\documentclass[aps,prc,superscriptaddress,showpacs,nofootinbib,twocolumn] 
{revtex4-1}

\usepackage{graphicx,amssymb,amsmath,gensymb,aas_macros,mathtools,xcolor}

\usepackage{dcolumn}
\newcolumntype{d}[1]{D{.}{.}{#1}}

\usepackage[normalem]{ulem}

\newcommand{\unit}[1]{\ensuremath{\,\mathrm{#1}}}

\newcommand{\etal}{\textit{et al.} }
\newcommand{\etaln}{\textit{et al.}}
\newcommand{\ie}{\textit{i.e.}, }
\newcommand{\eg}{\textit{e.g.}, }
\newcommand{\rLDP}{\ensuremath{\mathrm{LDP}}}
\newcommand{\LS}{\textit{LS} }
\newcommand{\LSn}{\textit{LS}}

\newcommand{\rsat}{{\rm sat}}
\newcommand{\rsym}{{\rm sym}}

\newcommand{\rmax}{{\rm max}}
\newcommand{\rbulk}{{\rm bulk}}
\newcommand{\tr}{{\rm tr}}

\begin{document}

% \preprint{APS/123-QED}

\title{The APR equation of state for simulations of supernovae, neutron stars and \\
binary mergers}% Force line breaks with \\

\author{A.~S. Schneider}
\email{andre.schneider@astro.su.se}
\affiliation{Department of Astronomy and the Oskar Klein Centre, Stockholm University, AlbaNova, SE-106 91 Stockholm, Sweden}
\affiliation{TAPIR, Walter Burke Institute for Theoretical Physics, MC 350-17, 
California Institute of Technology, Pasadena, CA 91125, USA}

\author{C. Constantinou}
\email{cconstaa5@kent.edu }
\affiliation{Department of Physics, Kent State University, Kent, OH 
442242}
\affiliation{Department of Physics and Astronomy, Ohio University, Athens,  
OH 45701}

\author{B. Muccioli}
\email{brian.muccioli@gmail.com}
\affiliation{BAE Systems Inc., 21 Continental Blvd, Merrimack, NH 03054}
\affiliation{Department of Physics and Astronomy, Ohio University, Athens, OH 
45701}

\author{M. Prakash}
\email{prakash@ohio.edu}
\affiliation{Department of Physics and Astronomy, Ohio University, Athens, OH 
45701}

\date{\today}% It is always \today, today,
             % but any date may be explicitly specified

\begin{abstract}

 Differences in the equation of state (EOS) of dense matter translate into 
 differences in astrophysical simulations and their multi-messenger signatures. 
Thus, extending the number of EOSs for astrophysical simulations allows us to probe the effect of different aspects of the EOS in astrophysical phenomena. 
In this work, we construct the EOS of hot and dense matter based on the Akmal, Pandharipande, and Ravenhall (APR) model and thereby extend the open-source SROEOS code which computes EOSs of hot dense matter for Skyrme-type parametrizations of the nuclear forces. Unlike Skrme-type models, 
in which parameters of the interaction are fit to reproduce the energy density of nuclear matter and/or properties of heavy nuclei, the EOS of APR  is obtained from potentials resulting from fits to nucleon-nucleon scattering and properties of light nuclei.  
In addition, this EOS features a phase transition to a neutral pion condensate at supra-nuclear  densities.
We show that differences in the effective masses between EOSs  have consequences for the properties of nuclei in the sub-nuclear inhomogeneous phase of matter. We also test the new EOS of APR in spherically symmetric core-collapse of massive stars with $15M_\odot$ and $40M_\odot$, respectively. 
We find that the phase transition in the EOS of APR  speeds up the collapse of the star. 
However, this phase transition does not generate a second shock wave or another neutrino burst as reported for the hadron-to-quark phase transition. The reason for this difference is that the onset of the phase transition in the EOS of APR  occurs at larger densities than for the quark-to-hadron transition employed earlier which results in a significantly smaller softening of the high density EOS. \\

\noindent Keywords: Supernova matter, potential models, thermal effects.

\end{abstract}

% PACS numbers
% 26.30.-k Nucleosynthesis in novae, supernovae, and other explosive 
% environments
% 26.60.-c Nuclear matter aspects of neutron stars
% 26.60.Dd Neutron stars cores
% 97.60.Jd Neutron stars (Late stage evolution)
% 26.50.+x Nuclear physics aspects of Supernovae
% 21.65.Mn Equations of State of nuclear matter
% 26.60.Kp Neutron stars equations of state

\pacs{21.65.Mn,26.50.+x,26.60.Kp}

\maketitle

% \tableofcontents

\section{Introduction}

Extreme conditions of temperatures, densities, and isospin asymmetries (excess of neutrons over protons) are found in various places across the Universe. 
Matter may be compressed beyond several times nuclear saturation density, heated up to dozens or even hundreds of MeV, and driven to highly neutron rich conditions by nuclear reactions
inside neutron stars (NSs), during compact object mergers as well as in core-collapse supernovae events, which lead to the formation of proto-NSs and black holes. 
A complete comprehension of these astrophysical environments and phenomena depends on our ability to understand 
 the phases of matter and its  equation of state (EOS) over a wide range of conditions. 
As some of these conditions are not accessible to laboratory experiments, knowledge must be deduced from a combination of theoretical and computational efforts and astronomical observations. 

Recently, the extent to which we can probe into hot and dense matter has been extended significantly by the detection of gravitational (GW) waves in the NS merger event GW170817 \cite{abbott:17a}. 
The subsequent observation of the same event in the electromagnetic spectrum \cite{abbott:17b} has 
shed much light on, \eg  synthesis  of heavy elements through rapid capture of neutrons, and the origin of some gamma-ray bursts, {\it cf.} Refs. \cite{soumi:18, most:18}. 
From future events, such as galactic core-collapse supernovae \cite{gossan:16}, we expect that combined observations of gravitational waves (GWs), electromagnetic (EM) signals, and neutrinos will further enhance  our understanding of the equation of state (EOS) of dense matter \cite{richers:17a, morozova:18}.

Despite ongoing progress, there are  many uncertainties in the EOS of dense matter which prevents accurate prediction of outcomes for astrophysical phenomena.  The foremost question is  what is the final state of core-collapse supernovae, and of NS mergers and their GW, neutrino, and EM signals \cite{hempel:12, morozova:18}? 
Many different approaches are used to study the EOS of dense matter. 
A recent review of EOSs used in studies of supernovae and compact stars is presented by Oertel \etal in Ref. \cite{oertel:17}. 
EOSs are usually provided to the astrophysical community in a tabular form that covers a wide range of densities, temperatures, and proton fractions. 
To construct these EOS tables, one first choses the degrees of freedom in the various phases to be considered. 
For simplicity, we choose to work solely with nucleons, nuclei, electrons, positrons, and photons in  this work. 
Extensions to include muons and anti-muons \cite{bollig:17}, hyperons \cite{banik:14}, and efforts to include quarks 
\cite{sagert:09,heinimann:16} also exist. 
We consider charge neutral matter in which the number density of electrons matches that of protons and positrons.  
%and that the pressure is large enough that 
Leptons and photons  are approximated as ideal relativistic gases and, thus, their EOSs decouple from the nuclear part. 
This procedure is commonly adopted in computations of dense matter EOSs. 

%\prak{Here!}

%To compute an EOS many other approximations are made to reflect the desired degrees of freedom of the system, our limited knowledge of nuclear interactions and of the strong force. 
%Computational constraints considerations also have to be accounted for. 
%For example, while some groups adopt 

In the construction of EOS tables, both 
non-relativistic potential model \cite{lattimer:91, schneider:17} and 
 %others work with fully 
 relativistic field-theoretical  \cite{shen:98a, shen:98b, shen:10b, hempel:10, hempel:12, steiner:13, banik:14, furusawa:11, furusawa:13, furusawa:17a, togashi:17} approaches have been employed. 
% \andre{Any others?} 
%There are also differences on the approach to compute the 
Differences also exist in the determination of inter-particle interactions in both approaches. In some cases, free space nucleon-nucleon interactions have guided the in-medium interactions, whereas in some others parameters of the chosen model are calibrated to fit empirical bulk nuclear matter properties.  
%, whether they are based on realistic potentials \cite{} or make use of effective fields \cite{}. 
Variations in the treatment of the sub-nuclear inhomogeneous phase, where light and heavy nuclei, pasta-like configurations, a gas of nucleons, electrons, and photons co-exist also exist. 
%, may be treated 
In the single nucleus approximation (SNA) \cite{lattimer:91, shen:98a, shen:98b, schneider:17},  a single representative nucleus describes the average thermodynamics of a nuclear ensemble.  An ensemble of nuclei in nuclear statistical equilibrium (NSE) \cite{shen:10a, shen:10b, shen:11, hempel:10, hempel:12, steiner:13, furusawa:11, furusawa:13, furusawa:17a, furusawa:17b, togashi:17, lalit:18} is used at very low densities when inter-nuclear interactions can be deemed small. Fully coupled reaction networks that change from dozens to a few thousand nuclear species have also been used \cite{lippuner:17,mosta:18,goni:18}. 
Generally, neutrinos and anti-neutrinos are not included in the EOS  because  simulations of supernovae and mergers of binary neutron stars treat neutrino transport separately from the EOS by incorporating all relevant neutrino scattering and absorption processes.  The time dependence of their properties is automatically included in the neutrino transport scheme coupled with hydrodynamics.  In proto-neutron star evolution, however, effects of neutrinos and antineutrinos are included in the  EOS (as free Fermion gases) as neutrino transport is treated in the diffusion regime.

%Finally, neutrino physics also has to be considered and many choices related to neutrino matter interactions also have to be considered \cite{}. 

%Amongst the many EOSs available one of 
The %most 
widely used EOS of Lattimer and Swesty (LS) \cite{lattimer:91} is based on the Lattimer, Lamb, Pethick, Ravenhall (LLPR) compressible liquid droplet model of nuclei \cite{lattimer:85}. 
%In the LS EOS, %the nuclear potential and 
Here, the mean-field interactions between nucleons are modeled using a Skyrme-type parametrization of the nuclear forces. 
%The LS EOS itself was modeled after the Lattimer, Lamb, Pethick, Ravenhall (LLPR) compressible liquid droplet model of nuclei \cite{lattimer:85}. 
%Nucleon properties take into account effects of their interactions and degeneracy, while 
The composition of heavy nuclei are determined 
%obtained treated 
in the SNA,  whereas light nuclei are represented  by alpha-particles  treated in the excluded volume approach. 
%nuclear deformations and 
The phase transition to the nuclear pasta phase considers various configurations that can exist due to competition between surface and coulomb effects. 
Although the SNA 
%is accurate enough to 
adequately describes the thermodynamics of the system \cite{burrows:84}, a full ensemble of nuclei is required to properly account for neutrino-matter interactions that are sensitive to the mass, charge numbers and abundances of the various nuclei present in addition to the most probable one. Extensions to include multiple nuclei in the NSE approach can be found in Refs. 
%extensions to transition to a NSE approach are available 
\cite{shen:10b, schneider:17, furusawa:17b, grams:18}.
% as neutrino-matter interactions are sensitive to the specific sizes and abundances of nuclei. 
At the time of the publication of the LS EOS, the bulk incompressibility $K_\rsat$ of nuclear matter was poorly constrained; thus, three different parametrizations of the EOS with $K_\rsat=180,\,220$ and  $375\unit{MeV}$ were made available. 
%Now we know that 
Subsequent studies have determined that
$K_\rsat\simeq230\pm20\unit{MeV}$ \cite{khan:12,margueron:18a} prompting most astrophysical studies to use the EOS with $K_\rsat=220\unit{MeV}$ (often referred to as LS220).
%Nevertheless, 
However, recent studies have  shown that the LS220  does not obey current nuclear physics constraints that correlate the symmetry energy at saturation density $J$ and its slope $L$ \cite{tews:17}.

Recently, Schneider \etal \cite{schneider:17} published an open-source code, SROEOS, which extends the LS approach in  many ways. 
The improvements made included 
(1) extra terms in the Skyrme parametrization of the nuclear force used by LS so as to fit results of more microscopic calculations, 
(2) a self-consistent treatment to determine the mass and charge numbers of heavy nuclei, and 
(3) the ability to compute the nuclear surface tension at finite temperature for the chosen Skyrme parametrizations. 
Additionally, density-dependent nucleon masses which control thermal effects in important ways were also included in their code. 
Although effects of density-dependent  effective masses were considered in the work of LS, it was not implemented in their open-source code.

The primary objective of this work is to construct an 
EOS for astrophysical simulations based on the potential model EOS of Akmal, Pandharipande, and Ravenhall (APR) \cite{akmal:98}.  At $T=0$, the  EOS of APR is fit to reproduce the variational calculations of Akmal and Pandharipande (AP) \cite{akmal:97} for symmetric nuclear matter (SNM) and pure neutron matter (PNM).  The nuclear interactions in these calculations are based on (1) the Argonne $v_{18}$ two-nucleon interaction \cite{wiringa:95} fit to nucleon phase shift data,  (2) the Urbana IX three-nucleon interaction that reproduces properties of light nuclei \cite{carlson:83, pudliner:95}, and 
(3) a relativistic boost interaction $\delta v$ \cite{forest:95, akmal:97, akmal:98}.  The EOS of APR reproduces the accepted values of empirical SNM properties such as the binding energy at the correct saturation density and incompressibility as well as the  
symmetry energy and its slope at the SNM  saturation density.  A characteristic feature of the EOSs of AP and APR is the phase transition to a neutral pion condensate at supra-nuclear densities. Although this induces softening at high densities, 
%when APR is extrapolated to intermediate proton fractions between SNM and PNM, 
the EOS predicts cold beta-equilibrated NS masses and radii that are in agreement with current observations \cite{antoniadis:13, fonseca:16, most:18, nattila:16, soumi:18}. 

%incompressibility, as well as fits to heavy-ion collision data of Danielewicz \etal \cite{danielewicz:02}. 

 Constantinou \etal \cite{constantinou:14} have calculated the thermal properties of the bulk homogeneous phase of supernova matter based on the EOS of APR.  However, properties of the sub-nuclear inhomogeneous phases based on the EOS of APR have not been investigated yet so that a full EOS based on the APR model is not yet available for use in astrophysical applications.  In this work, we take advantage of the structure of the SROEOS code to include inhomogeneous phases of sub-nuclear density matter using the EOS of APR within the LS formalism. 
 
The inhomogeneous phase has been incorporated into EOS models using techniques of differing complexity. 
The work of Negele and Vautherin \cite{negele:73} employed Hartree-Fock calculations for a single nucleus distributed in unit cells at zero-temperature. 
Bonche and  Vautherin \cite{bonche:81}, and, later Wolff \cite{wolff:83} extended this type of approach to finite temperatures. 
Alternately, a Thomas-Fermi calculation in which the nuclear wave functions are solved after appropriate approximations was undertaken in the works by \cite{marcos:82, ogasawara:83, bancel:84, shen:98b} (note this list is representative not exhaustive). 
These approaches treat nuclei in a realistic manner but are computationally slow. 

In this work, as was the case in Ref. \cite{schneider:17}, we follow the Lattimer and Swesty prescription \cite{lattimer:91} who developed a simplified version of the earlier work by Lattimer \etal \cite{lattimer:85}. 
In these approaches, nuclei are treated using the finite temperature compressible liquid-drop model which yields close agreement with results of more microscopic approaches. 
This approach is significantly faster than the previous approaches as it yields a system of equilibrium equations which is readily solved. 
It also utilizes the SNA in which the system is considered to consist of a single type of heavy nucleus plus alpha particles representing light nuclei. 
In principle different types of light nuclei should be considered (e.g., deuterons, tritons etc.) but these nuclei have significantly smaller binding energies than the alpha particle and thus, to leading order do not contribute to the thermodynamics of the system. 
Furthermore, it was shown in Ref. \cite{burrows:84} that the SNA gives an adequate representation of the thermodynamics of the system.  
However, in applications involving neutrino-nucleus, electron-nucleus scattering and capture processes, use of the full ensemble of nuclei is warranted. Several improvements to this first stage of our EOS calculation to be undertaken in later works will be noted in the concluding section.

This paper is organized as follows. 
In Sec \ref{sec:apr_model}, we review the bulk matter EOS of APR and discuss its main differences compared to the Skyrme EOSs. This is followed by a description of how we determine the nuclear surface contributions using the EOS of APR. 
Results for the sub- and supra-nuclear phases of stellar matter are presented in Sec. \ref{sec:dense} beginning with discussions of cold neutron star properties and nucleon effective masses. Thereafter, an in depth discussion of the finite temperature EOS of APR  along with detailed comparisons to two Skyrme-type models is provided.  
Temperature dependent nuclear surface tension and the composition of the system at sub-nuclear densities in the EOS of APR are also detailed in this section.  The EOS of APR is then used to simulate spherically symmetric collapse of massive stars in Sec. \ref{sec:ccsn}. Our conclusions are in Sec. \ref{sec:conclusions}. 
Appendices \ref{app:free} through \ref {app:B} contain formulas that are helpful in constructing the full EOS. 
The open-source APR EOS code is available at \url{https://bitbucket.org/andschn/sroeos/}.

\section{Equation of State Models}
\label{sec:apr_model}

The goal of this work is to present an equation of state (EOS) based on the  
potential model of Akmal, Pandharipande, and Ravenhall (APR) \cite{akmal:98}. 
The methodology used is similar to that used for the SRO EOS of Schneider \etal 
(SRO) \cite{schneider:17}, which was  based on the model of Lattimer and Swesty 
(LS) \cite{lattimer:91}. 
In these models, the nuclear EOS is decoupled from the EOS of leptons and 
photons, the later two forming  background uniform gases. The nuclear part 
takes into account nucleons, protons and neutrons, and alpha particles. 
Nucleons are free to cluster and form massive nuclei if the conditions are 
favorable.
The system is assumed to be charge neutral and in thermal equilibrium. 
Alpha particles are treated via an excluded volume (EV) approach so that their 
mass fraction vanishes at densities above $n\simeq0.1\unit{fm}^{-3}$. 
Recently, Lalit \etal have extended the EV model to include other light 
clusters ($^2$H, $^3$H, and $^3$He) and discussed the limitations of such models 
\cite{lalit:18}. 
These upgrades will be taken up in a future study.

If both density and temperature of the system are low enough, 
nucleon number density $n\lesssim0.1\unit{fm}^{-3}$ and temperature 
$T\lesssim1-16\unit{MeV}$,  
the nucleons can separate into a dense phase (heavy nuclei) and a dilute phase with 
nucleons and light nuclear clusters represented by  alpha particles here. 
The total free energy of the system is the sum of free energies of its 
individual components:
\begin{equation}\label{eq:F}
 F=F_o+F_\alpha+F_h+F_e+F_\gamma.
\end{equation}
Above, $F_o$, $F_\alpha$, $F_h$, $F_e$ and $F_\gamma$ are, respectively, the 
free energy density of the nucleons outside heavy nuclei, alpha particles, heavy 
nuclei, leptons, and photons. 
Leptons and photons are treated as relativistic gases of appropriate 
degeneracy following the EOS of Timmes \& Arnett \cite{timmes:99}.
As in LS and SRO, we determine the composition of the system by minimizing its 
free energy for a given baryon density $n$, temperature $T$, and proton fraction 
$y$.

Heavy nuclei are treated in the single nucleus approximation (SNA) and their 
bulk interiors considered to have a uniform density. 
The treatment of nuclear surface is discussed in Sec. \ref{ssec:apr_surface} 
below. 
The free energy density $F_i$ of nucleons in the bulk (inside) of heavy nuclei 
is treated with the same model as nucleons in the dilute gas around heavy 
nuclei. 
Other contributions to the free energy density $F_h$ of heavy nuclei are the 
surface, $F_S$, Coulomb, $F_C$, and translational,  $F_T$ terms, \ie 
\begin{equation}\label{eq:Fh}
 F_h=F_i+F_S+F_C+F_T.
\end{equation}
A refined model has been developed by Gramms \etal to include multiple nuclear 
species and effects of nuclear shell structure and realistic nuclear mass tables 
\cite{grams:18}.
Such improvements are not implemented in this work, but will be taken up in 
future studies as neutrino transport near the neutrino-sphere can be sensitive 
to nuclear composition \cite{yoshida:08,hempel:10,balasi:15,nakazato:18}.

%\andre{
A full description of the terms in Eqs. \eqref{eq:F} and \eqref{eq:Fh},  
and details of how to compute the thermodynamical properties of the nucleon system 
are given in the Appendices. 
In the remainder of this section, we describe differences between the APR and Skyrme  models and 
the computation of the surface properties of heavy nuclei.

\subsection{Bulk Matter}
\label{ssec:bulk}

We consider a general Hamiltonian density for bulk nucleonic matter of the form
\begin{equation}\label{eq:Hamiltonian}
\mathcal{H}(n,y,T)=\sum_t\frac{\hbar^2}{2m_t^\star(n,y)}\tau_t(n,y,T) + \mathcal{U}(n,y) \,, 
\end{equation}
where $n=n_n+n_p$ is the baryon density, with $n_n$ ($n_p$) denoting the neutron 
(proton) density, $y=n_n/n$ the proton fraction, $T$ the temperature of the 
system, and $t$  the nucleon isospin ($t=n$ or $p$). 
In Eq. (\ref{eq:Hamiltonian}), the effective masses $m_t^\star$ and nuclear potential $\mathcal{U}$ depend solely on the nucleon densities. 
In Skyrme-type models, the effective mass and nuclear potential are parametrized to reproduce properties of bulk nuclear matter and/or finite nuclei \cite{dutra:12, margueron:18a}. 
The APR model  Hamiltonian density is a parametric fit to the microscopic model calculations of Akmal and  Pandharipande (AP) \cite{akmal:97}. 
In the AP model, nucleon-nucleon interactions are modeled by the Argonne V18 potential \cite{wiringa:95}, the Urbana UIX three-body potential \cite{carlson:83, pudliner:95}, and a relativistic boost potential $\delta v$ \cite{forest:95, akmal:97, akmal:98}. 
As we show below, the density dependence of both the effective masses $m_t$ and nuclear potential $\mathcal{U}$ are  more complex for APR-type models than for the Skyrme-type ones. 
Note that neither APR nor Skyrme-type models have temperature dependent nucleon effective masses as in 
non relativistic EOSs based on finite-range forces \cite{} and 
relativistic EOSs \cite{hempel:12, steiner:13, furusawa:17a}.

From now on, except where explicitly needed, we omit the dependences of functions on $n$, $y$, and $T$. 
The effective masses $m_t^\star$ are defined through 
\begin{equation}\label{eq:meff}
\frac{\hbar^2}{2m_t^\star} = \frac{\hbar^2}{2m_t} + \mathcal{M}_{t}(n,y) \,,
\end{equation}
where $m_t$ are the vacuum nucleon masses and $\mathcal{M}_{t}$ are functions of the nucleonic densities. Note that for any function $F\equiv F(n,y)=F(n_n,n_p)$. 
The nucleon number densities $n_t$ and kinetic energy densities $\tau_t$  are 
\begin{align}
n_t & = \frac{1}{2\pi^2}\left(\frac{2m_t^\star T}{\hbar^2}\right)^{3/2}\mathcal{F}_{1/2} (\eta_t) 
\label{eq:nt} \\
\tau_t & = \frac{1}{2\pi^2}\left(\frac{2m_t^\star T}{\hbar^2}\right)^{5/2}\mathcal{F}_{3/2} (\eta_t) \,,
\label{eq:taut}
\end{align}
where the Fermi integrals are given by
\begin{equation}
\mathcal{F}_k(\eta)=\int\frac{u^k du}{1+\exp(u-\eta)} \,.
\end{equation} 
The degeneracy parameters $\eta_t$ are related to the chemical potentials $\mu_t$ through 
\begin{equation}\label{eq:etat}
\eta_t=\frac{\mu_t-\mathcal{V}_t}{T} \,,
\end{equation}
where the interaction potentials $\mathcal{V}_t$ are obtained from the functional derivatives
\begin{equation}
\mathcal{V}_t\equiv\left.\frac{\delta\mathcal{H}}{\delta n_t}\right|_{n_{-t},\tau_{\pm t}} \,.
\end{equation}
We note that the temperature dependence of the system is fully contained in the nucleon kinetic density terms $\tau_t$. 
Differences in the treatment of bulk matter for APR and Skyrme-type models appear only in the forms of the functions 
 $\mathcal{U}$, Eq. \eqref{eq:Hamiltonian} and 
$\mathcal{M}_t$, Eq. \eqref{eq:meff}.

\subsubsection{The EOS of APR}
\label{sssec:apr}

In the APR model, the interaction potential is parametrized by 
\begin{equation}\label{eq:uapr}
 \mathcal{U}(n,y) = g_1(n)\left[1-\delta^2(y)\right] + g_2(n)\delta^2(y),
\end{equation}
where $g_1(n)$ and $g_2(n)$ are functions of the baryon density $n$ and the isospin asymmetry $\delta(y) = (1 - 2y)$. 
The model exhibits a transition from a low density phase (LDP), where the only hadrons present are nucleons, to a high density phase (HDP), where a neutral pion condensate appears.
Owing to this transition, the potential energy density functions $g_1$ and $g_2$ have different forms below and above the transition density $n_\tr(y)$. 

For the low density phase (LDP), \ie for densities below those for which a neutral pion condensate forms, 
\begin{equation}\label{eq:uaprl}
\mathcal{U}\rightarrow \mathcal{U}_L= g_{1L}\left[1-\delta^2\right] + g_{2L}\delta^2 \,,
\end{equation}
where the functions $g_{iL}$ are parametrized by 
\begin{subequations}\label{eq:ldp}
\begin{align}
 -\frac{g_{1L}}{n^2} &= 
\left[p_1+p_2n+p_6n^2+(p_{10}+p_{11}n)e^{-p_9^2n^2}\right] \\
 -\frac{g_{2L}}{n^2} &= 
\left[\frac{p_{12}}{n}+p_7+p_8n+p_{13}e^{-p_9^2n^2}\right].
\end{align} 
\end{subequations}
In the high density phase (HDP) $\mathcal{U}\rightarrow\mathcal{U}_{H}$, where $g_{iH}$ are related to $g_{iL}$ by 
\begin{subequations}\label{eq:hdp}
\begin{align}
 \frac{g_{1L}-g_{1H}}{n^2} &= 
\left[p_{17}(n-p_{19})+p_{21}(n-p_{19})^2\right]e^{p_{18}(n-p_{19})} \\
 \frac{g_{2L}-g_{2H}}{n^2} &= 
\left[p_{15}(n-p_{20})+p_{14}(n-p_{20})^2\right]e^{p_{16}(n-p_{20})}.
\end{align}
\end{subequations}

Besides the interaction potential density, the Hamiltonian density is a function of the effective masses $m^\star_t$
which depend on the functions $\mathcal{M}_t(n,y)$, with $t=n$ or $p$, 
%of the baryon density $n$ and proton fraction $x$, 
[see Eq. (\eqref{eq:meff})]. 
In the APR model,
\begin{equation}\label{eq:meffapr}
 \mathcal{M}_{t}(n,y) = (p_3 n + p_5 n_t)e^{-p_4n}, 
\end{equation}
where neutron and proton densities are $n_n=n(1-y)$ and $n_p=ny$.

The parameters $p_i$ ($i=1,\hdots,21$) fully define the APR parametrization of the nuclear Hamiltonian density \cite{akmal:98}.  
These parameters are presented in Table \ref{tab:apr}.

\begin{table}[h]
\caption{\label{tab:apr} Parameters $p_i$ of the EOS of APR \cite{akmal:98}.
Values for $i=1,\hdots,13$ are for the LDP, whereas $i=14,\hdots,21$ refers to the HDP.} 
\begin{ruledtabular}
\begin{tabular}{c D{.}{.}{4.3} r c D{.}{.}{4.3} r}
\multicolumn{1}{c}{$p_i$} &
\multicolumn{1}{c}{Value} &
\multicolumn{1}{c}{units} &
\multicolumn{1}{c}{$p_i$} &
\multicolumn{1}{c}{Value} &
\multicolumn{1}{c}{Units} \\
\hline
$p_1$ &  337.2   & MeV\,fm$^3$ & $p_{14}$ &    0.    & MeV\,fm$^6$ \\
$p_2$ & -382.0   & MeV\,fm$^6$ & $p_{15}$ &  287.0   & MeV\,fm$^3$ \\
$p_3$ &   89.8   & MeV\,fm$^5$ & $p_{16}$ &   -1.54  &      fm$^3$ \\
$p_4$ &    0.457 &      fm$^3$ & $p_{17}$ &  175.0   & MeV\,fm$^3$ \\
$p_5$ &  -59.0   & MeV\,fm$^5$ & $p_{18}$ &   -1.45  &      fm$^3$ \\
$p_6$ &  -19.1   & MeV\,fm$^9$ & $p_{19}$ &    0.32  &      fm$^{-3}$ \\
$p_7$ &  214.6   & MeV\,fm$^3$ & $p_{20}$ &    0.195 &      fm$^{-3}$ \\
$p_8$ & -384.0   & MeV\,fm$^6$ & $p_{21}$ &    0.    & MeV\,fm$^6$ \\
$p_{ 9}$ &   6.4  &      fm$^3$ \\
$p_{10}$ &  69.0  & MeV\,fm$^3$ \\
$p_{11}$ & -33.0  & MeV\,fm$^6$ \\
$p_{12}$ &   0.35 & MeV  \\
$p_{13}$ &   0.   & MeV\,fm$^3$ \\
\end{tabular}
\end{ruledtabular}
\end{table}

The potentials in the LDP and HDP are temperature independent. 
Thus, the transition from one phase to the other occurs when their energies are the same. 
As noted by Constantinou \etal \cite{constantinou:14}, the transition density is well approximated by
\begin{align}\label{eq:ntr}
 n_\tr(y)=& 0.1956    + 0.3389y   + 0.2918y^2 \nonumber\\
          &-1.2614y^3 + 0.6307y^4.
\end{align} 
A mixed-phase region is determined via a Maxwell construction following the details laid out in Sec. VI of Ref. \cite{constantinou:14}. 

%\prak{Here!}

\subsubsection{The Skyrme EOS}
\label{sssec:skyrme}

%As in the APR EOS case,  
The Skyrme Hamiltonian density also has the generic form of Eq. (\ref{eq:Hamiltonian}).
However, the functions $\mathcal{M}_t$ and $\mathcal{U}$ that define, respectively, the effective masses $m_t^\star$ and the interaction potential  $\mathcal{V}_t$ have density and proton fraction dependences that are simpler than those of the APR Hamiltonian density.
The effective mass takes the form 
\begin{equation}\label{eq:meffsk}
 \mathcal{M}_t=\alpha_1 n_t + \alpha_2 n_{-t} \,,
\end{equation}
where if $t=n$ then $-t=p$, and vice versa. The potential energy density $\mathcal{U}$ may be written in the form
\begin{equation}\label{eq:usk}
 \mathcal{U}(n,x)=\sum_{i=0}^{N}\left[a_i+4b_iy(1-y)\right]n^{\delta_i}.
\end{equation}
In Eqs. \eqref{eq:meffsk} and \eqref{eq:usk}, the parameters  $\alpha_1$, $\alpha_2$, $a_i$, $b_i$, and $\delta_i$ are specific to each Skyrme model. 
These parameters are related to the often employed Skyrme parameters $x_i$, $t_i$, and $\sigma_i$ for $i=0,\hdots,3$ by Eqs. (14a-g) in SRO.

Dutra \etal analyzed 240 Skyrme parametrizations available in the literature and found that only 16 of those fully agreed with 11 well determined nuclear matter constraints and few that did not match only one of the constraints \cite{dutra:12}. 
Nevertheless, the equation of state obtained for most of these 16 parametrizations is unable to  support neutron stars (NSs) as massive as the ones observed by Antoniadis \etaln, PSR J0348+0432 with $M=2.01\pm0.04$ \cite{antoniadis:13} or by Fonseca \etaln, PSR J1614-2230 with $M=1.93\pm0.02$ \cite{fonseca:16}. 
Amongst those parametrizations that satisfy both the nuclear physics constraints and the lower limit of a neutron star's maximum mass is the NRAPR parametrization. 
The coefficients of the NRAPR parametrization were computed by Steiner \etal to match as closely as possible the effective masses of the APR equation of state as well as the charge radii and binding energies of a few selected nuclei \cite{steiner:05}. 
However, it is impossible to completely reproduce the effective mass behavior of APR with a Skyrme-type parametrization due to the more complex behavior of the former; compare Eqs. \eqref{eq:meffapr} and \eqref{eq:meffsk}.

Besides the EOS of APR and its non-relativistic version NRAPR developed by Steiner \etaln, we develop another Skyrme EOS to fit APR and term it as SkAPR. 
In SkAPR, unlike NRAPR which is fit to reproduce the effective masses and properties of finite nuclei computed with APR, we compute the parameters $\alpha_1$, $\alpha_2$, $a_i$, $b_i$, and $\delta_i$ ($i=0,\hdots,3$) to reproduce (1) the empirical parameters of the APR EOS up to second order, see Eq. \eqref{eq:expansion} below, (2) the pressure of symmetric nuclear matter (SNM) and pure neutron matter (PNM) at $4n_\rsat$, (3) the effective mass of neutrons at saturation density for SNM,  $m_n^\star(n_\rsat,y=1/2)$, and (4) the splitting between neutron and proton effective masses at saturation density for PNM, $\Delta m^\star = m_n^\star(n_\rsat,0) - m_p^\star(n_\rsat,0)$. 

%\prak{Here!}

\subsubsection{Comparison of APR and Skyrme EOSs}
\label{sssec:comparison}

To very good approximation, 
the energy density $\epsilon_B(n,y)$ of isospin asymmetric matter can be expanded around the nuclear saturation density, 
$n_\rsat$, for symmetric nuclear matter, $y=1/2$, \ie
\begin{equation}\label{eq:expansion}
 \epsilon_B(n,y)=\epsilon_{\rm is}(x)+\delta^2\epsilon_{\rm iv}(x) \,,
\end{equation}
where $x=(n-n_\rsat)/(3n_\rsat)$ and $\delta=1-2y$ is the isospin asymmetry. 
The isoscalar (is) and isovector (iv) expansion terms are functions of the 
nuclear empirical parameters \cite{margueron:18a, piekarewicz:09}
\begin{subequations}\label{eq:emp}
\begin{align}
\label{eq:is}
 \epsilon_{\rm is}(x) &=\epsilon_\rsat+\frac{1}{2!}K_\rsat x^2
          +\frac{1}{3!}Q_\rsat x^3+\hdots\,,\\
\label{eq:iv}
 \epsilon_{\rm iv}(x) &=\epsilon_\rsym+L_\rsym x+\frac{1}{2!}K_\rsym x^2
\nonumber\\ &\quad+\frac{1}{3!}Q_\rsym x^3+\hdots\,,
\end{align}
\end{subequations}
shown here explicitly up to third order in $x$.   Terms involving $\delta^4$ and higher give very small contributions. 
Comparisons between the values of observables for the EOSs of APR, NRAPR, and SkAPR are shown in Table \ref{tab:obs}. 
A description of the methods used to compute  $\alpha_1$, $\alpha_2$, $a_i$, $b_i$, and $\delta_i$ ($i=0,\hdots,3$) will be discussed in a forthcoming manuscript by Schneider \etal \cite{schneider:18}.

\begin{table*}[htb]
\caption{\label{tab:obs} Characteristic properties of the EOSs of APR \cite{akmal:98}, NRAPR \cite{steiner:05}, and SkAPR. For a description of the properties listed, see text. 
Values quoted as ``Experimental''  are averages over experimental and theoretical values drawn from many sources, and compiled by Margueron \etal \cite{margueron:18a}.} 
\begin{ruledtabular}
\begin{tabular}{c D{.}{.}{4.3} D{.}{.}{4.3} D{.}{.}{4.3} D{+}{\,\pm\,}{5+5} c}
\multicolumn{1}{c}{Property} &
\multicolumn{1}{c}{APR} &
\multicolumn{1}{c}{SkAPR} &
\multicolumn{1}{c}{NRAPR} &
\multicolumn{1}{c}{Experimental} &
\multicolumn{1}{c}{Units} \\
\hline
$n_\rsat$           &    0.160 &    0.160 &   0.161  &  0.155+0.005 & fm$^{-3}$ \\ 
$\epsilon_\rsat$    & - 16.00  & - 16.00  & -15.85   &  -15.8+0.03  & MeV \\ 
$K_\rsat$           &  266.0   &  266.0   & 225.6    &    230+20    & MeV \\ 
$Q_\rsat$           & -1054.1  & -348.3   & -362.5   &    300+400   & MeV \\ 
$\epsilon_\rsym$    &   32.59  &   32.59  &  32.78   &     32+2     & MeV \\ 
$L_\rsym$           &   58.47  &   58.47  &  59.63   &     60+15    & MeV \\ 
$K_\rsym$           & -102.63  & -102.63  & -123.32  &   -100+100   & MeV \\ 
$Q_\rsym$           & 1216.8   &  420.02  &  311.6   &      0+400   & MeV \\ 
$P^{(4)}_{\rm SNM}$ &  133.2   &  133.2   & 125.0    &    100+ 50   & MeV\,fm$^{-3}$ \\ 
$P^{(4)}_{\rm PNM}$ &  167.43  &  167.43  & 127.8    &    160+ 80   & MeV\,fm$^{-3}$ \\ 
$m^\star$           &    0.698 &    0.698 &  0.694   &   0.75+0.10  & $m_n$ \\ 
$\Delta m^\star$    &    0.211 &    0.211 &  0.214   &   0.10+0.10  & $m_n$ \\ 
\end{tabular}
\end{ruledtabular}
\end{table*}

%\prak{Here!}

\subsection{The Nuclear Surface}
\label{ssec:apr_surface}

If the density and/or temperature of the system is low enough, nuclear matter  separates into a dense phase of nucleons (heavy nuclei) surrounded by a dilute gas of nucleons and alpha particles (in general, light nuclear clusters) in thermal equilibrium. 
The free energy of heavy nuclei has contributions from the bulk nucleons that form it as well as from surface, Coulomb, and translational terms. 
The bulk term is treated with the Hamiltonian density  in Eq. \eqref{eq:Hamiltonian}. 
Coulomb and translational terms are discussed in detail in Appendix \ref{app:heavy}.

As in Refs. \cite{schneider:17,lattimer:91,lim:12}, the surface free energy density is taken to be
\begin{equation}\label{eq:FS}
 F_S=\frac{3s(u)}{r}\sigma(y_i,T)\,.
\end{equation}
Above $s(u)=u(1-u)$ is a shape function that depends on the volume $u$ occupied by the heavy nuclei with generalized radius $r$ within the Wigner-Seitz cell. 
More details are discussed in Sec. II B of SRO \cite{schneider:17}, in Sec. 2.6 of \LS \cite{lattimer:91}, and are reviewed in Appendices \ref{app:surface} and \ref{app:heavy}.
The surface tension $\sigma$ (energy per unit area) is a function of the proton fraction $y_i$ of the bulk phase and the temperature $T$ of the system, and  is parametrized by \cite{lattimer:91}
\begin{equation}\label{eq:sigma}
 \sigma(y_i,T)=\sigma_s h\left(y_i,T\right)
 \frac{2\cdot2^{\lambda}+q}{y_i^{-\lambda}+q+(1-y_i)^{-\lambda}} \,,
\end{equation} 
where $\sigma_s\equiv\sigma(0.5,0)$. 
The function $h(y_i,T)$ contains the temperature dependence of the surface tension:
\begin{equation}\label{eq:h}
 h\left(y_i,T\right)=
 \begin{dcases}
    [1-({T}/{T_c(y_i)})^2]^p\,,& \mathrm{if\, } T\leq T_c(y_i)\,;\\
    0\,,              & \mathrm{otherwise}\quad.
\end{dcases}
\end{equation}
In Eqs. \eqref{eq:sigma} and \eqref{eq:h}, $\lambda$, $q$, and $p$ are parameters to be determined, while $T_c(y_i)$ is the critical temperature for which the dense and the dilute phases coexist. 
We fit $T_c(y_i)$ using the same polynomial form used in SRO, \ie 
\begin{equation}\label{eq:Tc}
 T_c(y) = T_{c0}(a_c+b_c\delta(y)^2+c_c\delta(y)^4+d_c\delta(y)^6)
\end{equation} 
where $T_{c0}\equiv T_{c}(y=0.5)$ is the critical temperature for
symmetric nuclear matter and $\delta(y)=1-2y$ is the neutron excess.

The bulk nucleons inside heavy nuclei are assumed to have density $n_i$ and proton fraction $y_i$ while the dilute gas has density $n_o\leq n_i$ and proton fraction $y_o$. 
The parameters $\lambda$, $q$, and $p$, are obtained as in Sec. II B of SRO \cite{schneider:17}, and shown in Appendix \ref{app:surface} for completeness.  
However, the Hamiltonian density for the EOS of APR has a different functional form than that of the Skyrme EOS and so does its gradient term. 
The gradient part of the Hamiltonian density is used to obtain the surface tension  $\sigma(y_i,T)$.
For semi-infinite nucleonic matter
\begin{align}\label{eq:HS}
 E_S(z)=\frac{1}{2}\bigg[&q_{nn}\left(\boldsymbol{\nabla}n_n\right)^2
 +q_{np}\boldsymbol{\nabla}n_n\cdot\boldsymbol{\nabla}n_p\nonumber\\
 &+q_{pn}\boldsymbol{\nabla}n_p\cdot\boldsymbol{\nabla}n_n
 +q_{pp}\left(\boldsymbol{\nabla}n_p\right)^2\bigg]\,,
\end{align}
For Skyrme-type parametrizations $q_{tt'}$ are computed from the Skyrme parameters $x_1$, $x_2$, $t_1$, and $t_2$, see Eqs. (27a-b) in SRO, and satisfy the relations $q_{nn}=q_{pp}$ and $q_{np}=q_{pn}$. 
In the APR model, however, we obtain, following Pethick \etal \cite{pethick:95} and Steiner \etal \cite{steiner:05}, 
\begin{subequations}\label{eq:qtt}
\begin{align}
q_{nn}&=-\tfrac{1}{4}e^{-p_4n}\left[6p_5+p_4(p_3-2p_5)(n_n+2n_p)\right],\\
q_{pp}&=-\tfrac{1}{4}e^{-p_4n}\left[6p_5+p_4(p_3-2p_5)(n_p+2n_n)\right],\\
q_{np}&=q_{pn}=\tfrac{1}{8}e^{-p_4n}\left[4(p_3-4p_5)-3p_4(p_3-2p_5)n\right] \,.
\end{align}
\end{subequations}
This implies that $q_{nn}=q_{pp}$ only for SNM, \ie for $y = 0.5$
\footnote{There is a straightforward connection between the nucleon effective 
masses $m_t^\star$ and the coefficients $q_{tt'}$  for APR and Skyrme
EOSs in the $p_4\rightarrow0$ limit. 
However, using Eq. \eqref{eq:qtt} this connection is possible if and 
only if the Skyrme parameters $x_1=x_2=0$ \cite{pethick:95}.
Assumptions about the form of $q_{tt'}$ coefficients for the EOS of APR
should be relaxed in future works.}. 
Apart from the form of coefficients $q_{tt'}$, the method to compute the 
parameters $\lambda$, $q$ and $p$ of the surface tension $\sigma(y_i,T)$ 
is the same for the APR and Skyrme-type EOSs. 
Details are discussed in that work and in Appendix \ref{app:surface} 
for completeness.

\section{Results for Sub- and Supra-Nuclear Phases}
\label{sec:dense}

For the EOSs listed in Table \ref{tab:obs}, our calculations for astrophysical applications are performed using the single nucleus approximation (SNA).
We consider two forms of the EOS of APR, namely, APR and APR$_{\rLDP}$. 
In the latter, we ignore the transition to the high density phase of the nuclear potential $\mathcal{U}$ and set $\mathcal{U}\rightarrow\mathcal{U}_L$ for all $n$, see Eqs. \eqref{eq:uapr} and \eqref{eq:uaprl}. 
In addition to these EOSs of APR, we use the SRO EOS code \cite{schneider:17} to compute two other EOS tables (for Skyrme-type models), namely, NRAPR \cite{steiner:05} and SkAPR. 
In SkAPR, the parameters of the Skyrme interaction in Eqs. \eqref{eq:meffsk} and \eqref{eq:usk} are chosen to reproduce the nuclear empirical parameters up to second order, Eqs. \eqref{eq:emp}, the nucleon effective mass at nuclear saturation density and its isospin splitting for pure neutron matter, as well as the pressure of symmetric nuclear matter (SNM) and pure neutron matter (PNM) at $4n_\rsat$.

\subsection{Equation of State at $T=0$}
\label{ssec:t=0}

\begin{figure}[!htb]
\centering
\includegraphics[trim=0 40 0 0, clip, width = 0.50\textwidth]{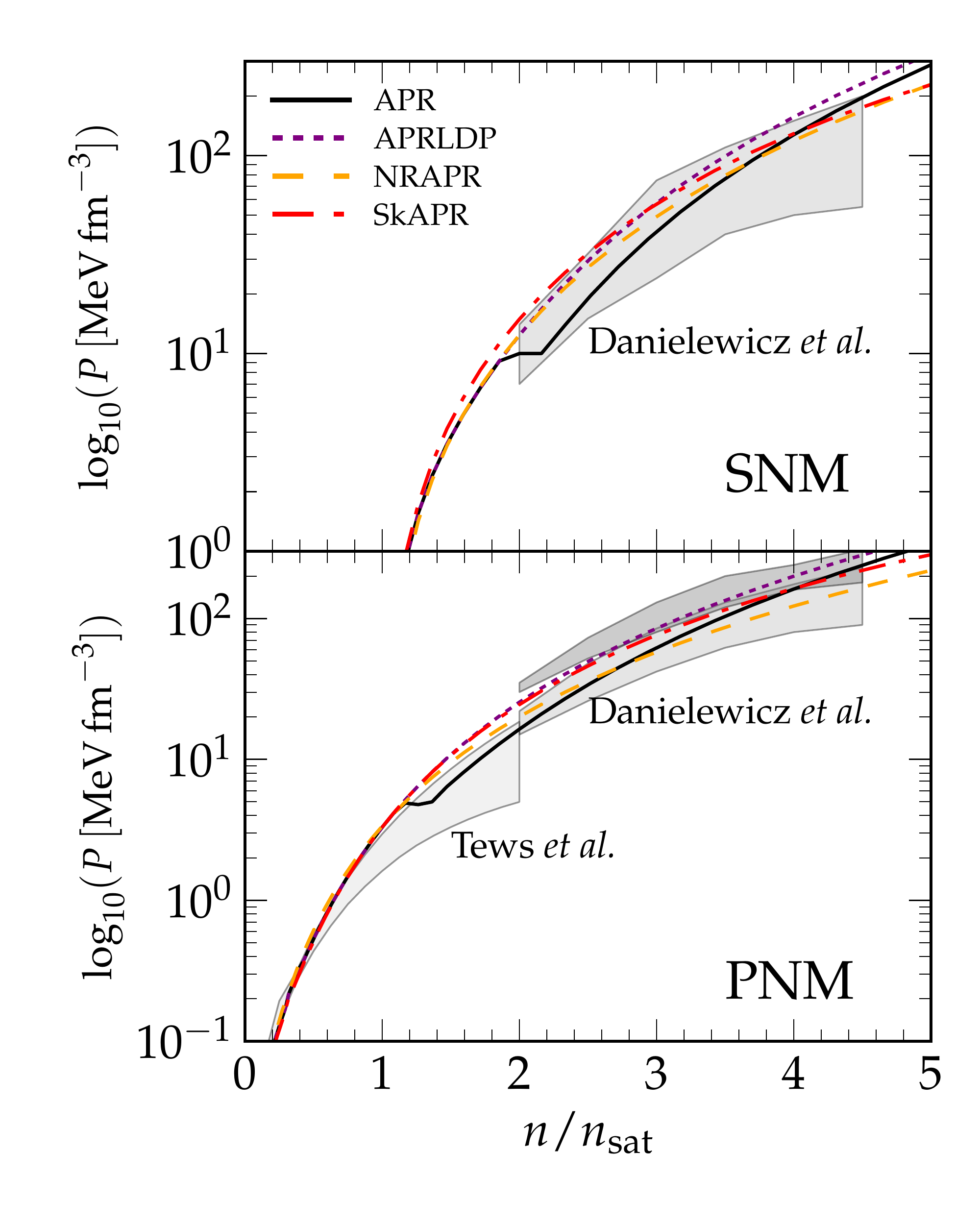}
\caption{\label{fig:pres} Pressure of SNM (top) and PNM (bottom) for the EOS's of APR, 
APR$_{\rLDP}$, NRAPR, and SkAPR. The SNM and PNM results are compared to the 
pressure of nuclear matter deduced from analysis of heavy ion collision experiments by
Danielewicz \etal \cite{danielewicz:02}. The PNM results are also compared to results from chiral 
effective field theory supplemented by piece-wise polynomials by Tews \etal \cite{tews:18}. }
\end{figure}

In Fig. \ref{fig:pres}, we show the pressures of SNM and PNM at zero temperature for each of the four EOSs: APR, APR$_{\rLDP}$, NRAPR, and SkAPR. 
The pressures as a function of density for all EOSs are mostly within the bands computed by Danielewicz \etal from analysis of collective flow in heavy ion collision experiments \cite{danielewicz:02} and the chiral effective theory results of Tews \etal \cite{tews:18}. Note that results from microscopic calculations from the latter source are limited to about $2n_{sat}$, but are extended beyond using piece-wise polynomials that preserve causality.  Quantitative differences between predictions of the different EOSs become apparent with progressively increasing density.

\begin{figure}[!htb]
\centering
\includegraphics[trim=0 0 0 0, clip, width = 0.50\textwidth]{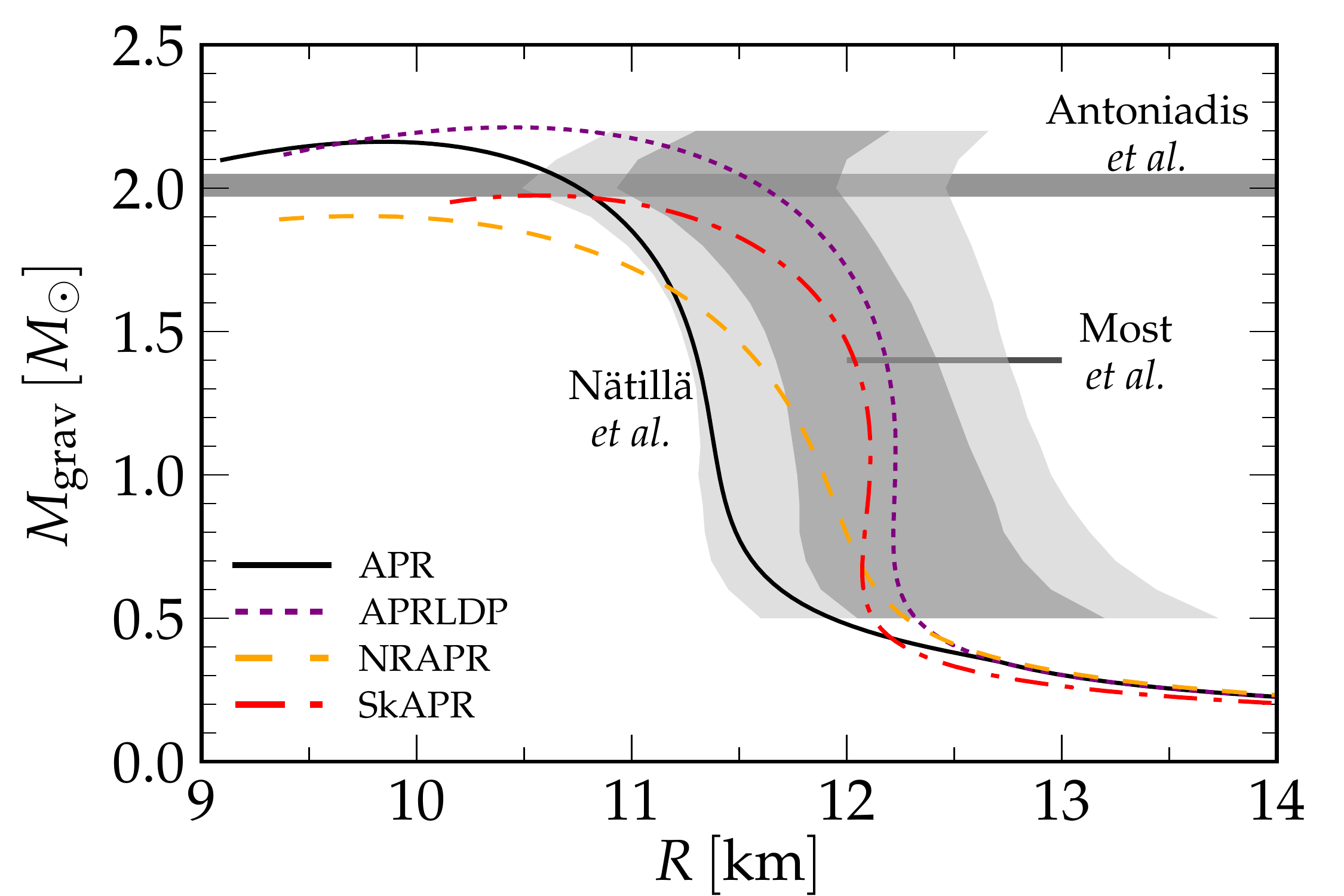}
\caption{\label{fig:mr} Mass radius relationships for the models of APR, APR$_{\rLDP}$, NRAPR, and SkAPR computed for charge neutral and temperature beta-equilibrated matter at zero temperature. 
Results are compared to the maximum NS mass observed by Antoniadis \etal \cite{antoniadis:13}, the mass radius relationship of N\"{a}ttil\"{a} \etal \cite{nattila:16}, and the radius of a $1.4M_\odot$ NS inferred by Most \etal \cite{most:18}.}
\end{figure}

\begin{table*}[htb]
\caption{\label{tab:NS} Properties of cold beta-equilibrated NSs with $M=1.4M_\odot$ and $M=M_\rmax$ for the four EOSs computed in this work, APR, APR$_\rLDP$, SkAPR, and NRAPR.  The compactness $\beta=(GM_\odot/c^2) (M/M_\odot)/R$. 
Conversion factors used were $1\unit{MeV\,fm}^{-3}=6.242\times10^{-34}\unit{erg\,g}^{-1}$ for the pressure and $1\unit{g\,cm}^{-3}=5.97\times10^{-16}\unit{fm}^{-3}$ for the density.} 
\begin{ruledtabular}
\begin{tabular}{c D{.}{.}{4.3} D{.}{.}{4.3} D{.}{.}{4.3} D{.}{.}{4.3} c}
\multicolumn{1}{c}{Property} &
\multicolumn{1}{c}{APR} &
\multicolumn{1}{c}{APR$_\rLDP$} &
\multicolumn{1}{c}{SkAPR} &
\multicolumn{1}{c}{NRAPR} &
\multicolumn{1}{c}{Units} \\
\hline
$R_{1.4}$      &  11.31  &  12.18  &  12.04  &  11.58  & km \\  
$\beta_{1.4}$ & 0.18 & 0.17 & 0.17 & 0.18 & \\
$n_{c,1.4}$    &   3.46  &   2.85  &   2.98  &   3.58  & $n_\rsat$      \\ 
$P_{c,1.4}$    &  96.2   &  67.9   &  72.1   &  94.2   & Mev\,fm$^{-3}$ \\  
$y_{c,1.4}$    &   0.089 &   0.110 &   0.094 &   0.079 &                \\ 
$R_{\rmax}$    &  10.41  &   9.88  &  10.59  &   9.77  & km \\  
$M_{\rmax}$    &   2.162 &   2.212 &  1.974  &   1.903 & $M_\odot$      \\
$\beta_{\rmax}$ & 0.31 & 0.33 & 0.28 & 0.29 & \\
$n_{c,\rmax}$  &   6.63  &   7.11  &   6.79  &   7.98  & $n_\rsat$      \\ 
$P_{c,\rmax}$  & 889     &1059     & 514     & 780     & MeV\,fm$^{-3}$ \\  
$y_{c,\rmax}$  &   0.160 &   0.129 &   0.146 &   0.039 &                \\ 
\end{tabular}
\end{ruledtabular}
\end{table*}

The mass radius relationship of cold non-rotating neutron stars (NSs) 
for each EOS is shown in Fig. \ref{fig:mr}. 
These relations are obtained solving the TOV equations for charge neutral and beta-equilibrated matter at zero temperature \cite{tolman:39}. 
For comparison the maximum NS mass observed to date, that of PSR J0348+0432 with $2.01\pm0.04 
M_\odot$ \cite{antoniadis:13} is also shown in this figure. 
A similar mass measurement, but for a different NS, PSR J1614-2230 with $M=1.93\pm0.02$ \cite{fonseca:16}, boosts our confidence that NSs with at least $2M_\odot$ exist in nature and, thus, any realistic EOS should reproduce this limit. 
While APR and APR$_{\rLDP}$ predict, respectively, maximum masses $M_{\rm max}=2.17M_\odot$ and $2.21 M_\odot$,  well above the  $2.01\pm0.04 M_\odot$ limit, SkAPR barely reaches the lower limit of the observation, $M_{\rm max}=1.97M_\odot$, and NRAPR is two standard deviations below the lower limit,  $M_{\rm max}=1.90M_\odot$.  
Properties of cold beta-equilibrated NSs are shown in Table \ref{tab:NS}. The compactness parameters $\beta=(GM_\odot/c^2) (M/M_\odot)/R$ are very nearly the same for both the $1.4M_\odot$ and maximum mass stars for all the EOSs listed  in this table.

NS radii are less constrained than their maximum masses. 
From the NS merger observation GW170817 \cite{abbott:17a, abbott:17b}, Most \etal predict that canonical NSs with mass $1.4M_\odot$ have radii in the range $12\unit{km} \leq R_{1.4}\leq 13.45 \unit{km}$. 
In contrast, De \etal constrain radii to be in the $8.9\unit{km} < \bar R < 13.2\unit{km}$ interval by analyzing of Love numbers from the observation of GW170817 \cite{soumi:18}. 
Although we show the constraint of Most \etal in our plot for comparison between EOSs, more observations are needed to confirm their result. 
Note that only APR$_{\rLDP}$ and SkAPR satisfy the constraint of Most \etal and predict, respectively, $R_{1.4}=12.2\unit{km}$ and $12.0\unit{km}$.
The APR and NRAPR EOSs, on the other hand, predict radii that are too small for a canonical NS \textit{when compared to the results of Most \etaln}, $R_{1.4}=11.2\unit{km}$ and $11.5\unit{km}$, respectively. 
However, all four EOSs are well within the bounds determined by De \etal \cite{soumi:18}.
Furthermore, except for the heaviest NSs in the SkAPR case, both APR$_{\rLDP}$ and SkAPR mass-radius relationships are within $1\sigma$ range of ``model A'' of N\"attil\"a \etal obtained from observations of x-ray bursts \cite{nattila:16} and also shown in our Fig. \ref{fig:mr}. 
APR (NRAPR) is within the $2\sigma$ range of the results of N\"attil\"a \etaln, except for the NSs above $2.1M_\odot$ ($1.7M_\odot$).

It is worthwhile to note here that combining electromagnetic \cite{abbott:17b} and gravitational wave information from the merger GW170817, Ref. \cite{margalit:17} provides constraints on the radius $R_{\rm ns}$ and maximum gravitational mass $M_{\rm max}^g$ of a neutron star:
\begin{eqnarray}
M_{\rm max}^g &\lesssim& 2.17 M_\odot \, \nonumber \\
R_{1.3} &\gtrsim& 3.1GM_{\rm max}^g \simeq 9.92~{\rm km} \,, 
\end{eqnarray}
where $R_{1.3}$ is the radius of a 1.3$M_\odot$ neutron star and its numerical value above corresponds to 
$M_{\rm max}^g=2.17~M_\odot$. 

\subsection{Effective Masses}
\label{ssec:meff}

\begin{figure}[!htb]
\centering
\includegraphics[trim=0 0 0 0, clip, width = 0.50\textwidth]{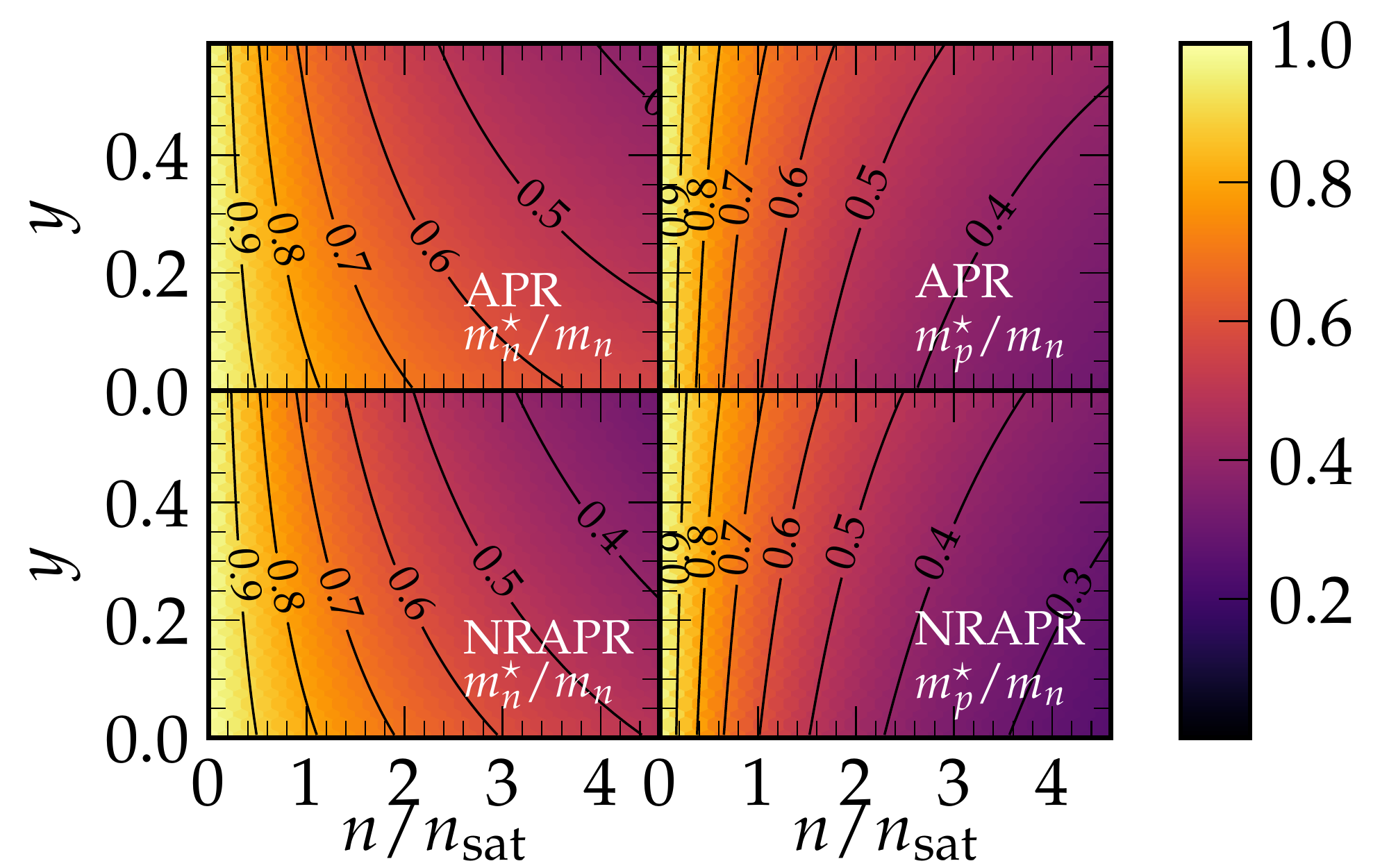}
\caption{\label{fig:meff} Neutron (left) and proton (right) effective masses $m_t$ normalized by the neutron vacuum mass $m_n$ as a function of density $n$ and proton fraction $y$ for the APR (top) and NRAPR (bottom) models.} 
\end{figure}

\begin{figure}[!htb]
\centering
\includegraphics[trim=0 0 0 0, clip, width = 0.50\textwidth]{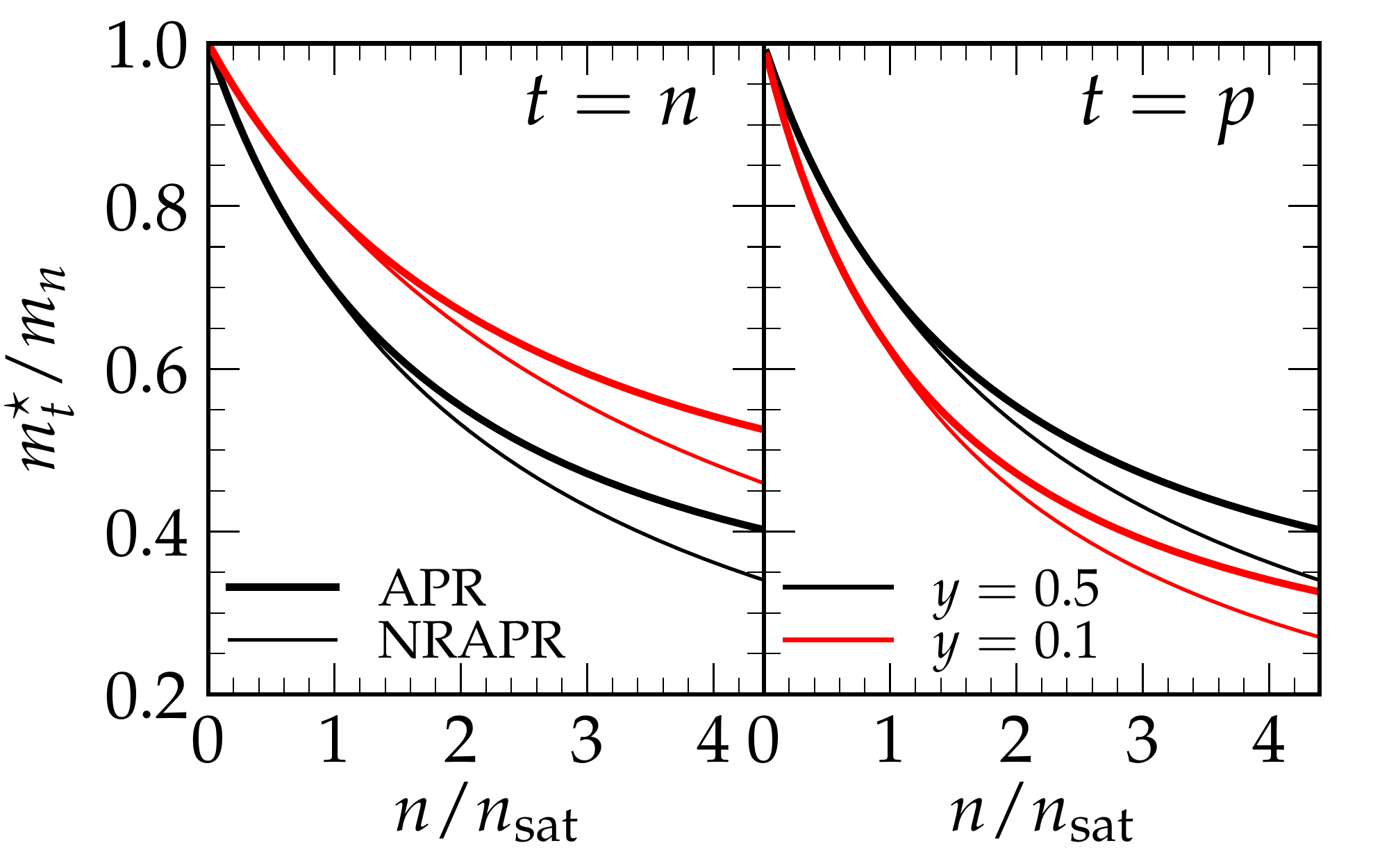}
\caption{\label{fig:meff1} Neutron (left) and proton (right) effective masses for the APR (thick) and NRAPR (thin) EOSs for proton fractions $y=0.10$ (red) and $0.50$ (black). } 
\end{figure}

Nucleon effective masses for the APR and NRAPR EOSs are compared in Figs. \ref{fig:meff} and \ref{fig:meff1}. 
Results for SkAPR are not shown as they are very similar to those of NRAPR in that $m^\star$ and $\Delta m^\star$ are nearly the same for the two models, see Tab. \ref{tab:obs}. 
The effective mass contributes directly to the thermal component of the EOS, see Eq. \eqref{eq:Hamiltonian}. 
Thus, differences in effective masses contribute to differences in the thermodynamical properties of dense matter at non-zero temperatures. 
While differences in effective masses between APR and NRAPR below nuclear saturation density $n_\rsat$ are negligible, 
they become significant  
with increasing as density.  Specifically, the decrease of the effective masses for the APR model is somewhat slower than those of the NRAPR model. 
A similar behavior has also been observed by Constantinou \etal \cite{constantinou:14} when comparing APR and Ska EOSs \cite{kohler:76}; see their Fig. 1.  Such differences can have consequences in astrophysical applications. 
As the stellar core compresses in core collapse supernovae simulations, we expect that for the same density the temperature will be larger the lower the effective mass is.

\subsection{Surface Properties of Nuclei at $T\neq 0$}
\label{ssec:surf}

Using the methods described in Appendix \ref{app:surface}, we compute surface properties of nuclei by minimizing the nuclear surface tension $\sigma'(y_i,T)$ between two slabs of semi-infinite matter: 
a dense slab with nucleon number density $n_i$ and proton fraction $y_i$, and a dilute one with density $n_o$ and proton fraction $y_o$. 
We determine the equilibrium configurations for a range of proton fractions $y_i$ in the densest phase and temperatures $T$ of the system.  
We then compute the parameters $\lambda$, $q$, and $p$ that define the fit $\sigma(y_i,T)$ in Eqs. \eqref{eq:sigma} and \eqref{eq:h} by  minimizing the difference between $\sigma$ and $\sigma'$. 
The values of the surface tension fit parameters as well as the surface level density $A_S$, surface symmetry energy $S_S$, and the parameters of the critical temperature fit for phase coexistence $T_{c}(y)$, Eq. \eqref{eq:Tc}, are shown in Table \ref{tab:surf}. 
Note that since APR and APR$_{\rLDP}$ only differ at densities larger than the ones of interest here their surface properties are exactly the same.  
For the SkAPR EOS we computed the values of the fit assuming that its $q_{tt'}$ coefficients match those of APR for SNM at saturation density.

\begin{table}[h]
\caption{\label{tab:surf} Nuclear surface tension $\sigma(y_i,T)$ fitting parameters, $\sigma_s$, $\lambda$, $q$, and $p$ in Eq. \eqref{eq:sigma} for the EOSs of APR, NRAPR, and SkAPR. 
We also show the nuclear surface symmetry energy $S_S$, the surface level density $A_S$, and the parameters of the critical temperature fit $T_c(y)$ for phase coexistence for two semi-infinite slabs of nuclear matter.} 
\begin{ruledtabular}
\begin{tabular}{c D{.}{.}{4.3}  D{.}{.}{4.3}  D{.}{.}{4.3}  c}
\multicolumn{1}{c}{Quantity} &
\multicolumn{1}{c}{APR} &
\multicolumn{1}{c}{SkAPR} &
\multicolumn{1}{c}{NRAPR} &
\multicolumn{1}{c}{Units} \\
\hline
$\lambda$  &  3.12  &  3.38  &  3.51  &     \\
$q$        & 11.2   & 21.1   & 13.6   &     \\
$p$        &  1.61  &  1.52  &  1.46  &     \\
$\sigma_s$ &  1.20  &   1.31 &  1.14  & MeV\,fm$^{-2}$ \\
$A_S$      &  0.978 &  1.30  &  1.31  & MeV\,fm$^{-1}$ \\
$S_S$      & 76.8   & 79.1   & 93.0   & MeV \\
$T_{c0}$   & 17.92  & 15.80  & 14.39  & MeV \\
$a_{c}$    &  1.004 &  1.004 &  1.002 & \\
$b_{c}$    & -1.025 & -1.053 & -1.152 & \\
$c_{c}$    &  0.697 &  0.771 &  0.470 & \\
$d_{c}$    & -1.400 & -1.456 & -0.993 & \\
\end{tabular}
\end{ruledtabular}
\end{table}

\begin{figure}[!htb]
\centering
\includegraphics[trim=0 70 0 0, clip, width = 0.50\textwidth]{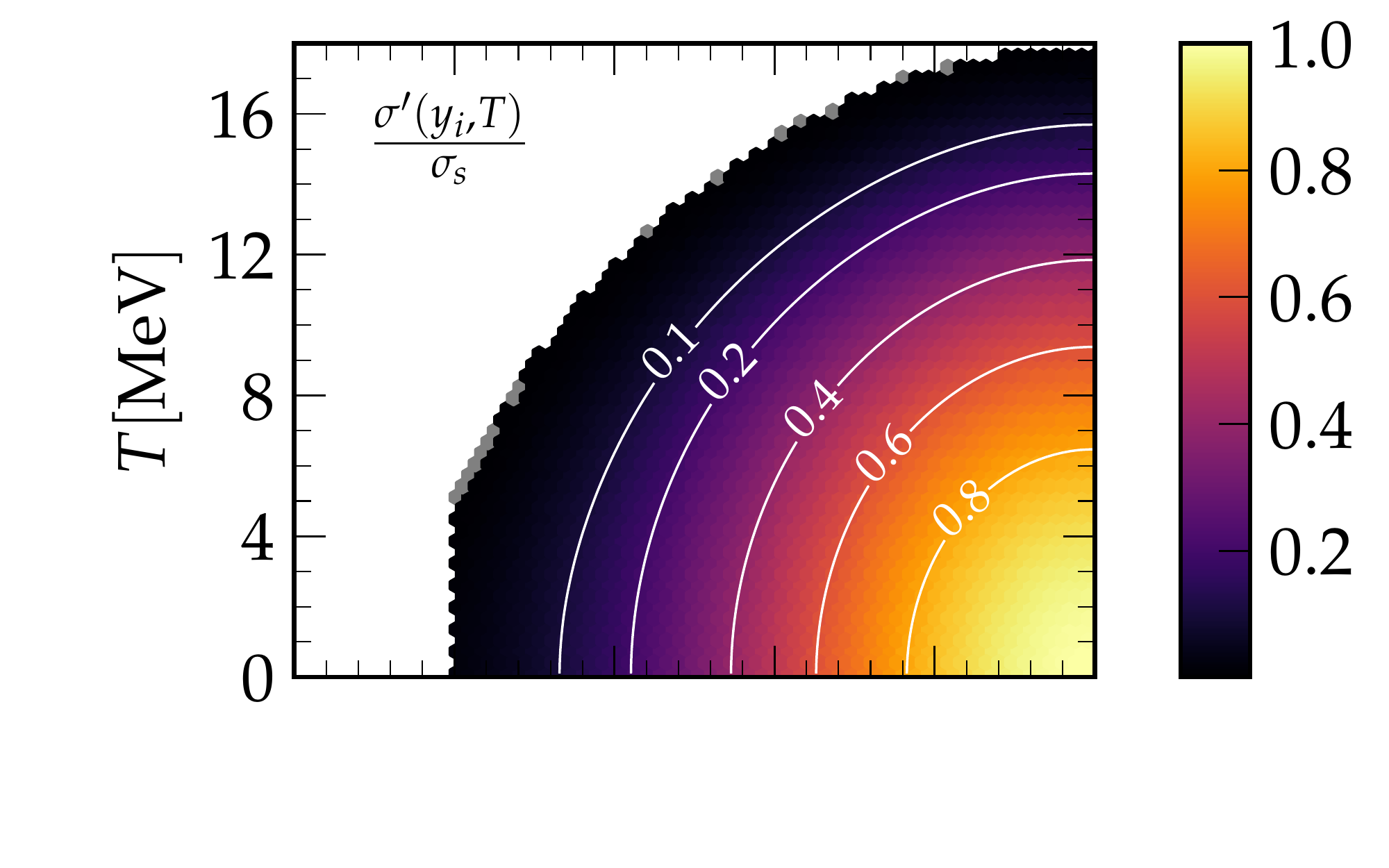}
\includegraphics[trim=0 70 0 0, clip, width = 0.50\textwidth]{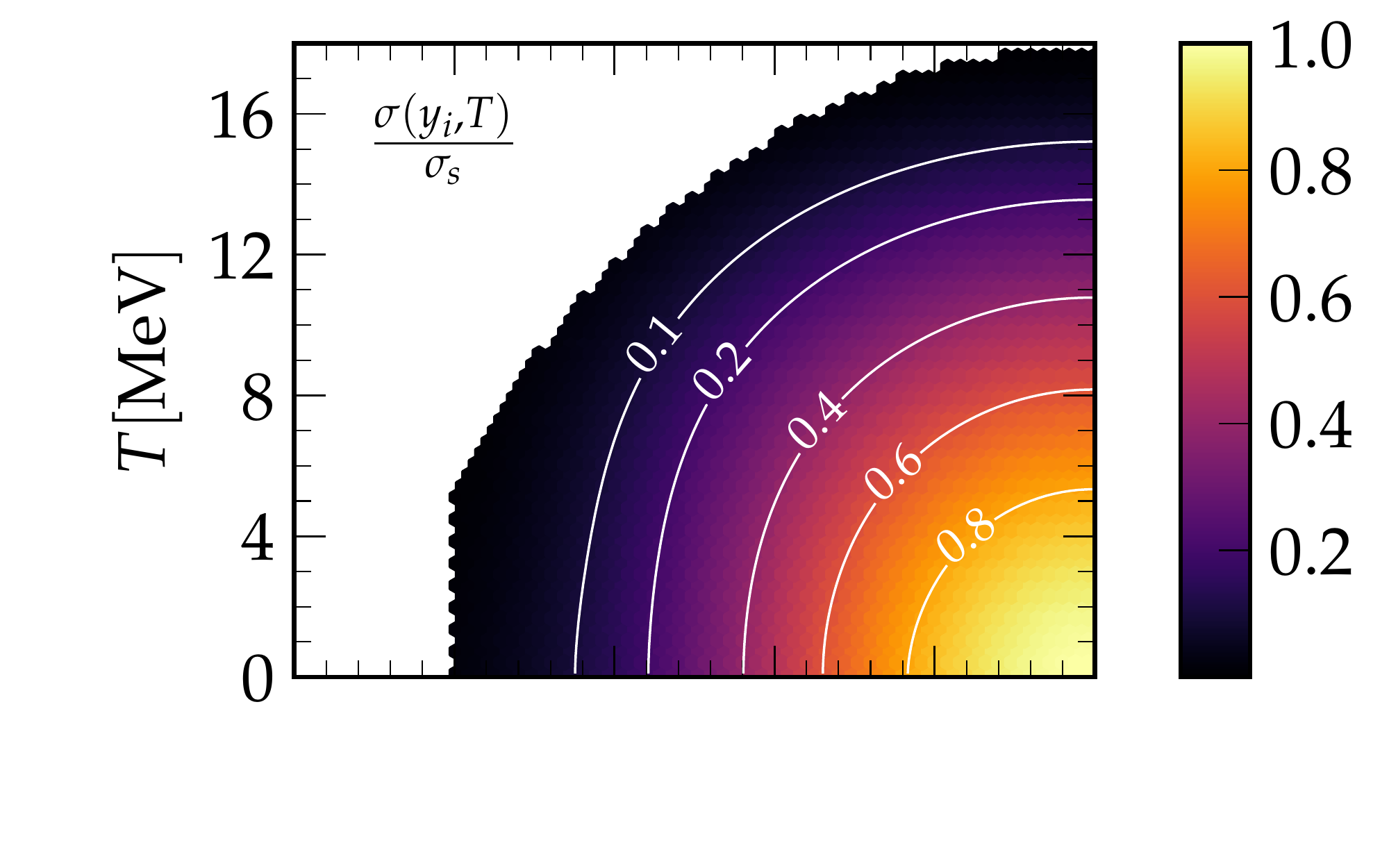}
\includegraphics[trim=0 20 0 0, clip, width = 0.50\textwidth]{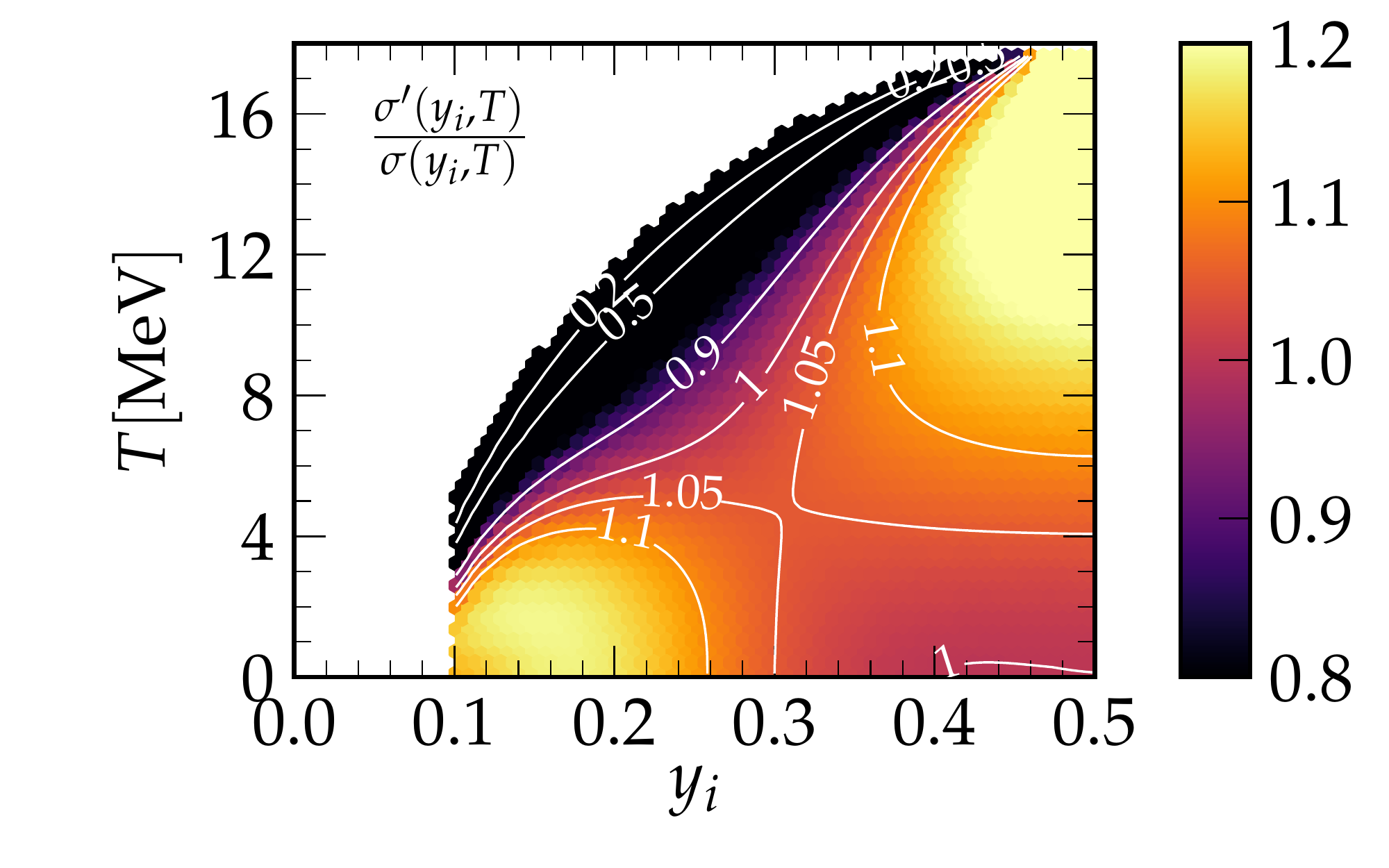}
\caption{\label{fig:surf_APR} Surface properties $\sigma'(y_i,T)$ computed for the APR model (top), its best fit $\sigma(y_i,T)$ using Eqs. \eqref{eq:sigma} and \eqref{eq:h} (center), and ratio between the computed and best fit $\sigma'(y_i,T)/\sigma(y_i,T)$ (bottom). 
White regions are places where matter is unstable against phase coexistence.} 
\end{figure}

In Fig. \ref{fig:surf_APR}, we plot the surface tension $\sigma'$ obtained for the EOS of APR and its best fit $\sigma$ following Eqs. \eqref{eq:sigma} and \eqref{eq:h}, and the ratio between the computed properties and its best fit $\sigma'/\sigma$.  
As expected, the surface tension $\sigma'$ is largest for symmetric matter at zero temperature and decreases as matter becomes neutron rich and/or as its temperature is increased. 
For temperatures above the critical temperature $T_c=17.9\unit{MeV}$, the system is unstable against phase coexistence. 
For very neutron rich matter, the surface tension fit $\sigma$ goes to zero for $y_i=0.06$, indicating that there is no equilibrium between coexisting phases if the proton fraction of the densest phase drops below this value. 
However, our algorithm is unable to find solutions for $y\lesssim0.10$ as the surface tension between the dense and dilute phases is too small and the surface extends over long distances.

The values for the surface tension $\sigma'$ and its fit $\sigma$ agree well in most of the parameter space as seen in the top and center plots of Fig. \ref{fig:surf_APR}. 
We note, however, that the ratios between $\sigma$ and $\sigma'$ differ by 10 to 20\% for symmetric matter at  high temperatures and for neutron rich matter at temperatures below $\sim3\unit{MeV}$. 
Furthermore, for regions of the $y_i$-$T$ phase space where $\sigma'/\sigma_s$ is below 0.3, the fitting function $\sigma$ overestimates the surface tension $\sigma'$ by as much as a factor of 5. 
Thus, a different fitting function may be needed in order to accurately probe this region. 
We defer this to future work.

\begin{figure}[!htb]
\centering
\includegraphics[trim=0 70 0 0, clip, width = 0.50\textwidth]{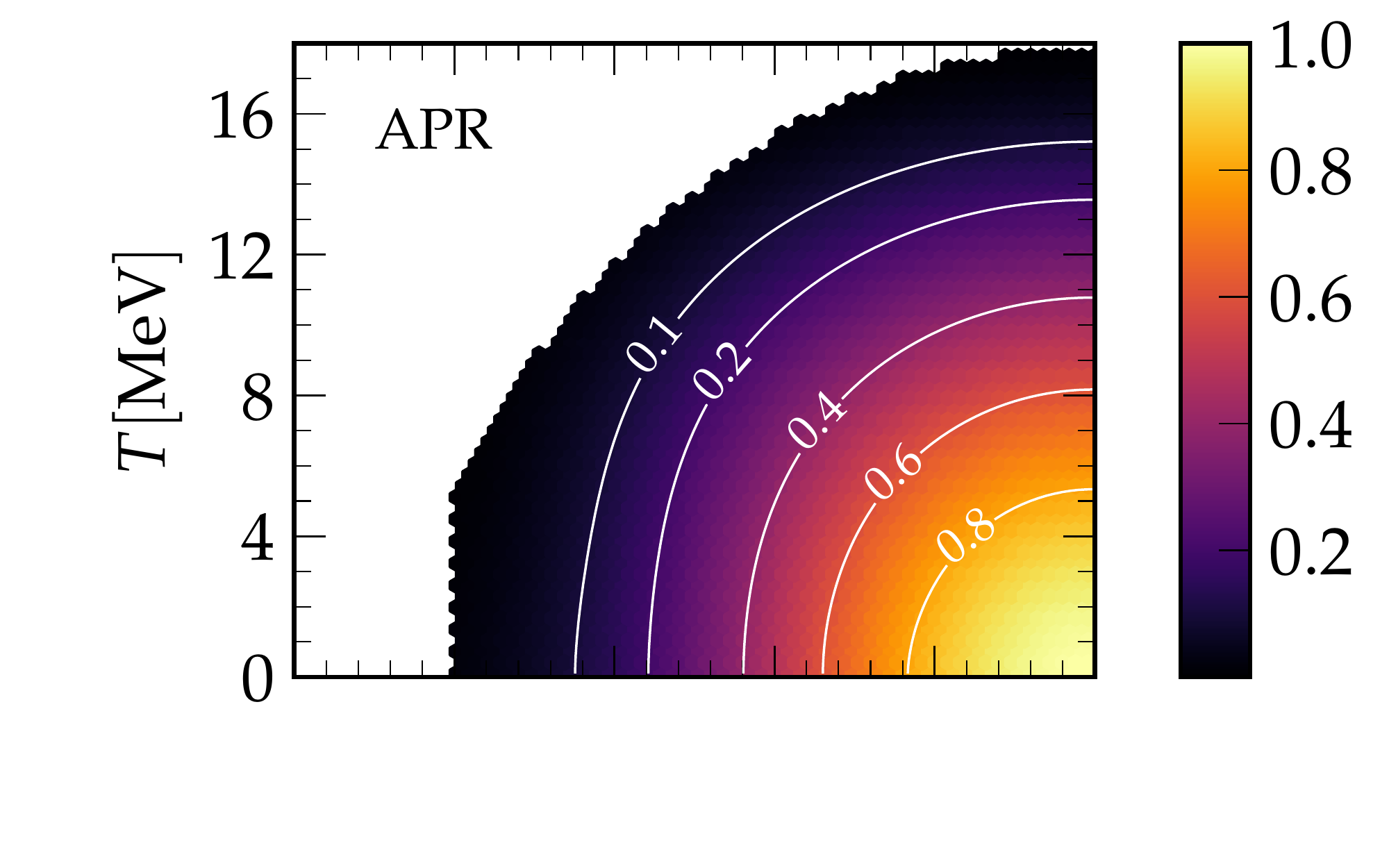}
\includegraphics[trim=0 70 0 0, clip, width = 0.50\textwidth]{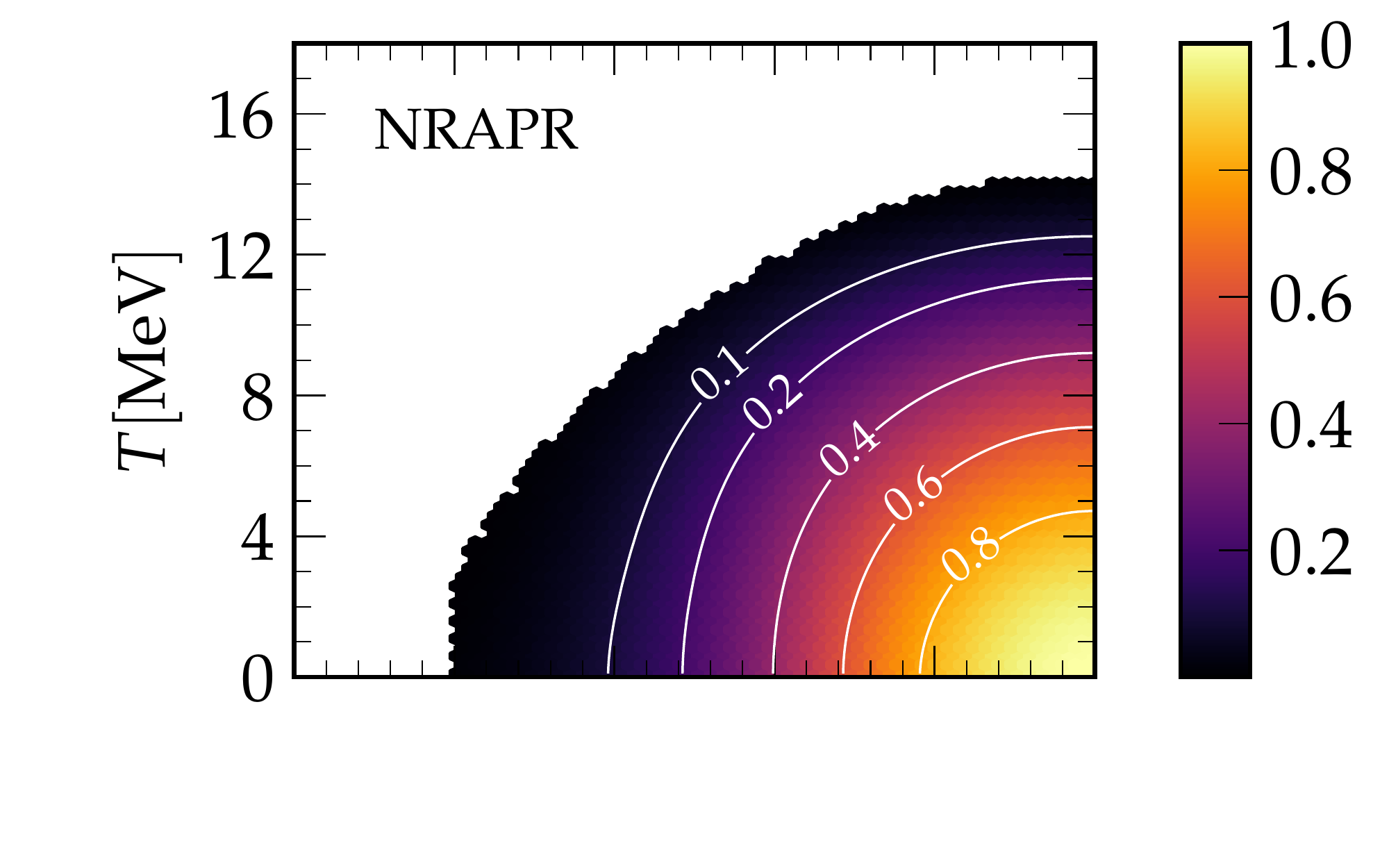}
\includegraphics[trim=0 20 0 0, clip, width = 0.50\textwidth]{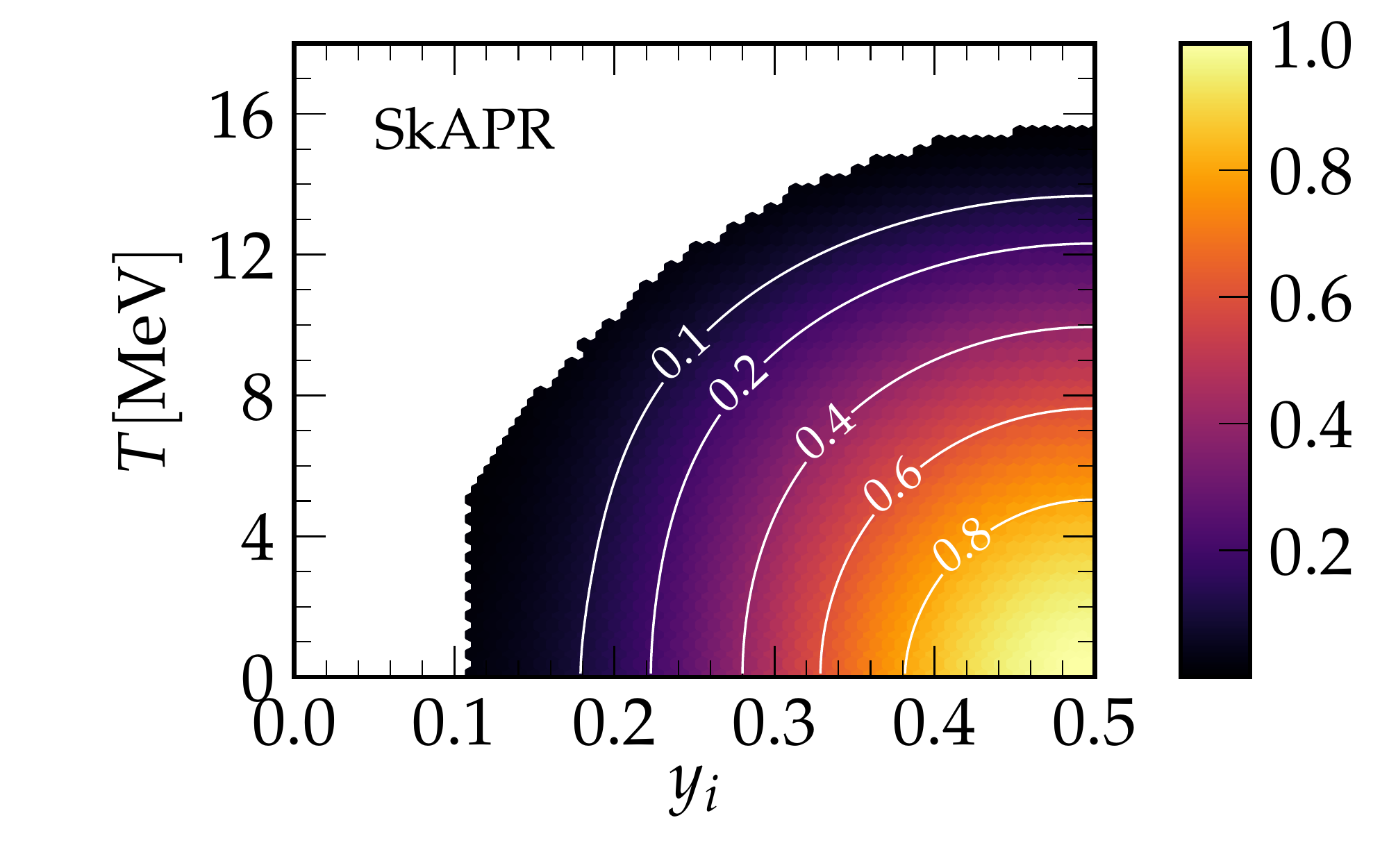}
\caption{\label{fig:surf_all} Plots of normalized surface tension $\sigma(y_i,T)/\sigma_s$ for APR (top), NRAPR (center), and SkAPR (bottom) EOSs. White regions indicate areas of the parameter space where there is no coexistence of matter with two different densities and proton fractions.}
\end{figure}

In Fig. \ref{fig:surf_all}, we plot the surface tension fit $\sigma(y_i,T)$ for the APR, NRAPR and SkAPR models. 
All three EOSs have the same qualitative behavior for $\sigma(y_i,T)$.
We notice that the APR model predicts coexistence of dense and dilute phases for symmetric nuclear matter for temperatures higher than the other two EOSs. 
The values of the critical temperatures for each EOS are presented in Tab. \ref{tab:surf}. 
As for the APR model, our algorithm to obtain $\sigma'$ fails to obtain coexisting phases for proton fractions lower than $y_i=0.10$ for NRAPR and SkAPR.  
However, we do not expect this failure to significantly alter the parameters of the fit function $\sigma$.

\begin{figure}[!h]
\centering
\includegraphics[trim=0 80 0 0, clip, width = 0.50\textwidth]{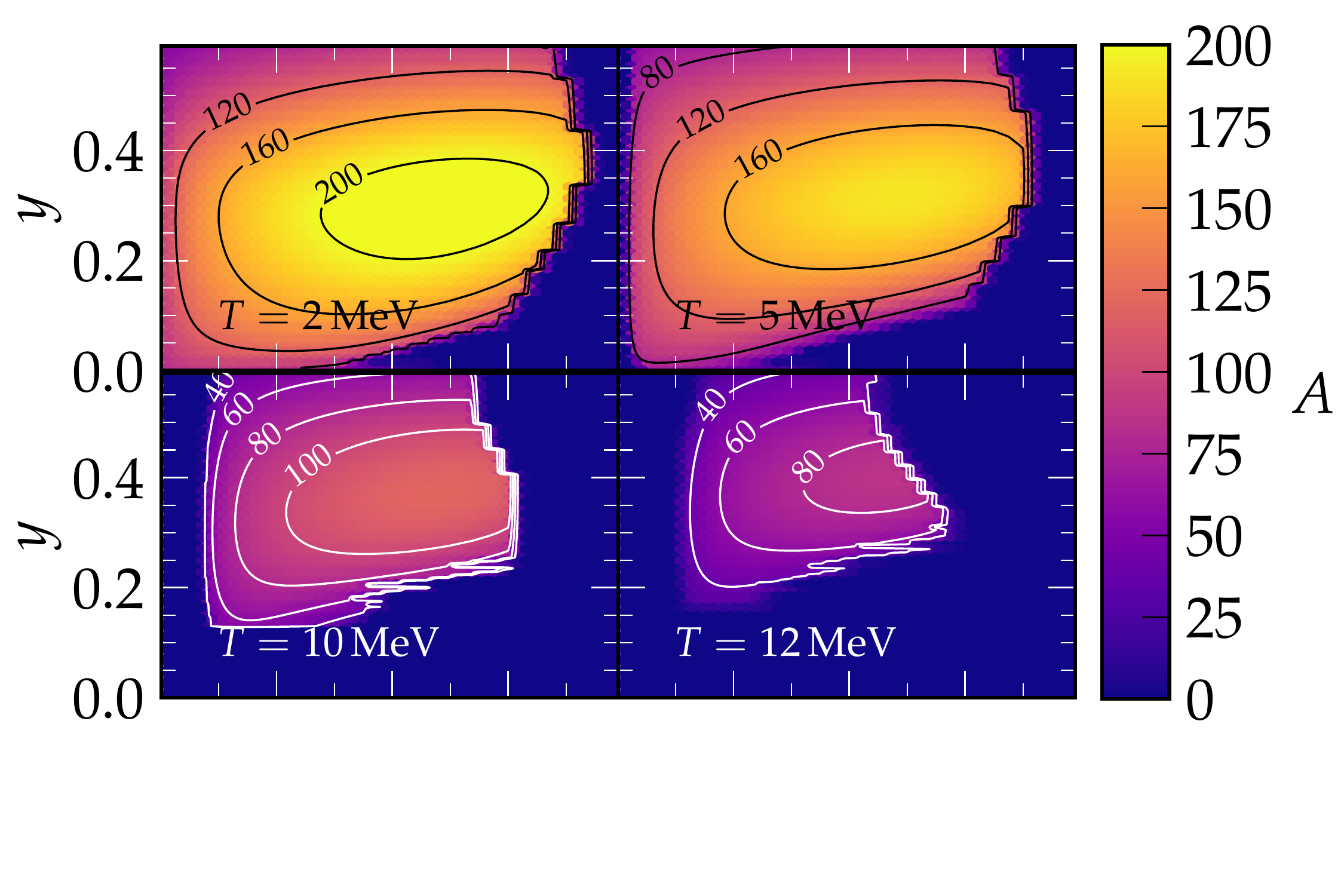}
\includegraphics[trim=0 80 0 0, clip, width = 0.50\textwidth]{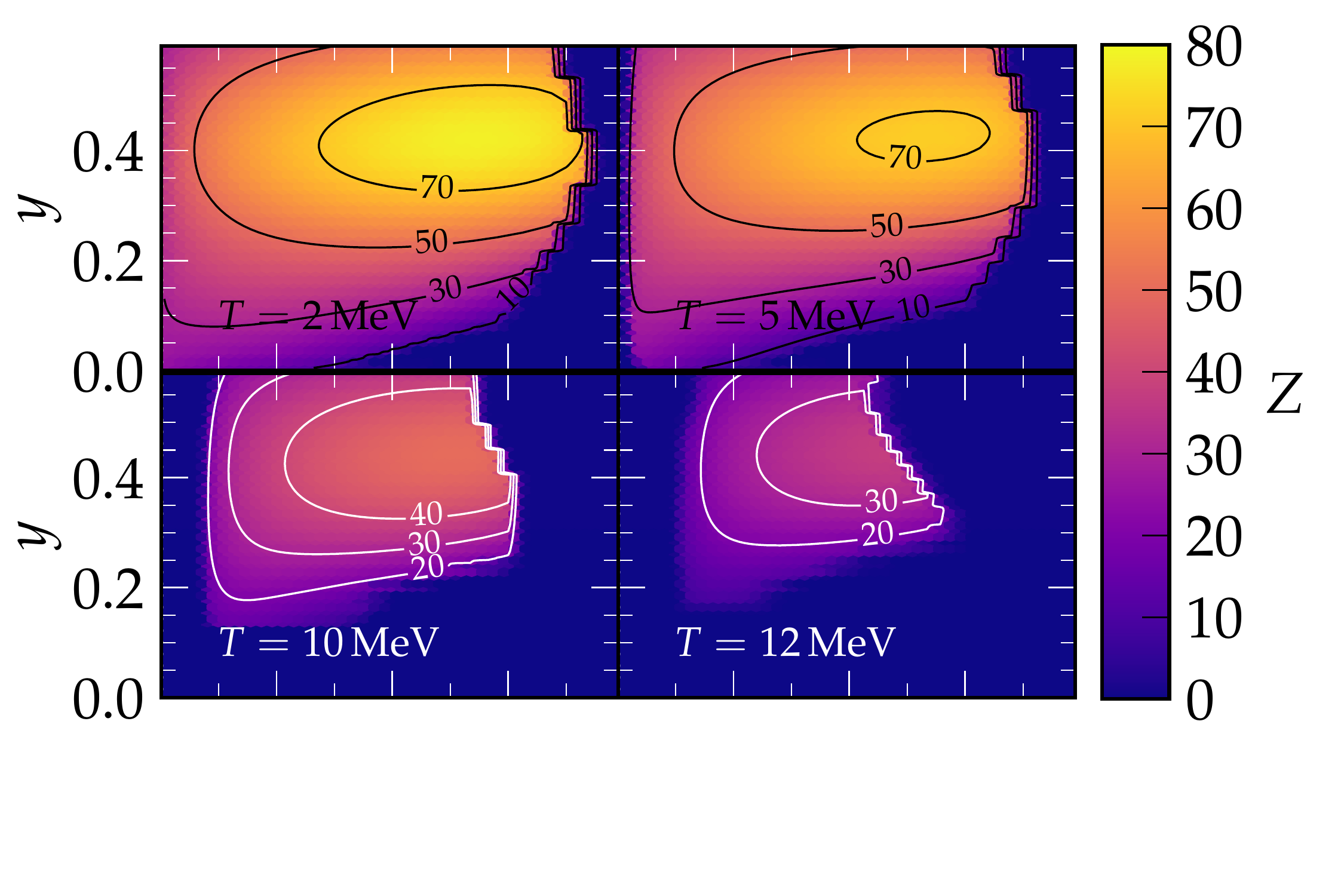}
\includegraphics[trim=0 30 0 0, clip, width = 0.50\textwidth]{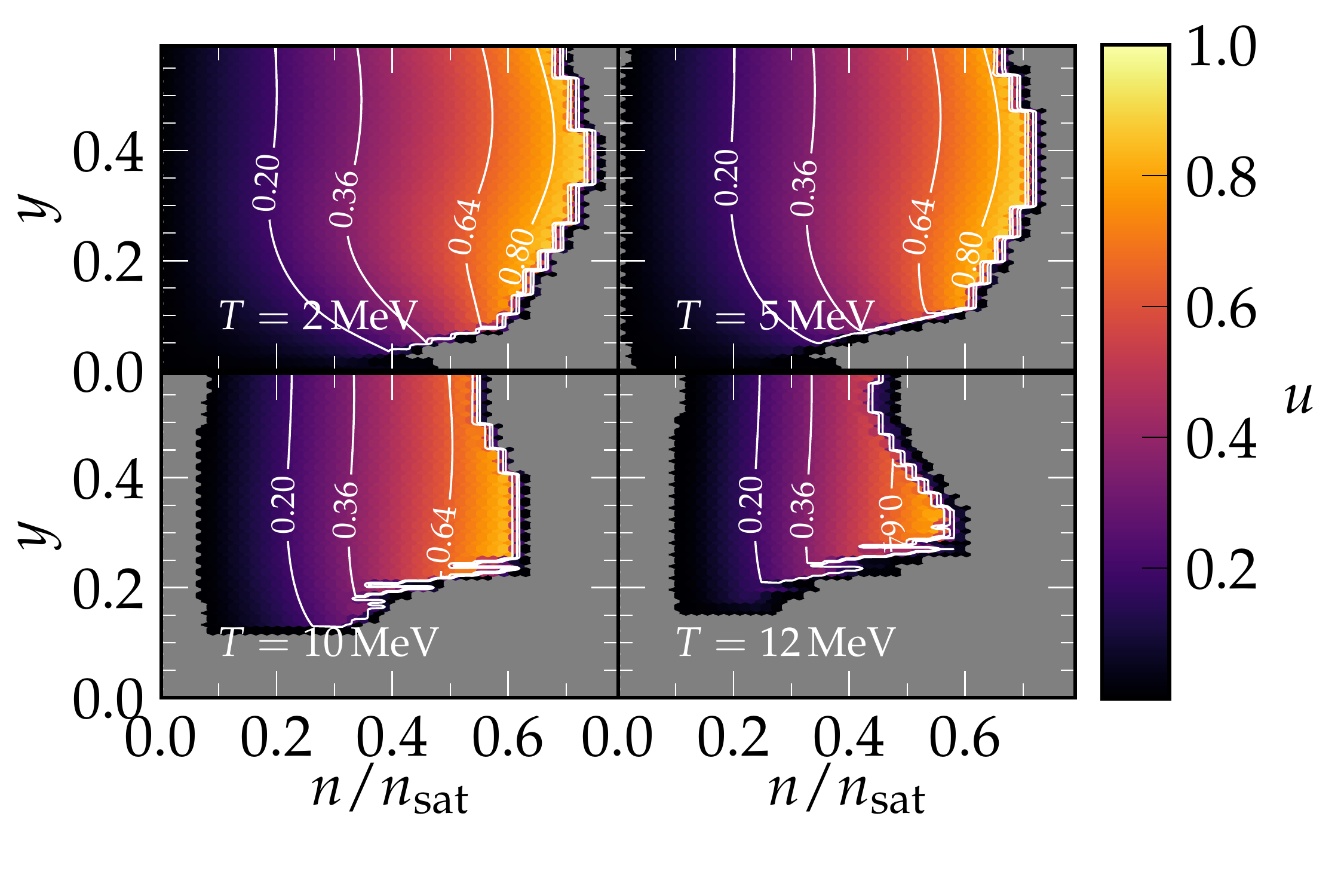}
\caption{\label{fig:u_APR} Nuclear mass number $A$, charge $Z$ and the volume fraction $u$ occupied by the dense phase for the APR model.}
\end{figure}

\begin{figure}[!h]
\centering
\includegraphics[trim=0 80 0 0, clip, width = 0.50\textwidth]{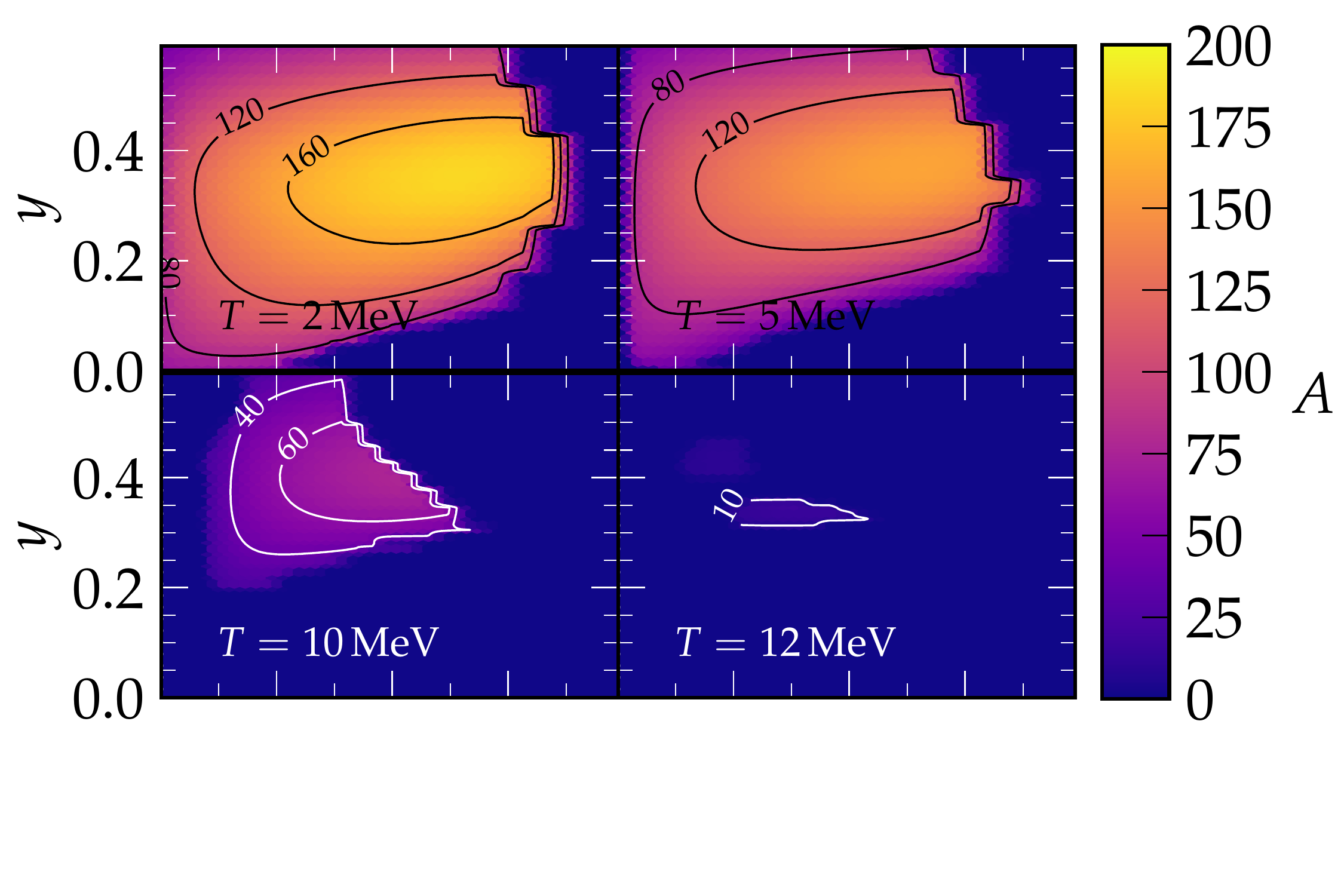}
\includegraphics[trim=0 80 0 0, clip, width = 0.50\textwidth]{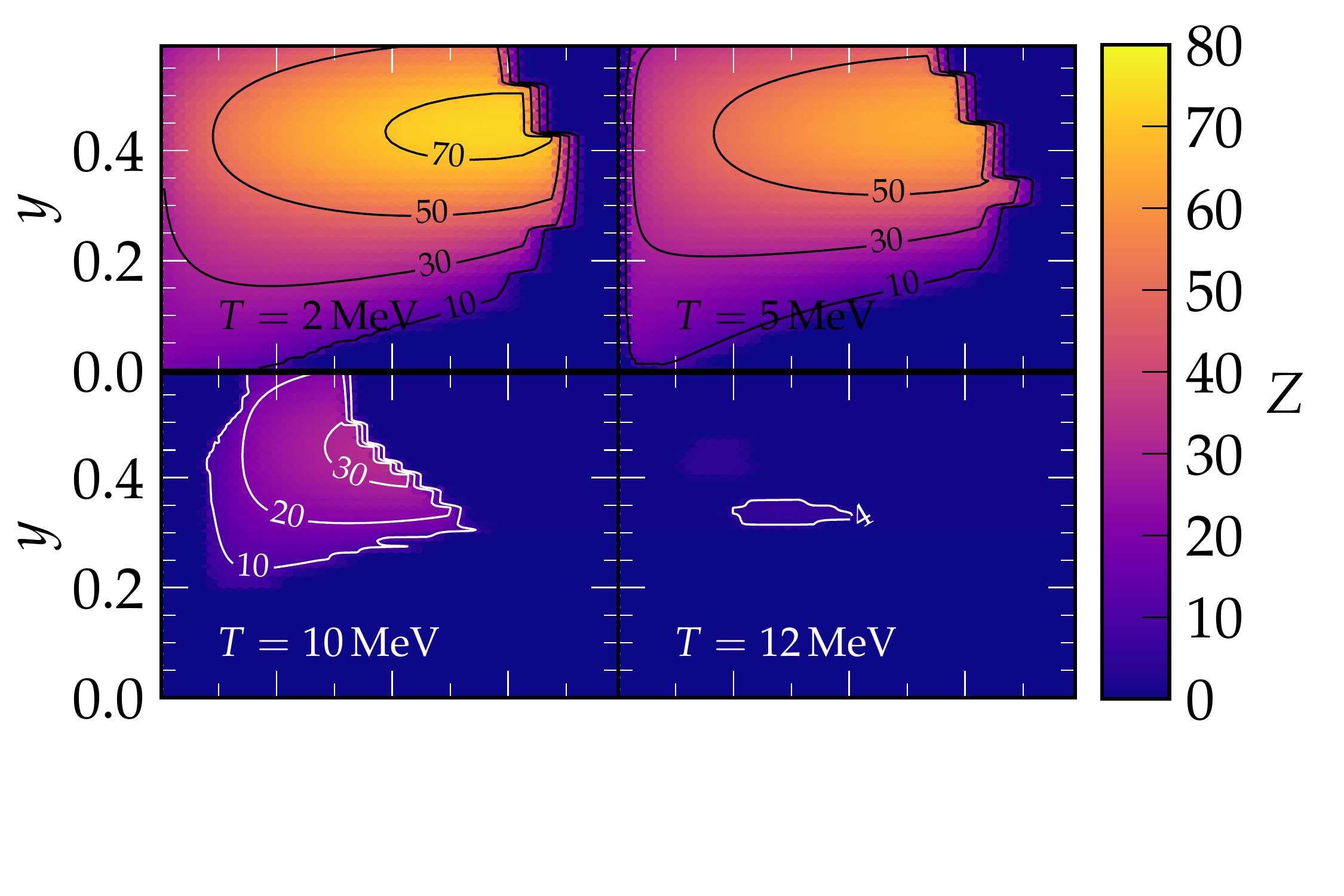}
\includegraphics[trim=0 30 0 0, clip, width = 0.50\textwidth]{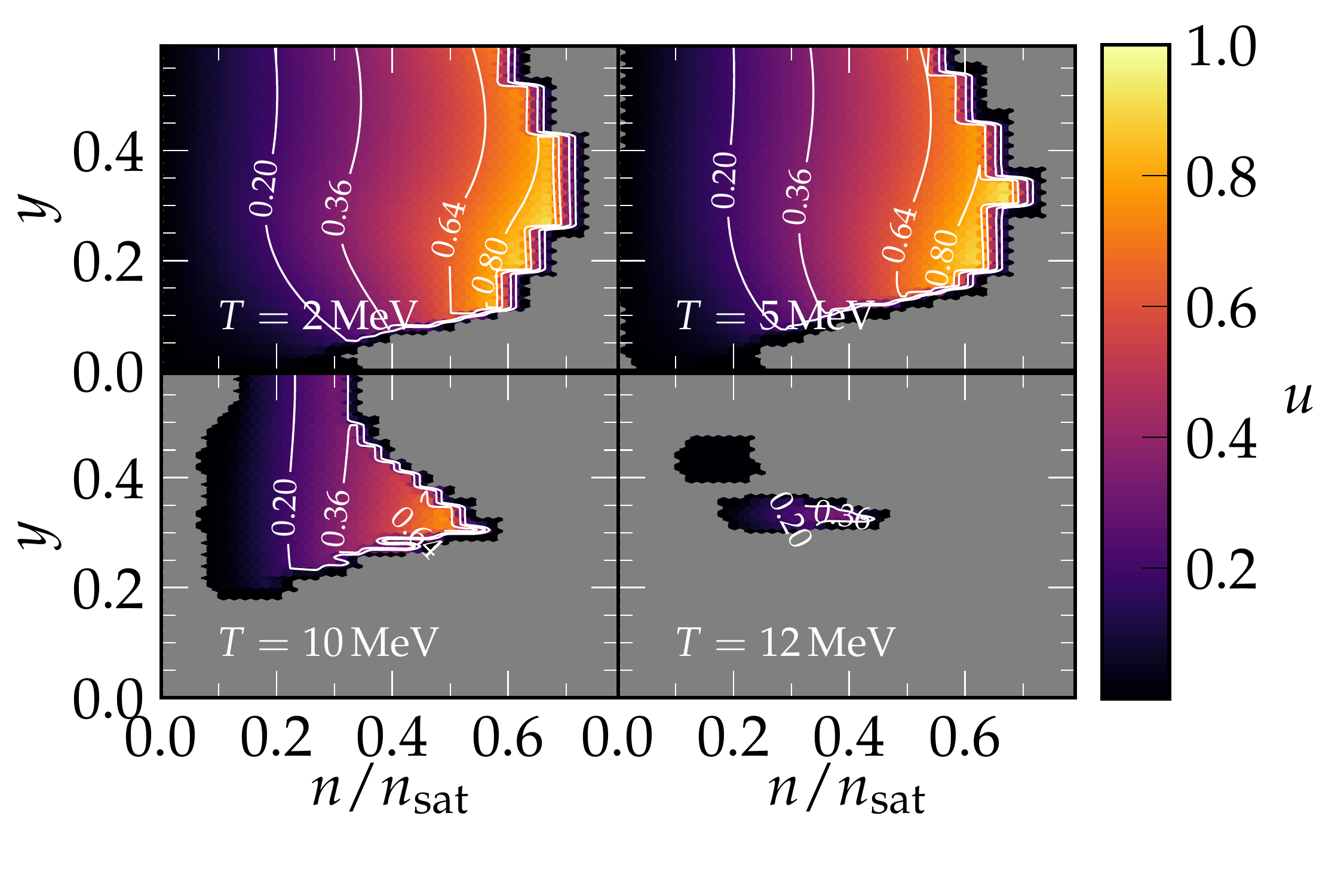}
\caption{\label{fig:u_NRAPR} Same as Fig. \ref{fig:u_APR}, but for  the NRAPR model.}
\end{figure}

\begin{figure}[!h]
\centering
\includegraphics[trim=0 80 0 0, clip, width = 0.50\textwidth]{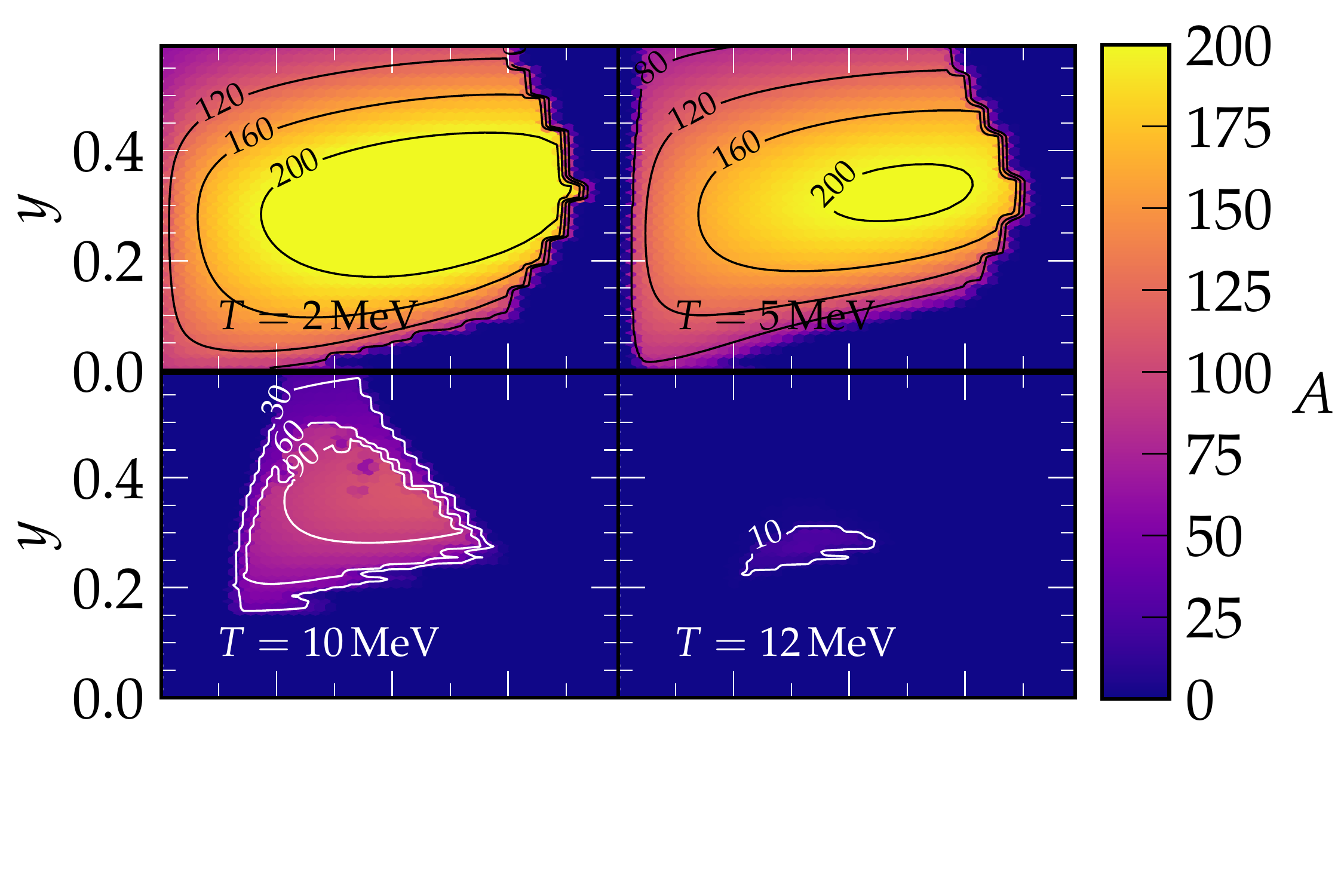}
\includegraphics[trim=0 80 0 0, clip, width = 0.50\textwidth]{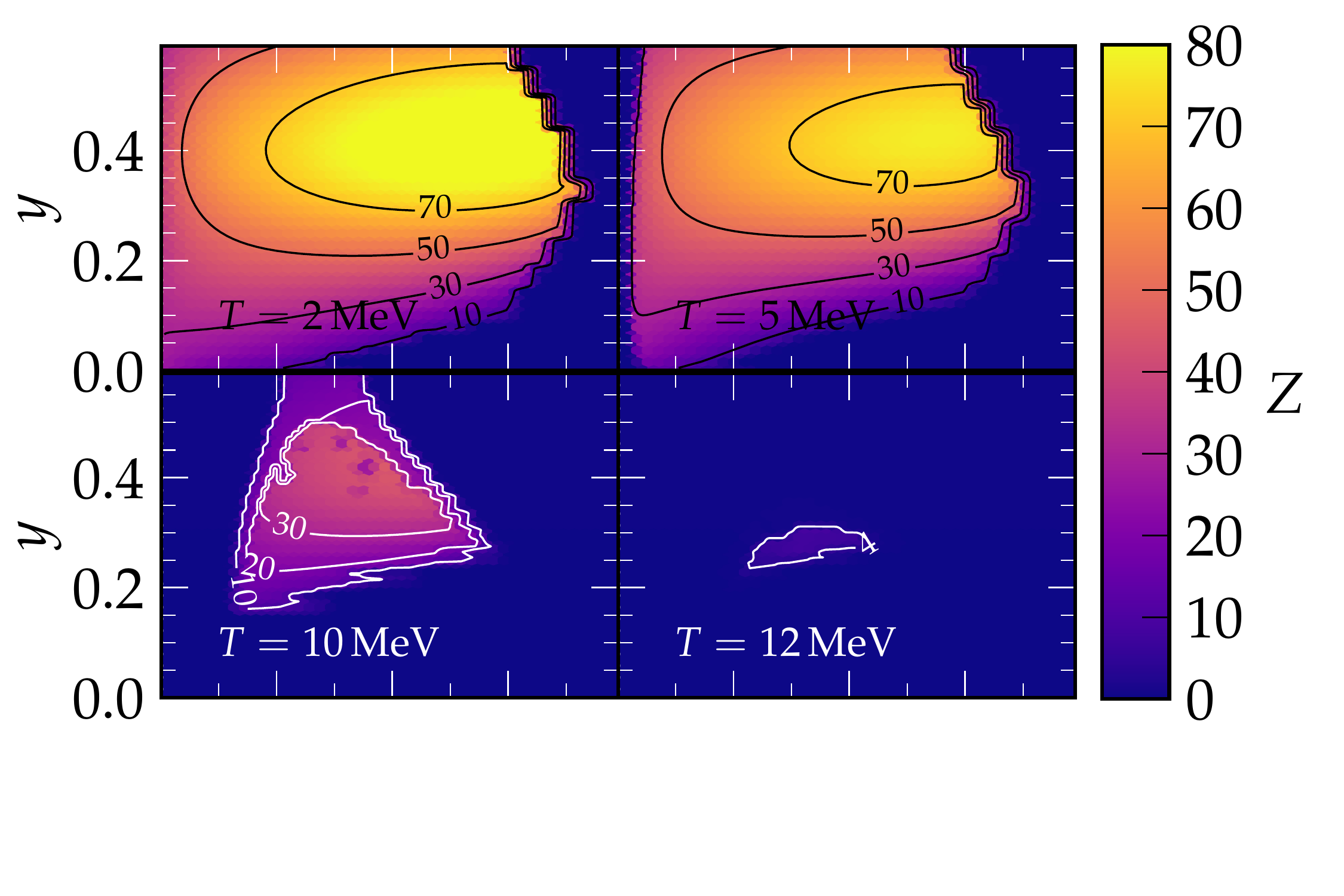}
\includegraphics[trim=0 30 0 0, clip, width = 0.50\textwidth]{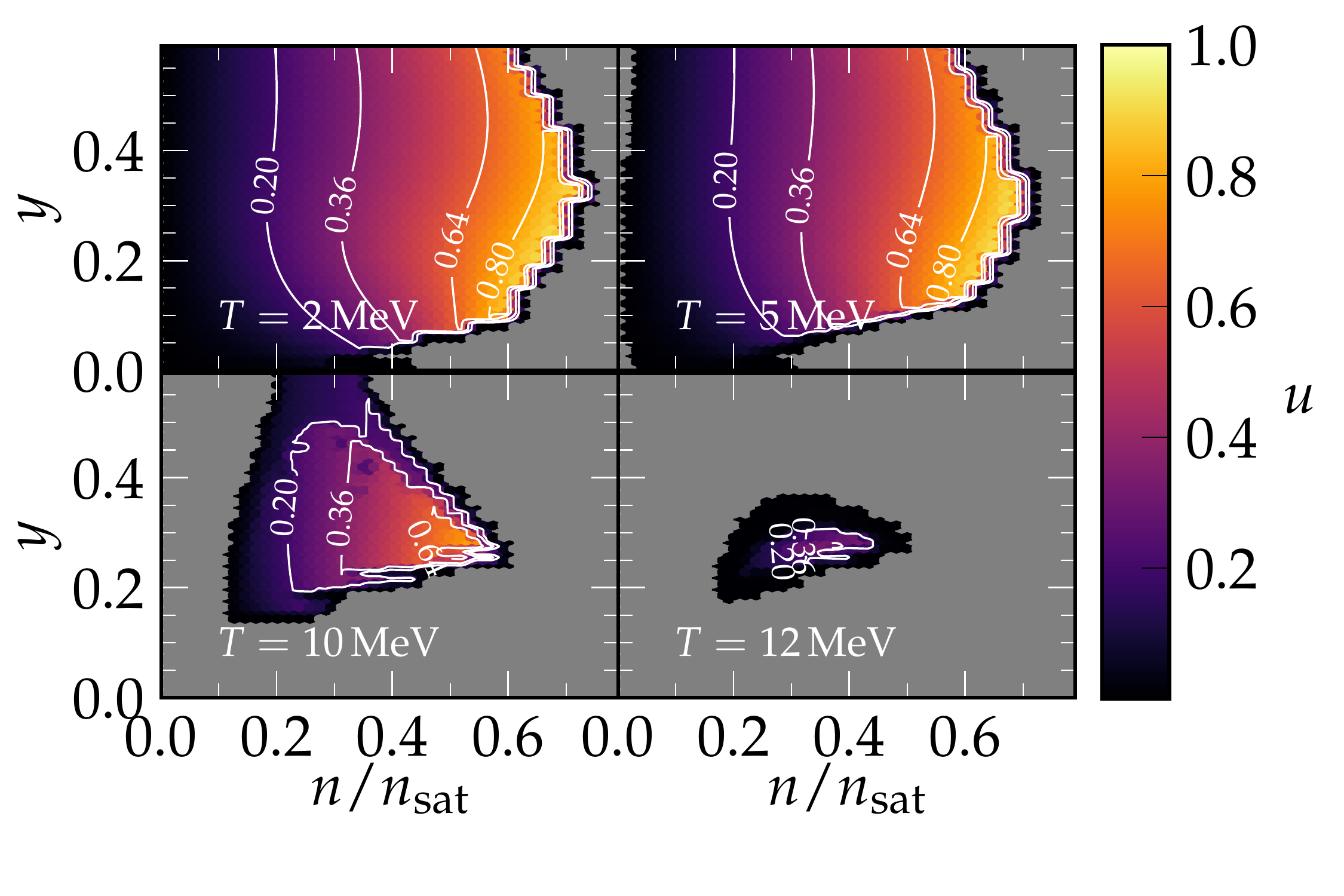}
\caption{\label{fig:u_SkAPR} Same as Fig. \ref{fig:u_APR}, but for  the SkAPR model.}
\end{figure}

Once the surface properties have been determined, we focus our attention of the subnuclear density region $0.1 \lesssim n/n_\rsat \lesssim 0.8$ of parameter space with temperatures lower than $T_c(y)$, \ie where the nuclear pasta is expected to occur. 
In Figs. \ref{fig:u_APR}, \ref{fig:u_NRAPR}, and \ref{fig:u_SkAPR} we plot, respectively,  the nuclear mass number $A$, its charge $Z$ as well as the volume fraction $u$ occupied by the dense phase in each Wigner-Seitz cell for the APR, NRAPR, and SkAPR models. Results shown are for four temperatures, $T=2,\,5,\,10,\,12\unit{MeV}$.

Within the formalism used, the volume fraction $u$ is directly related to the topological phase of nuclear matter. 
Following the procedure of Lattimer \& Swesty \cite{lattimer:91} and detailed in Fig. 4 of Lim \& Holt \cite{lim:17}, the occupied volume fraction of the dense phase describes (1) spherical nuclei for $u<0.20$, (2) cylindrical nuclei for $0.20\le u < 0.36$, (3) flat sheets for $0.36\le u < 0.64$, (4) cylindrical holes for $0.64 \le u < 0.80$, and (5) spherical holes for $u\ge 0.80$. 
In the Figs. \ref{fig:u_APR}, \ref{fig:u_NRAPR}, and \ref{fig:u_SkAPR}, the gray area represents regions where nuclear matter is in a uniform phase.

We note that SkAPR and APR produce nuclei with larger mass numbers than NRAPR owing to their higher compression moduli, $K_\rsat=266\unit{MeV}$ compared to $K_\rsat=226\unit{MeV}$ of NRAPR. 
As expected from the surface tension plot, Fig. \ref{fig:surf_all}, the APR model predicts nuclei that perist up to higher temperatures than for the Skyrme EOSs. 
This is likely due to the density dependence of the $q_{tt'}$ in the APR model, see Eqs. \eqref{eq:qtt}, which is absent in the Skyrme model.

As the temperature increases, uniform nuclear matter occupies larger and larger fraction of the $y-n$ parameter space. 
In all cases, spherical holes seem to disappear first followed by cylindrical holes. 
The last region to disappear for all EOSs (not shown for APR), is for proton fractions $0.2\lesssim y\lesssim 0.4$ at densities $0.2\lesssim n/n_\rsat\lesssim 0.4$.
This happens even though the surface tension is larger for SNM than for neutron rich matter. 
Similar results, albeit with small quantitative differences, are obtained in other works which use SNA near the transition to uniform nuclear matter \cite{lattimer:91, shen:98b, shen:11}. 
Relaxing the assumptions made therein to compute the free energy near the transition region, so that SNM melts at a higher temperature than neutron rich matter, will be taken up in future work.

\subsection{Composition of the System at $T\neq 0$}
\label{ssec:composition}

\begin{figure}[!h]
\centering
\includegraphics[trim=0 50 0 0, clip, width = 0.50\textwidth]{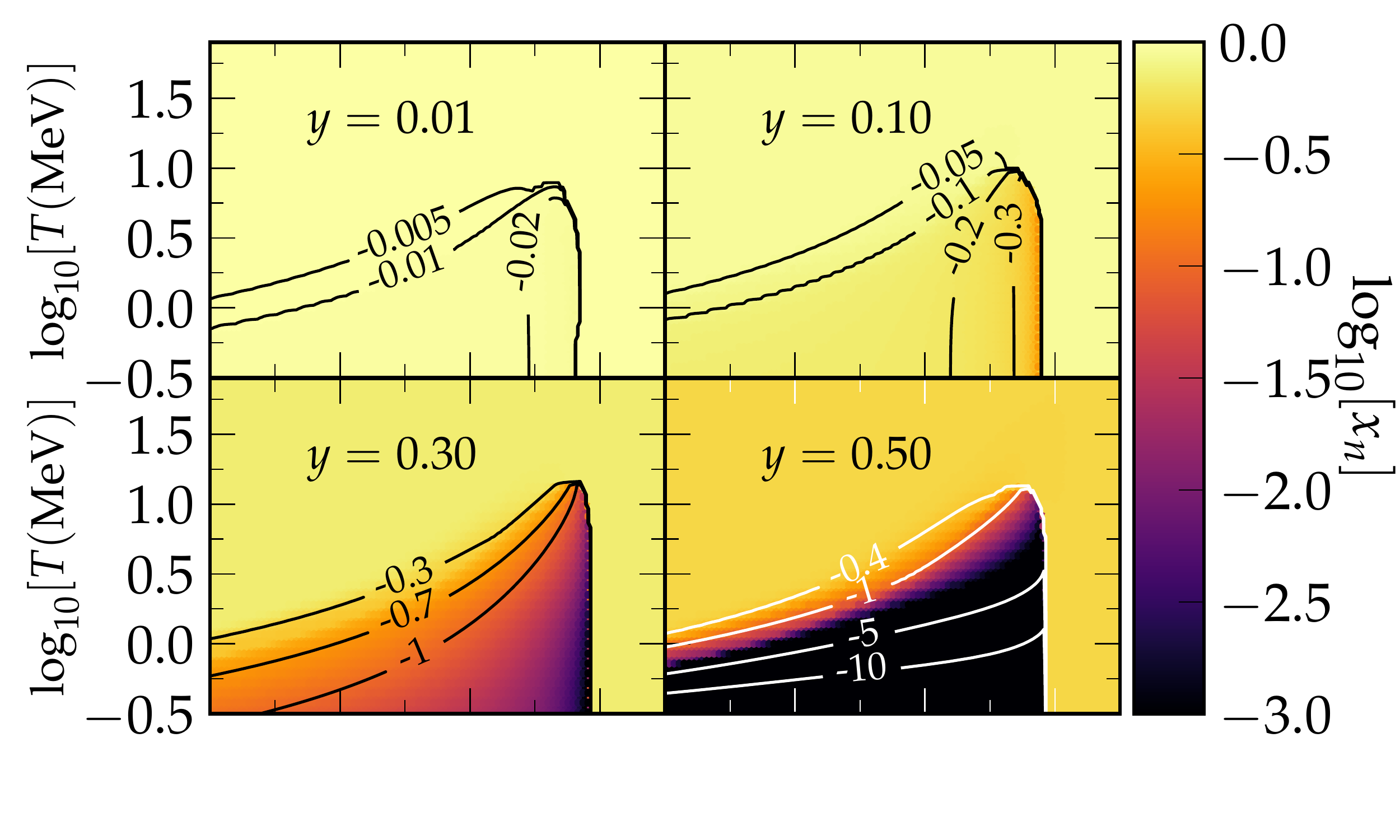}
\includegraphics[trim=0 50 0 0, clip, width = 0.50\textwidth]{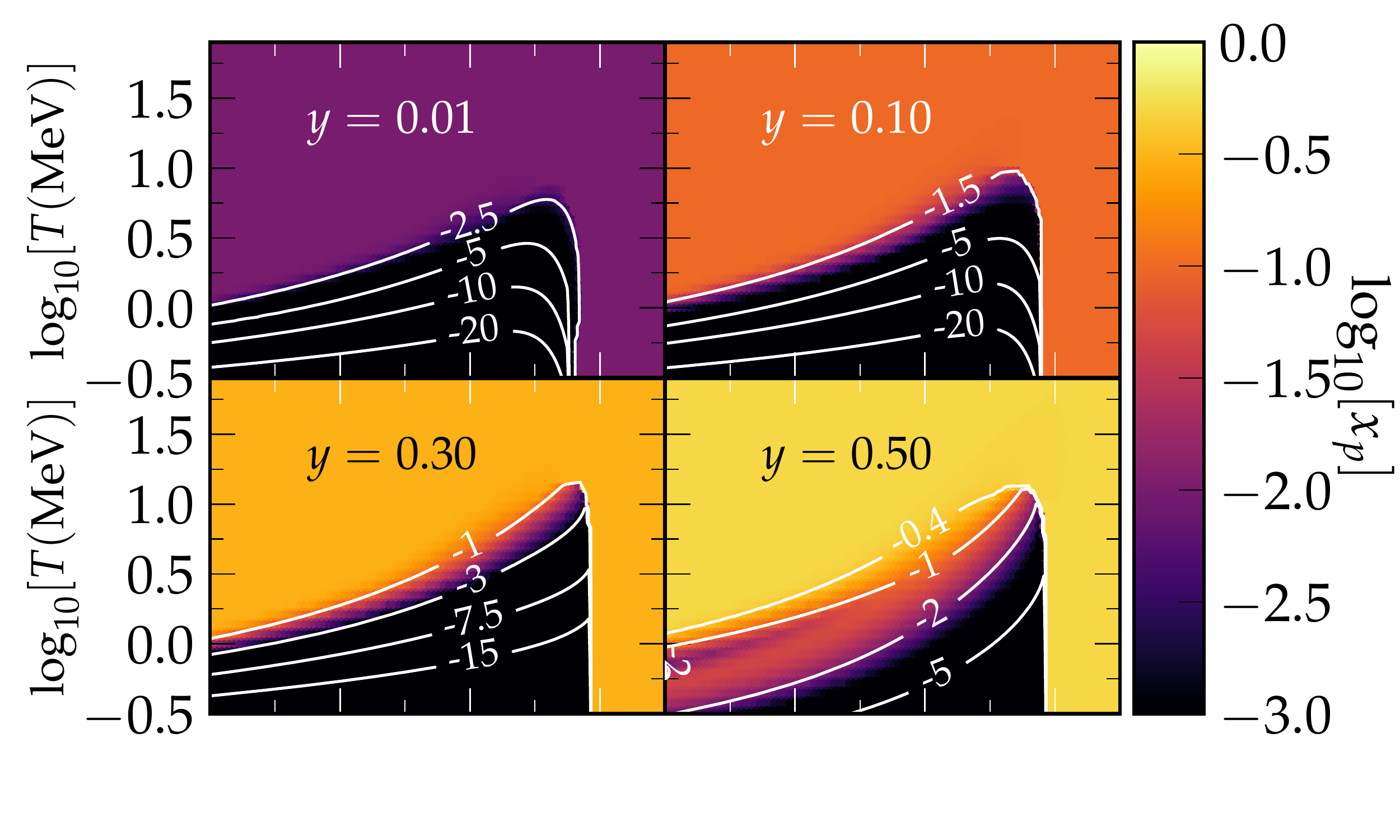}
\includegraphics[trim=0 50 0 0, clip, width = 0.50\textwidth]{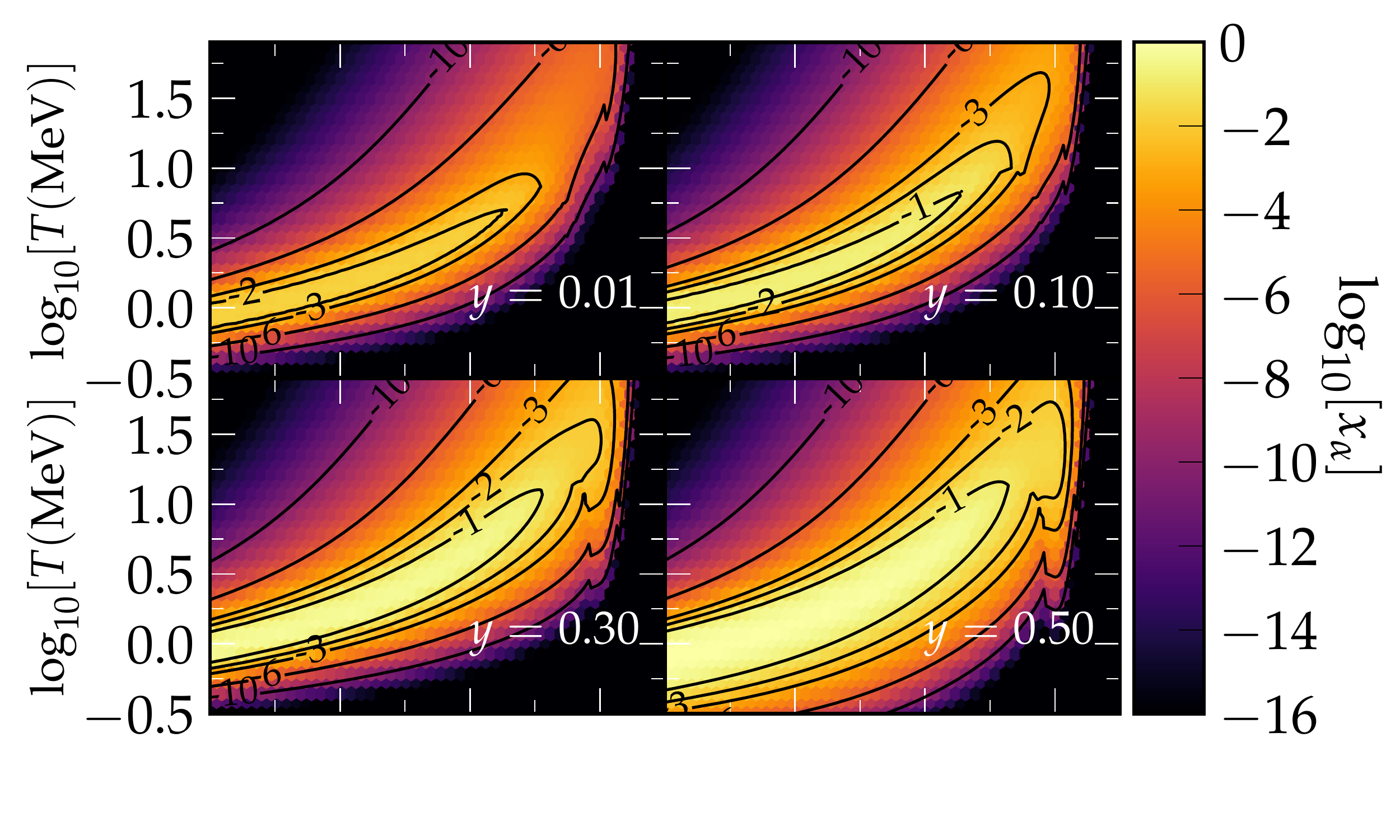}
\includegraphics[trim=0  0 0 0, clip, width = 0.50\textwidth]{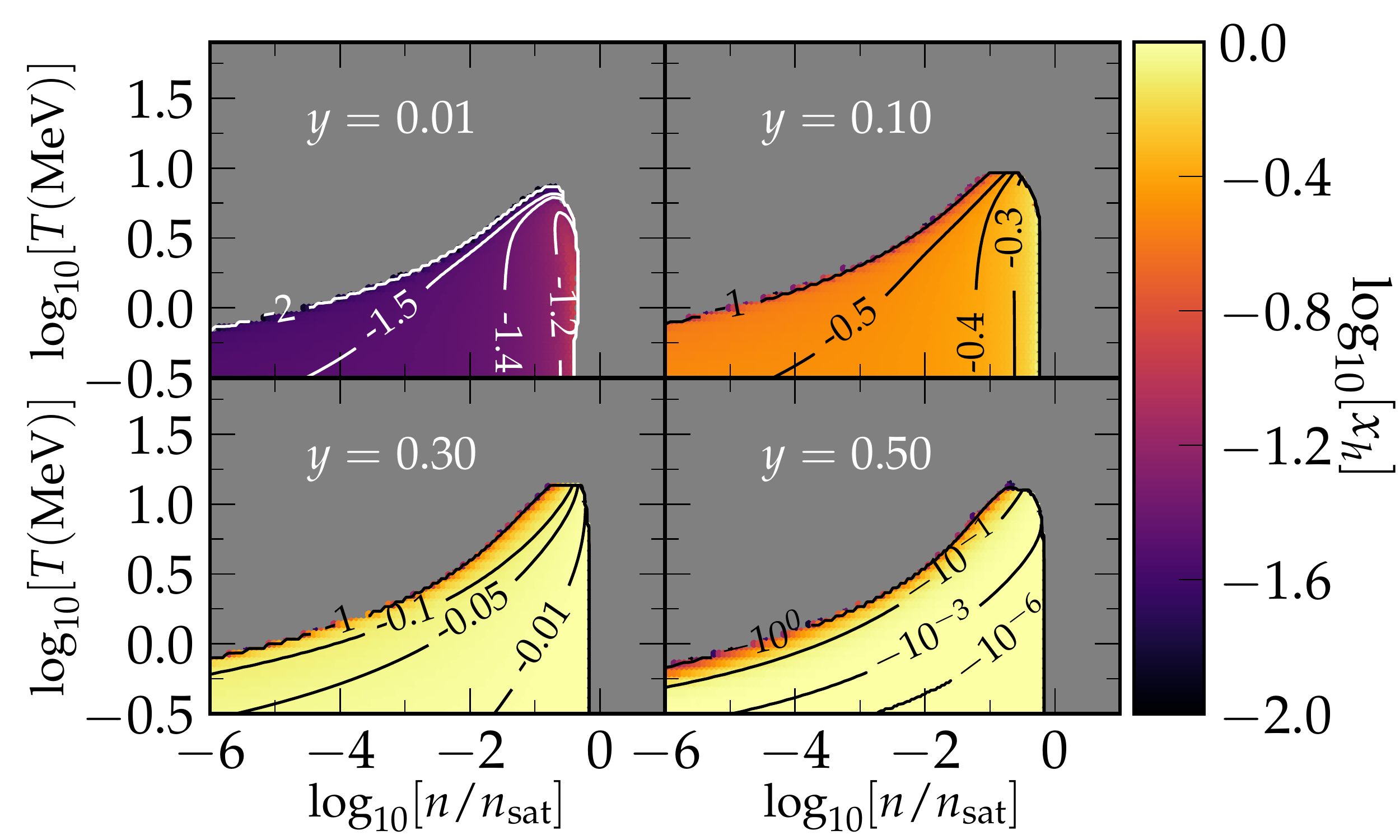}
\caption{\label{fig:APR_x} From top to bottom: number fraction of neutrons $x_n$, protons $x_p$, alpha particles $x_\alpha$, and heavy nuclei $x_h$ for proton fraction $y=0.01,\,0.10,\,0.30$ and $0.50$ for the APR EOS.}
\end{figure}

In Fig. \ref{fig:APR_x}, we display the composition of the system for the APR model. 
We plot neutron, proton, alpha particle, and heavy nuclei number fractions $x_n$, $x_p$, $x_\alpha$, and $x_h$, respectively. 
The qualitative behavior of the composition for the other EOSs is the same as for the APR EOS across all of parameter space. 
However, there are minor quantitative differences between the APR and the Skyrme EOSs, as for example, APR predicts that heavy nuclei melt at higher temperatures, especially at densities close to the nuclear saturation density.

We note that all expected qualitative behavior for the EOSs are fulfilled. 
For SNM at densities $n\lesssim0.10\unit{fm}^{-3}$ and temperatures $T\lesssim1\unit{MeV}$, most nucleons cluster into heavy nuclei, a few into alpha particles whereas a very small fraction is free due to temperature effects. 
As density increases and reaches $n\simeq0.10\unit{fm}^{-3}$ nucleons occupy all the space available to them and matter becomes uniform. 
As temperature is increased, heavy nuclei progressively breakup into alpha particles until at even alpha particles start to breakup and the system is driven closer to a uniform free nucleon gas. 
If, instead, proton fraction is decreased, neutrons drift out of heavy nuclei, alpha particles breakup, and the system as a whole becomes neutron rich.

\begin{figure}[!h]
\centering
\includegraphics[trim=0 50 0 0, clip, width = 0.50\textwidth]{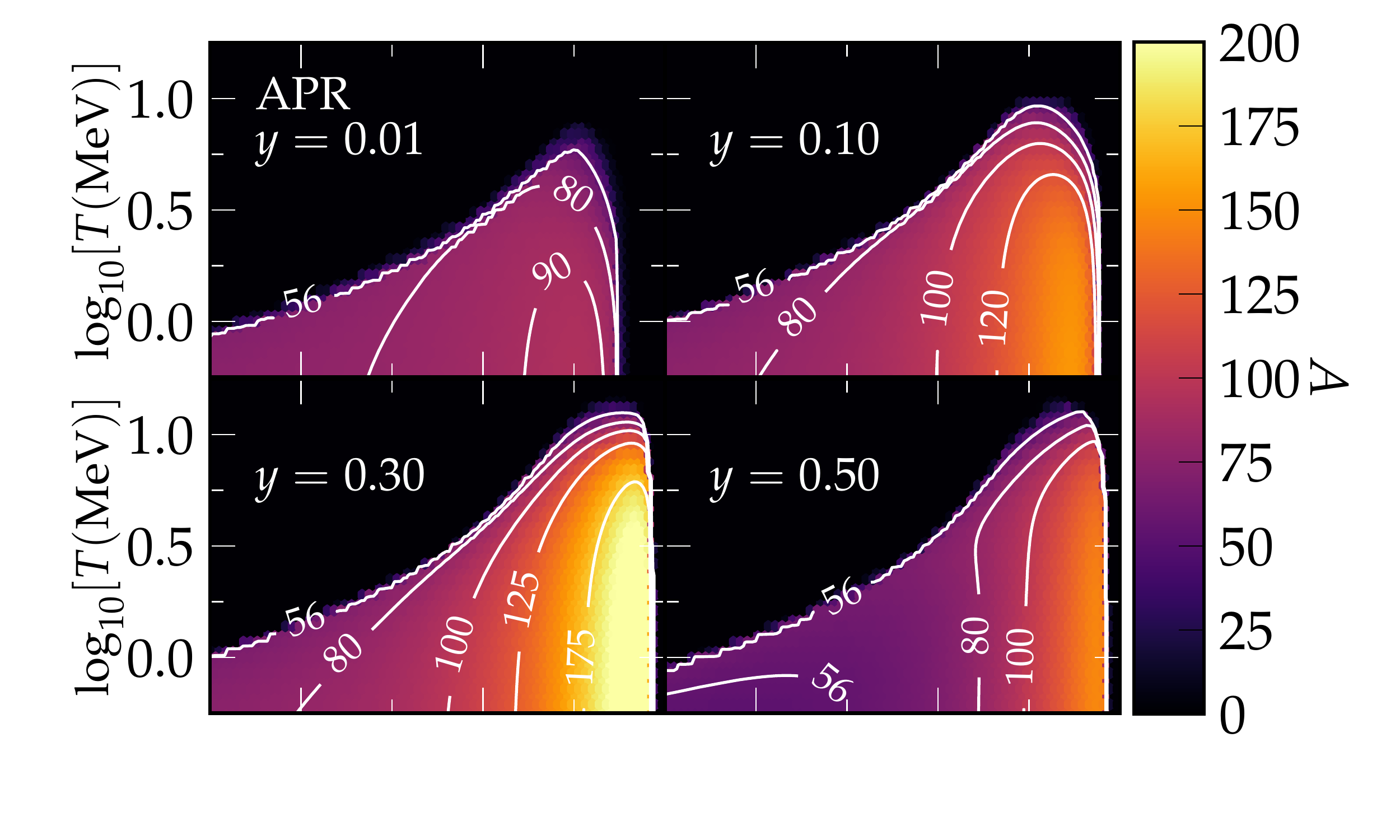}
\includegraphics[trim=0 50 0 0, clip, width = 0.50\textwidth]{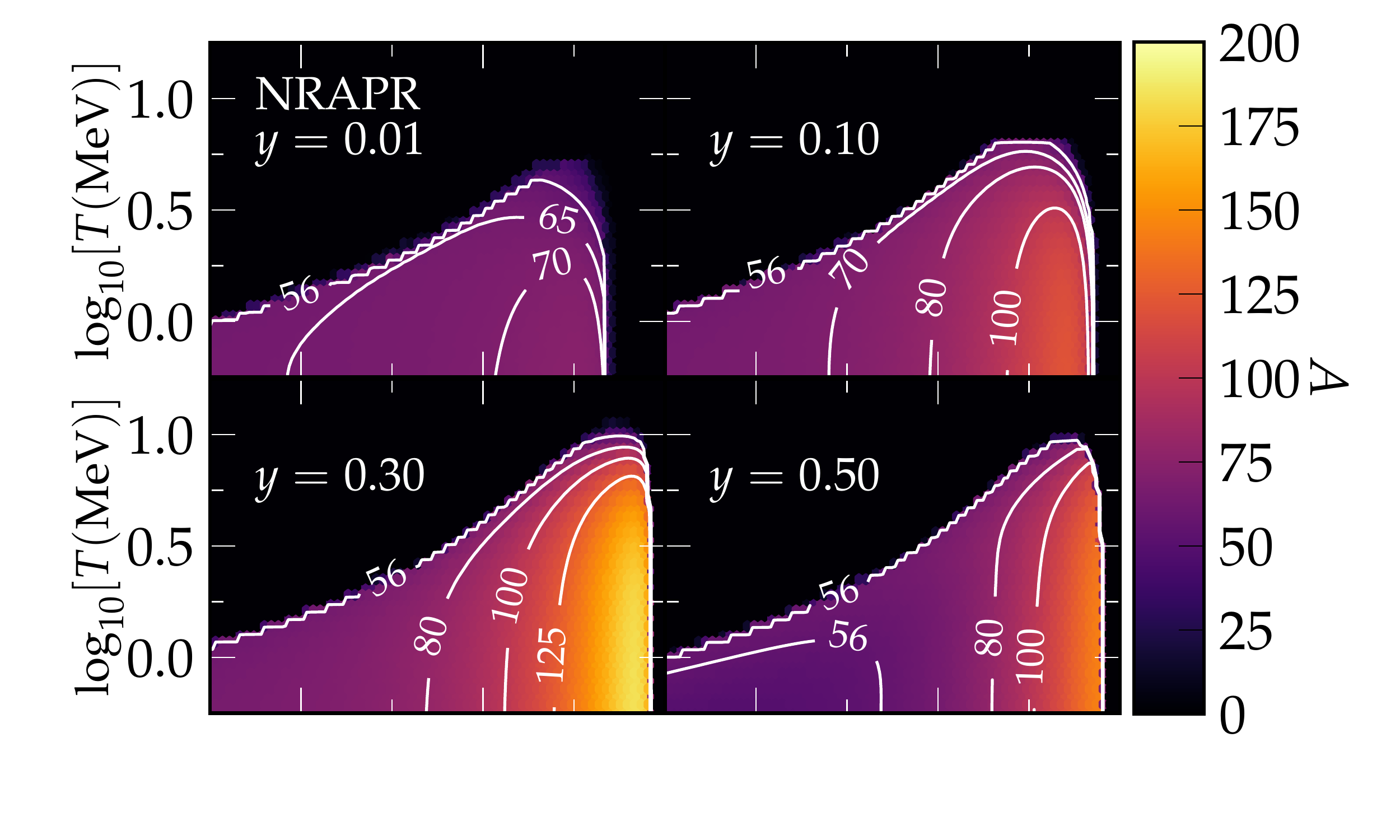}
\includegraphics[trim=0  0 0 0, clip, width = 
0.50\textwidth]{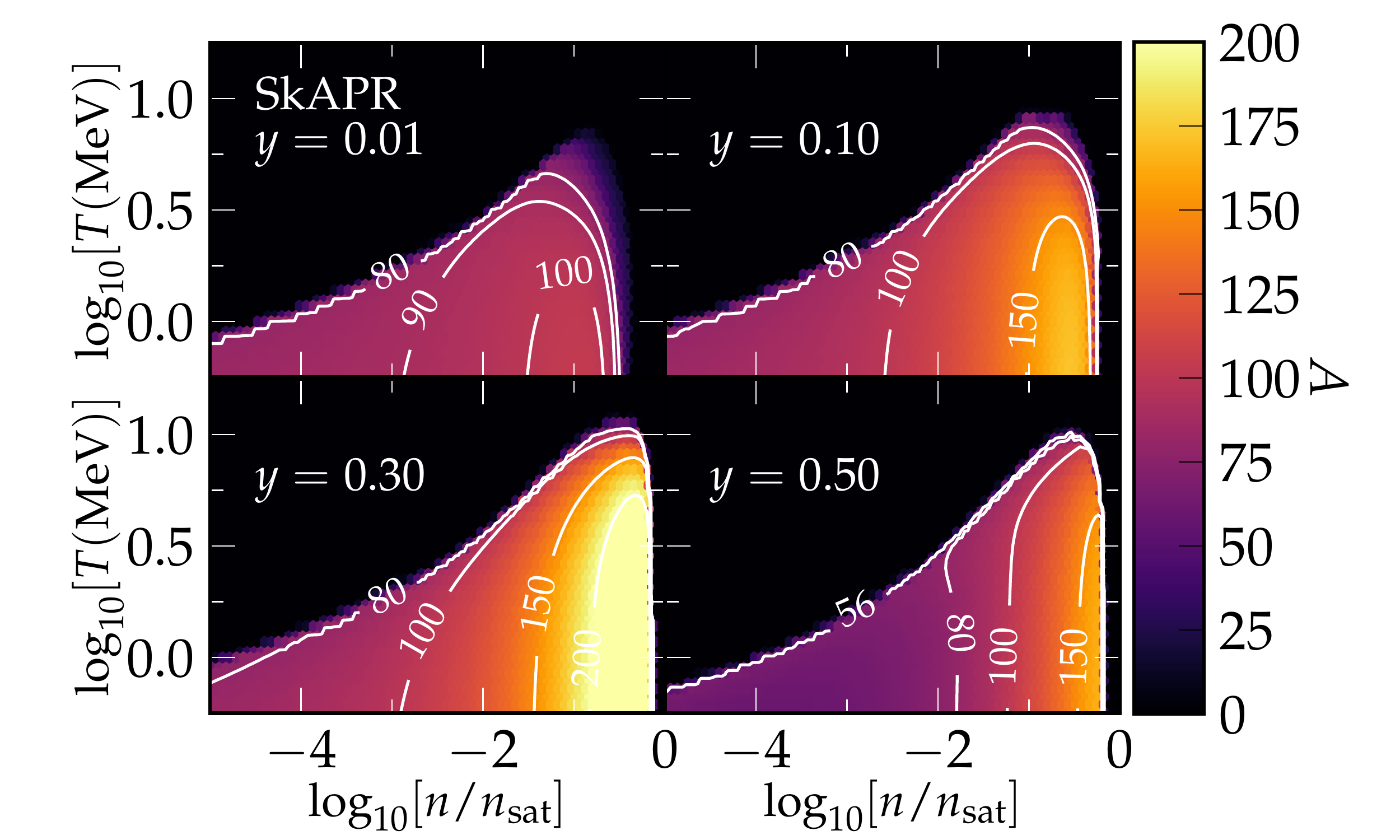}
\caption{\label{fig:A} From top to bottom: sizes of nuclei for the APR (top), NRAPR (center), SkAPR (bottom) EOSs for proton fraction $y=0.01,\,0.10,\,0.30$ and $0.50$.}
\end{figure}

In Fig. \ref{fig:A} we plot the mass numbers of nuclei for the different EOSs in the temperature and density plane. 
As was shown in Figs. \ref{fig:u_APR}, \ref{fig:u_NRAPR}, and \ref{fig:u_SkAPR}, the SkAPR EOS predicts the most massive nuclei, while APR often predicts higher melting temperatures for heavy nuclei and that nuclei survive up to larger isospin asymmetries. 
The black area represents regions where nuclear matter is in a uniform phase.

\subsection{High Density Phase Transition in APR}
\label{ssec:APR_phases}

The EOSs of AP, and thus APR, predicts that at high densities there is a phase transition from pure nucleonic matter to a phase that includes nucleons and a neutral pion condensation. 
In the APR formalism, this phase transition is taken into account by including extra potential terms in the high density phase (HDP) compared to the low density phase (LDP), see Eqs. \eqref{eq:ldp} and \eqref{eq:hdp}. 
The extra terms in the HDP soften the EOS at high densities and cause a discontinuity in the pressure and chemical potentials of the EOS of APR. 
In a self-consistent EOS for astrophysical simulations there must be no pressure discontinuities as well as no points where $dP/dn\vert_T<0$. 
To avoid such regions, we perform a Maxwell construction in the manner described in Sec. VI of Ref. \cite{constantinou:14}.
This results in a mixed phase for densities near $n\sim n_\tr(y)$, see Eq. \eqref{eq:nt}.

\begin{figure}[!h]
\centering
\includegraphics[trim=0 50 0 0, clip, width = 0.50\textwidth]{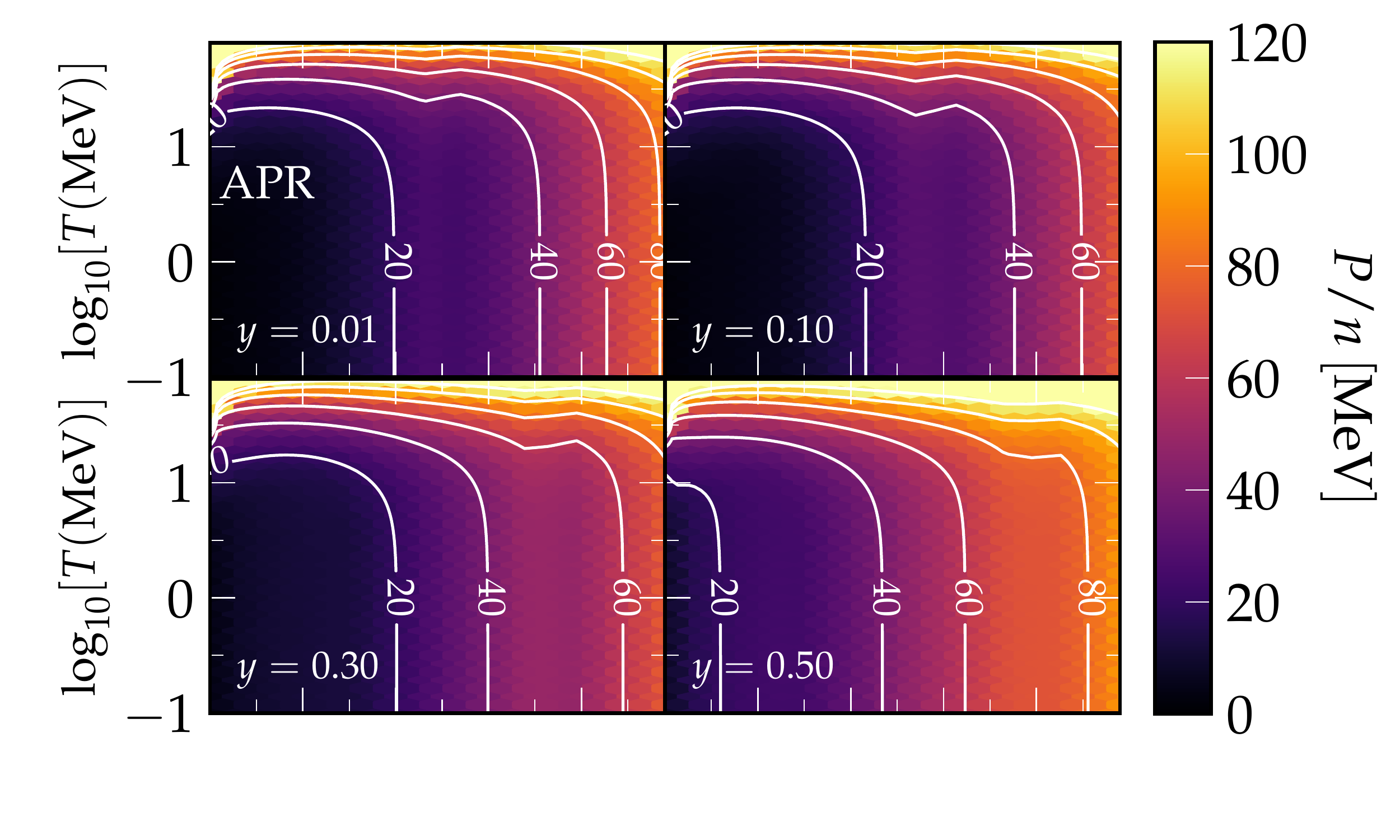}
\includegraphics[trim=0  0 0 0, clip, width = 0.50\textwidth]{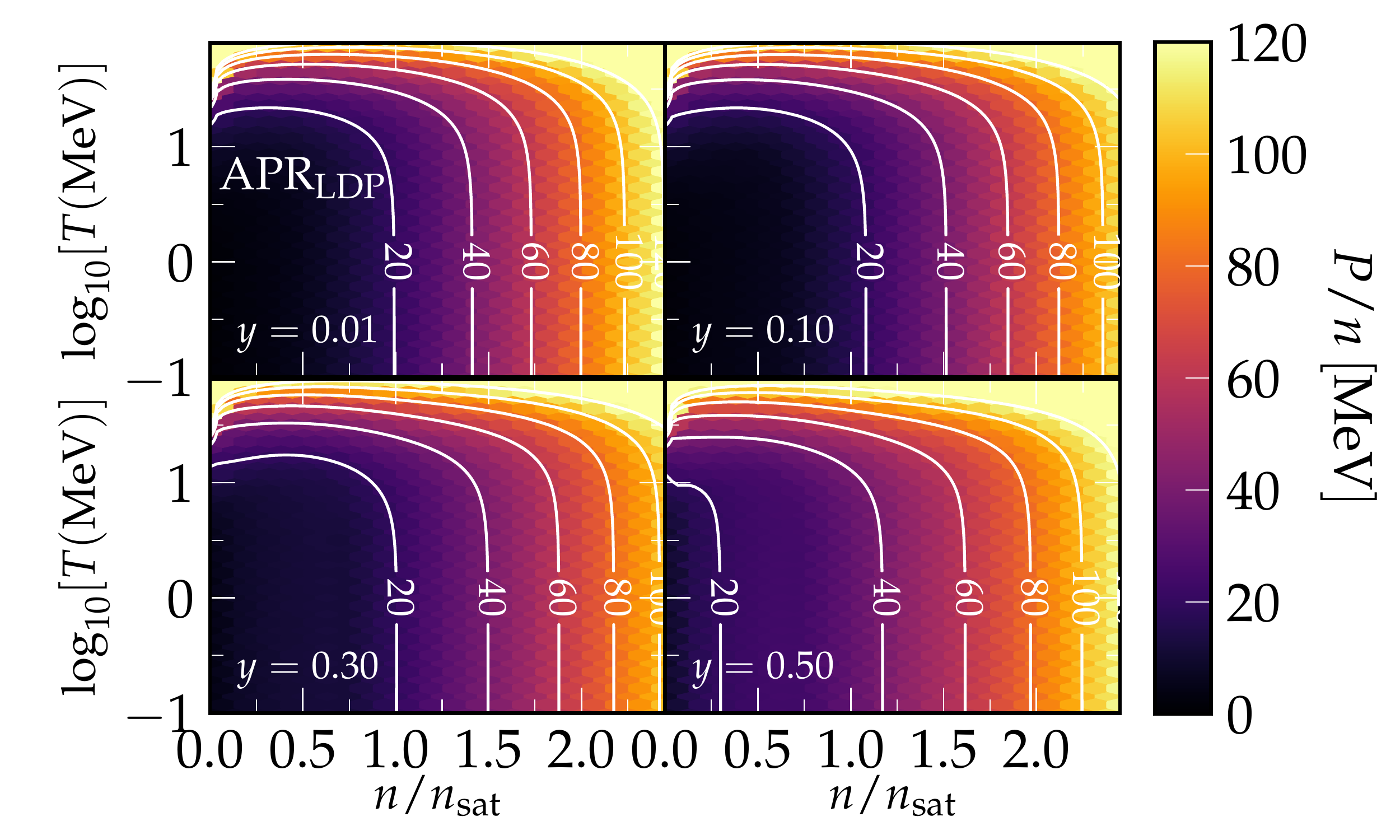}
\caption{\label{fig:APR_P} Pressure per baryon $P/n$ for APR EOS (top) and APRLDP (bottom) for proton fraction $y=0.01,\,0.10,\,0.30$ and $0.50$.}
\end{figure}

In Fig. \ref{fig:APR_P}, we compare the pressure per baryon $P/n$ of the APR EOS with its variant that only includes the stiffer LDP, APR$_{\rLDP}$. 
In regions of phase space near $n\simeq1.3n_\rsat$ for almost PNM, $y=0.01$, to $n\simeq2n_\rsat$ for SNM, $y=0.50$, the pressure per baryon remains constant as the baryon number density of the system increases at constant temperature. 
This is the region where our Maxwell construction finds a mixed phase of LDP and HDP, see also Fig. 32 of Constantinou \etal \cite{constantinou:14} for how the mixed phase changes with proton fraction and temperature. 
Notice that no such region exists for the APR$_{\rLDP}$ EOS and the pressure.

\begin{figure}[!h]
\centering
\includegraphics[trim=0 50 0 0, clip, width = 0.50\textwidth]{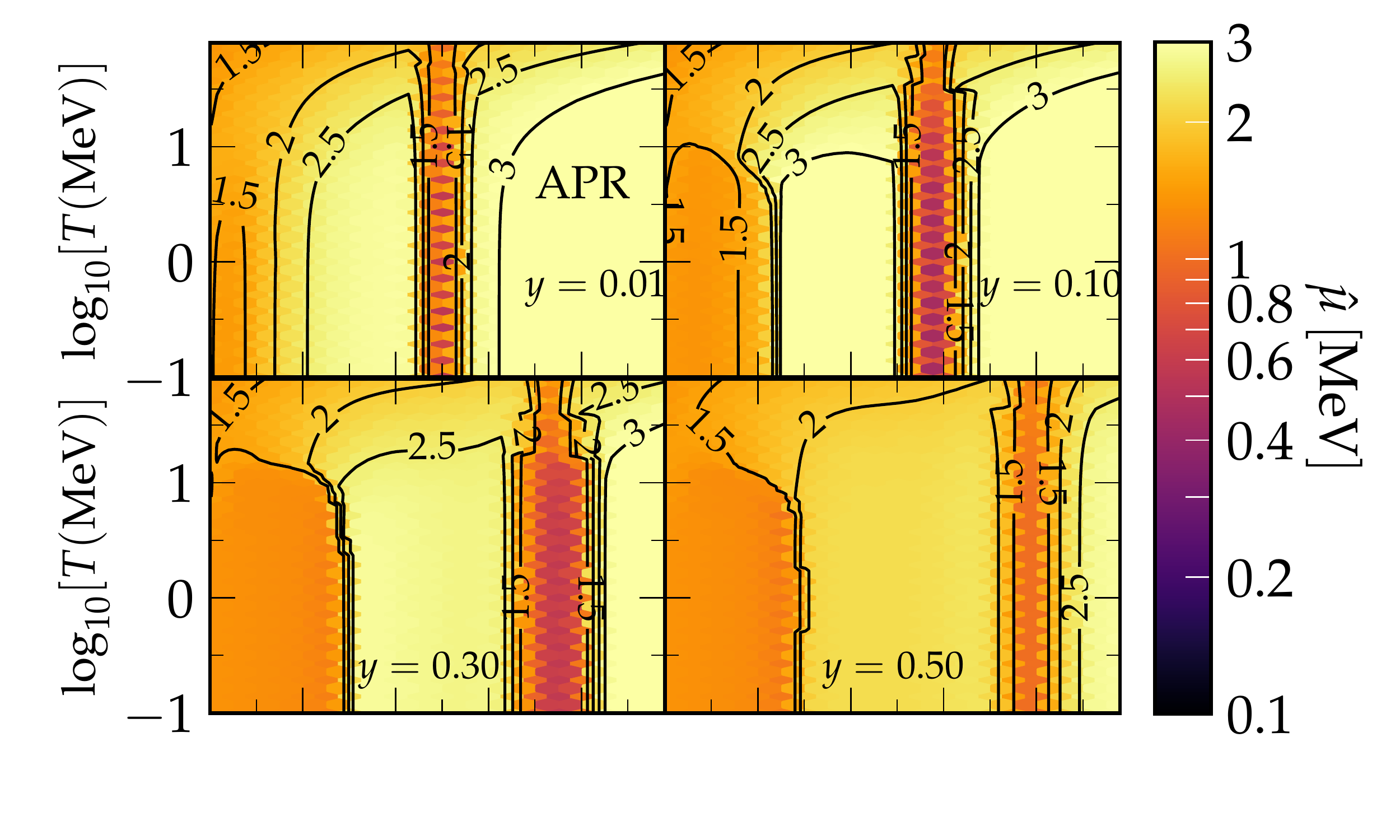}
\includegraphics[trim=0  0 0 0, clip, width = 0.50\textwidth]{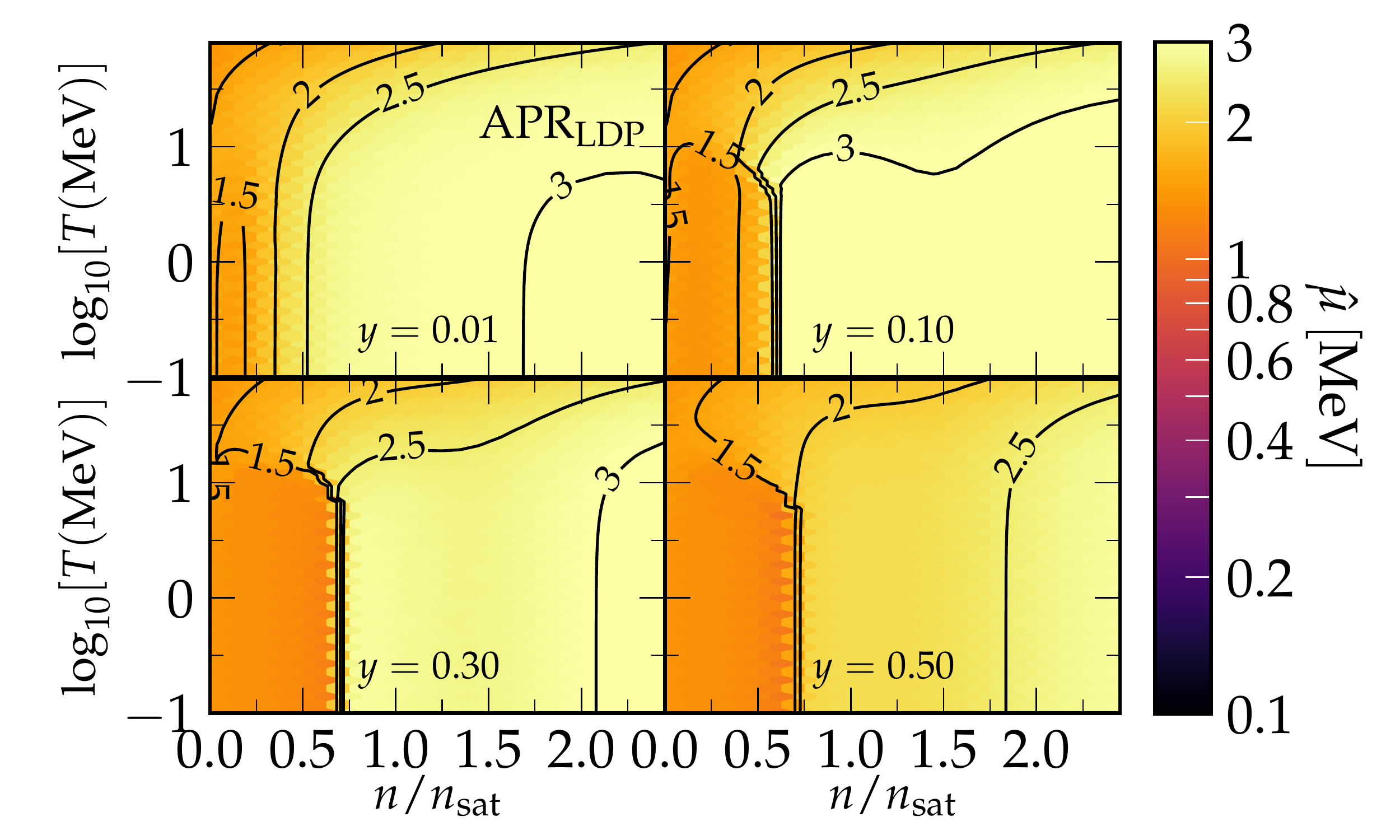}
\caption{\label{fig:APR_muh} Chemical potential splitting $\hat\mu=\mu_n-\mu_p$ for APR EOS (top) and APRLDP (bottom) for proton fraction $y=0.01,\,0.10,\,0.30$ and $0.50$.}
\end{figure}

We  show the chemical potential splitting $\hat\mu=\mu_n-\mu_p$ in Fig. \ref{fig:APR_muh}. 
Comparing the EOSs of APR and APR$_\rLDP$, we observe that $\hat\mu$ exhibits a sharp drop of about $1$ to $2\unit{MeV}$ in the mixed phase region.

\section{Core-collapse Supernovae}
\label{sec:ccsn}

We have carried out a set of example core-collapse and post bounce core-collapse 
supernovae (CCSNe) simulations in spherical symmetry. 
We investigated how the new APR EOSs compare to the Skyrme EOSs in this important 
astrophysical scenario and discuss the influence of the EOS on core-collapse post bounce 
evolution and black hole formation. 
For these simulations, the EOS for the low density phase below $10^{-3}~{\rm fm}^{-3}$ used was that of 3,335 nuclei in NSE. 
The match between the NSE and the single nucleus approximation (SNA) EOSs was 
performed using the simple merge function described in Sec. VII of SRO 
\cite{schneider:17} with the parameters $n_t=10^{-3}\unit{fm}^{-3}$ and 
$n_\delta=0.33$.

The CCSNe simulations were performed employing the open-source 
spherically-symmetric (1D) general-relativistic hydrodynamics code \texttt{GR1D} 
\cite{oconnor:10,oconnor:11,oconnor:13,oconnor:15}.  
Unlike in the SRO paper, we treat neutrino transport using the two-moment 
neutrino transport solver. 
This is achieved using the NuLib neutrino transport library which builds a 
database of energy dependent multi-species M1 neutrino transport properties 
\cite{oconnor:15}. 
We consider three neutrino species: $\nu_e$, $\nu_{\bar{e}}$, and $\nu_x 
=\nu_\mu=\nu_{\bar{\mu}} =\nu_\tau=\nu_{\bar{\tau}}$. 
The energy grid for each neutrino type has 24 logarithmically spaced groups. 
The first group is centered at 1\unit{MeV} and has a width of 2\unit{MeV}. 
The last group is centered at $\sim269\unit{MeV}$ and has a width of $\sim35\unit{MeV}$.

We simulated the core-collapse and post bounce evolution of two 
progenitors:
(1) a 15$M_\odot$ progenitor of Woosley, Heger, and Weaver \cite{woosley:02} 
and 
(2) a 40$M_\odot$ progenitor of Woosley and Heger \cite{woosley:07}.
While the former is expected to explode as a SN, at least in multi dimensional 
simulations, and leave a neutron star remnant the latter is very massive and has 
a high-compactness which favors  black hole (BH) formation \cite{oconnor:11}. 
For both progenitors we used a computational grid with $1\,500$ grid cells, 
constant cell size of $100\,\mathrm{m}$ out to a radius of $20\,\mathrm{km}$, 
and then geometrically increasing cell size to an outer radius of 
$10\,000\,\mathrm{km}$.

Stellar evolution codes, such as the ones that generate the two progenitors in 
our simulations, use reaction networks and, thus, the pre-collapse relationship 
between thermodynamical variables can differ substantially from the ones in the 
EOSs used in CCSN simulations. 
To start our simulations in a way that is as consistent as possible with the 
hydrodynamical structure of the progenitor models, we map the stellar rest-mass 
density $\rho$, proton fraction $y$, and pressure $P$ to \texttt{GR1D}, and then 
find the temperature $T$, specific internal energy $\epsilon$, entropy $s$, etc.,  
using our EOS tables. 
This approach for setting up the initial conditions results in differences 
between the original stellar profile and the \texttt{GR1D} initial conditions in 
all quantities except $\rho$, $y$, and $P$. 
This treatment differs from most CCSNe simulations which match 
$\rho$, $y$, and $T$ between pre-supernova progenitors and the core-collapse 
simulation.

\subsubsection{15$M_\odot$ Progenitor}
\label{ssec:15M}

We followed the collapse and post-bounce evolution of the $15M_\odot$ 
progenitor up to $1.0\unit{s}$ after bounce. 
Stars with such mass are expected to explode in nature and do so in some multi 
dimensional simulations \cite{yakunin:10,ott:18}, albeit for different 
pre-supernova progenitor models \cite{woosley:95, woosley:07}.  
However, we do not observe explosions in our \texttt{GR1D} simulations, which 
is consistent with other 1D simulations for this progenitor \cite{oconnor:10}.

\begin{figure}[htb]
\centering
\includegraphics[width=0.50\textwidth]{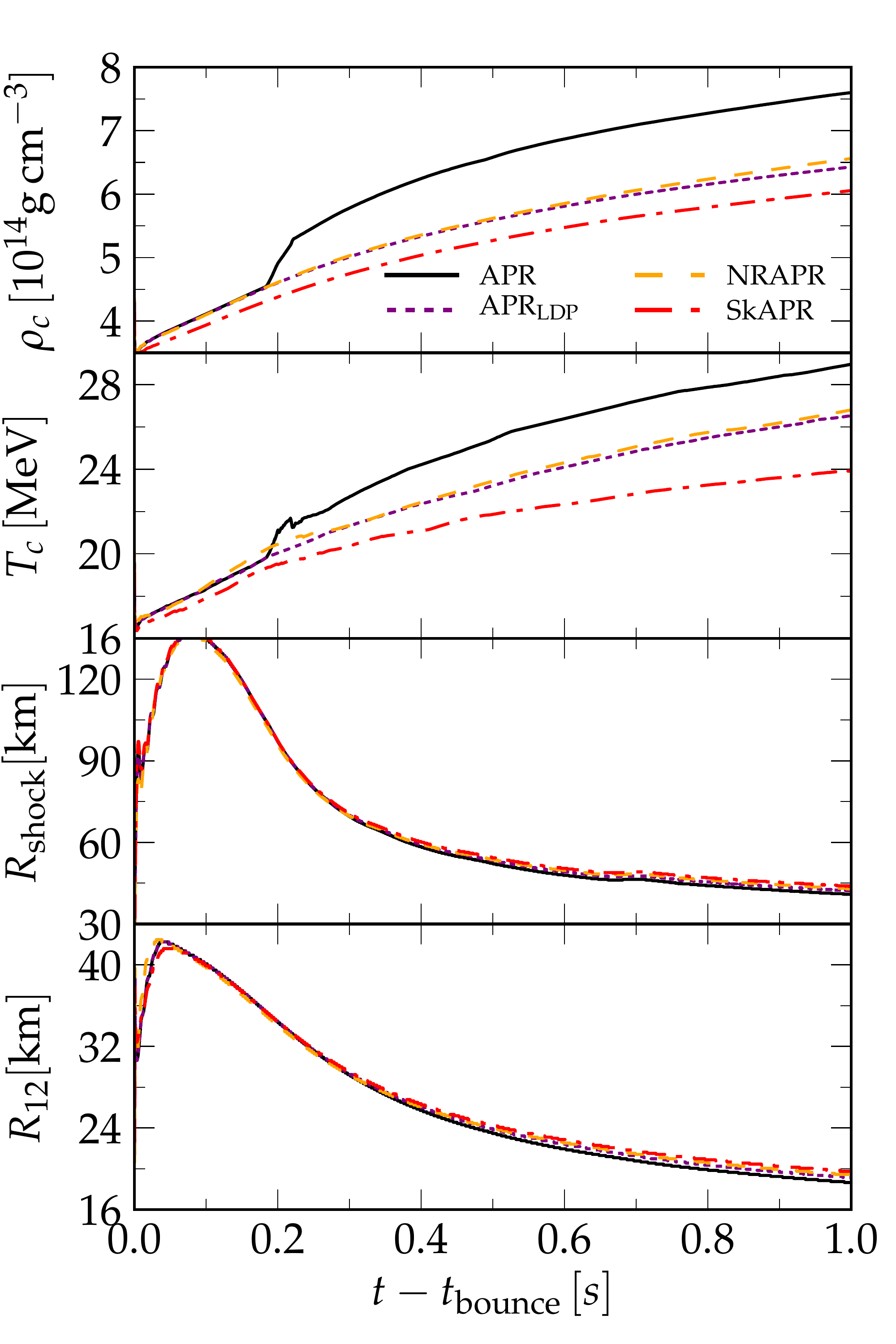}
\caption{\label{fig:15M_c} (Color online) From top to bottom: central density 
$\rho_c$, central temperature $T_c$, shock radius $R_{\rm{shock}}$, and PNS 
radius $R_{12}$ for the core-collapse of the $15\,M_\odot$ pre-supernova 
progenitor of Woosley, Heger, and Weaver \cite{woosley:02}.}
\end{figure}

In Fig. \ref{fig:15M_c}, we plot the central density and temperature temperature  as a 
function of time after bounce $t_{\rm bounce}=0.351\unit{s}$ as well as the 
shock radius and neutron star (NS) radius defined as the radius where the 
density is $\rho=10^{12}\unit{g\,cm}^{-3}$. 
We observe significant differences in the core density $\rho_c$ and its 
temperature $T_c$ between the APR EOS and its version APR$_\rLDP$ without the high density 
transition. 
Also, results for the NRAPR EOS of Steiner \etal \cite{steiner:05} agree better 
with those obtained using the APR$_\rLDP$ EOS than those from the SkAPR EOS. 
This happens even though the properties of SkAPR near saturation density match 
more closely those of APR$_\rLDP$ than NRAPR does, see Tab. \ref{tab:obs}, 
implying there is a trade-off between the different approaches to the EOS and 
exactly matching their observables.

\begin{figure*}[htb]
\centering
\includegraphics[width=0.9\textwidth]{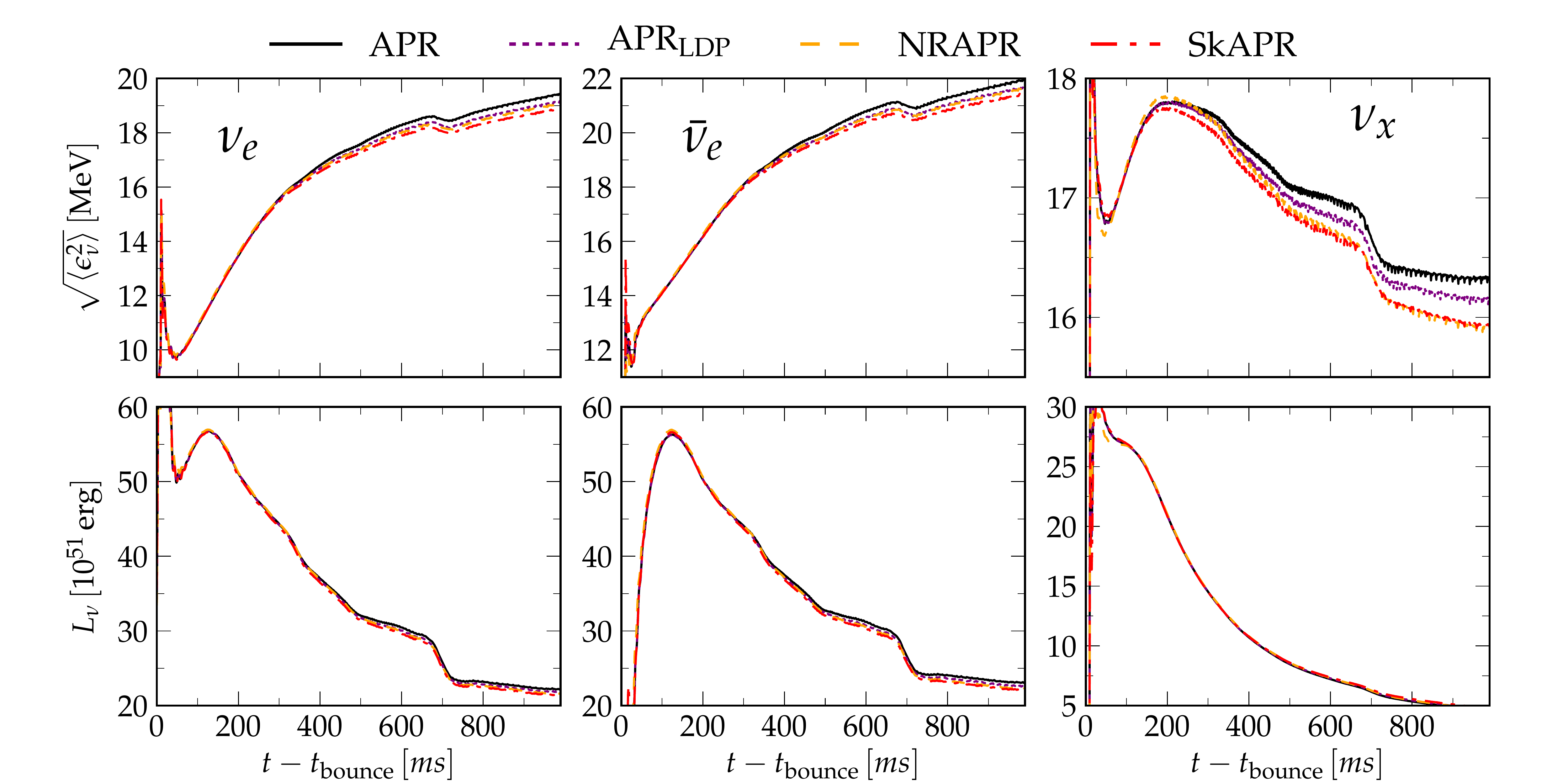}
\caption{\label{fig:15M_nu} (Color online) Root mean square energy 
$\sqrt{\langle\epsilon^2_\nu\rangle}$ (top) and luminosity $L_\nu$ (bottom) 
for electron neutrinos $\nu_e$ (left), electron anti-neutrinos $\nu_{\bar{e}}$ 
(center), and one of heavy neutrinos $\nu_x$ (right)  for the core-collapse 
of the $15\,M_\odot$ pre-supernova progenitor of Woosley, Heger, and Weaver 
\cite{woosley:02}. The results for the heavy neutrinos $\nu_x$ have been 
divided by 4 as it includes four neutrino types. We have added $2\times10^{52}\unit{erg}$ to the 
luminosity $L_{\nu_x}$  so that it fits with the same 
scale as the luminosity for the other two species.}
\end{figure*}

We see a shift in both the density and temperature at the core of the PNS once the 
core density is above the region where the pion condensate appears according to the EOS model.
However, the outer regions of the PNS and the shock front are only weakly 
affected by the phase transition. 
Although the PNS and shock radius contract faster due to the high density 
transition, this change is only of order a few \%. 
Furthermore, for this model and the APR EOS, we do not see a second spike in the 
neutrino signal triggered by the phase transition, as reported  by Sagert 
\etal for the case  a hadron-to-quark matter phase 
transition \cite{sagert:09}, see Fig. \ref{fig:15M_nu}.

The neutrino spectra, root mean square $\sqrt{\langle\epsilon^2_\nu\rangle}$ and 
luminosity $L_\nu$, for the different neutrino species are shown in Fig. \ref{fig:15M_nu}.
We note that at the time the core densities are large enough that there is a 
phase transition in the APR EOS, there is a short contraction in the PNS 
and shock radii. 
This contraction heats up slightly the neutrino-sphere and increases the energy 
and luminosity of neutrinos emitted. 
Nevertheless, this change is only of a few \% and of the same order as changes seen
between the two different Skyrme EOSs, NRAPR and SkAPR, for which observables 
such as the incompressibility $K_\rsat$ changed by a large amount.

\subsubsection{40$M_\odot$ Progenitor}
\label{ssec:40M}

We now follow the core-collapse and post-bounce evolution of the $40M_\odot$ 
progenitor of Woosley and Heger \cite{woosley:07} until a black hole (BH) 
forms. 
This progenitor is one of the many studied by O'Connor and Ott 
\cite{oconnor:11} using a neutrino leakage scheme transport and four different 
EOSs, the three Lattimer and Swesty (LS) variants \cite{lattimer:91} and the 
Shen EOS with the TM1 parametrization \cite{shen:98b}. 
O'Connor and Ott observed that larger incompressibilities lead to a faster 
collapse to BH, although effects of the effective mass, which are important 
for the temperature dependence of the EOS \cite{steiner:13, constantinou:14, 
schneider:18}, on the different BH formation time and its initial mass were not 
disentangled. 
The three LS EOSs, with incompressibility $K_\rsat=180$, $220$, and 
$375\unit{MeV}$ all have effective masses set to the nucleon vacuum mass. 
The Shen TM1 EOS has $K_\rsat=280\unit{MeV}$ and predicts an 
effective mass for symmetric nuclear matter at nuclear saturation density 
$m^\star=0.63m_n$. 
All else being equal for the zero temperature properties of nuclear matter, a lower 
effective mass will lead to higher thermal pressure and a slower collapse to 
BH \cite{schneider:18}.

\begin{figure}[htb]
\centering
\includegraphics[width=0.50\textwidth]{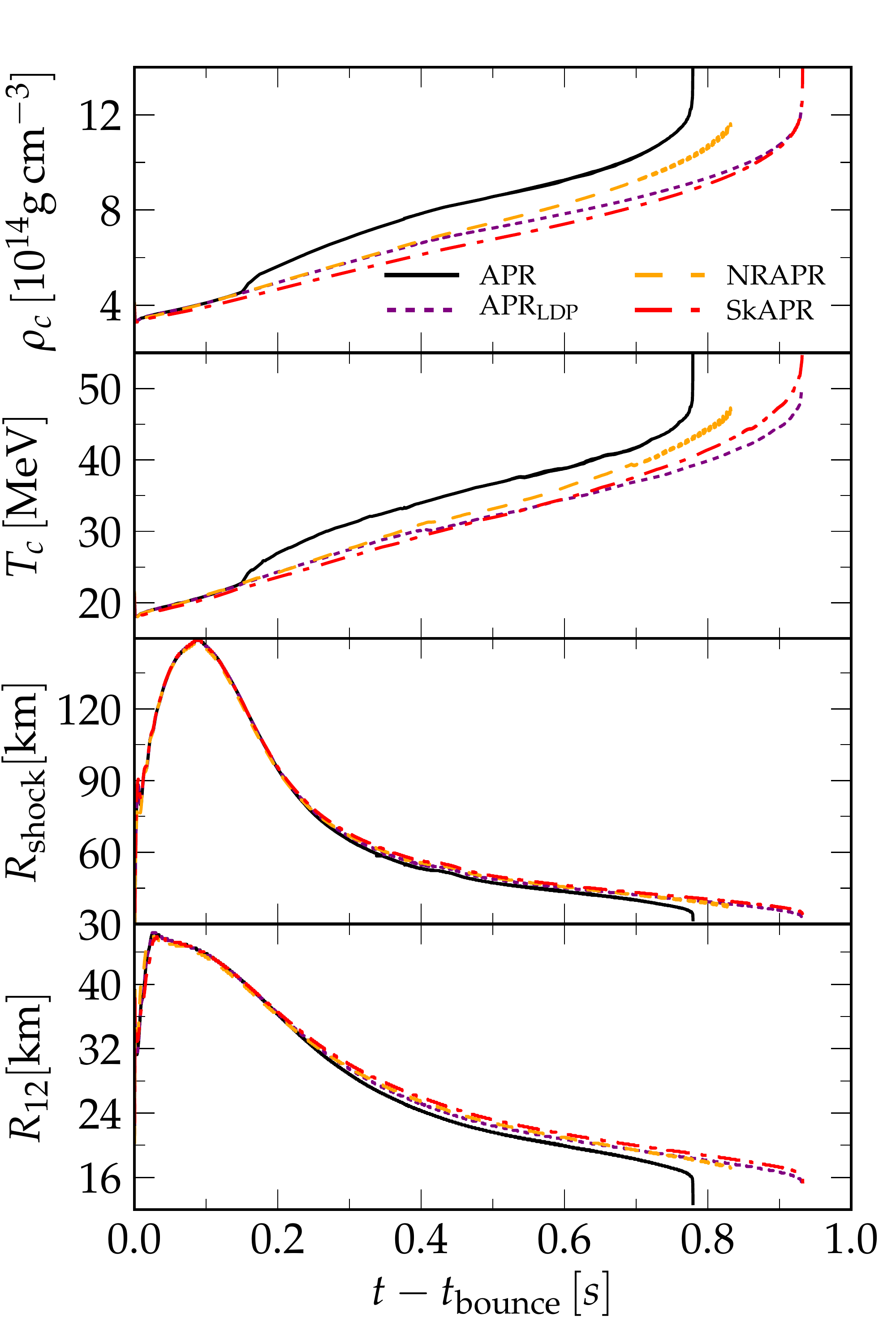}
\caption{\label{fig:40M_c} (Color online). Same as Fig. \ref{fig:15M_c}, but for the 40$M_\odot$ pre-supernova progenitor 
of Wooseley and Heger \cite{woosley:07}.}
\end{figure}

In the four EOSs studied here, the effective masses all have  very similar 
values at the saturation density, see Tab. \ref{tab:obs}. 
However, the effective masses for the Skyrme EOSs decrease faster at higher 
densities than for the APR EOSs, Fig. \ref{fig:meff}. 
Other main differences between these EOSs are the lower incompressibility $K_\rsat$ 
for NRAPR and the high density transition in APR. 
Thus, we expect that using the APR EOS will lead to a faster collapse to BH than 
for the other EOSs, due to the sharp phase transition discussed in Sec. 
\ref{sec:apr_model}. 
This is indeed the case as seen in Fig. \ref{fig:40M_c}.
We also expect the NRAPR EOS to predict a faster collapse than SkAPR and 
APR$_\rLDP$ due to its lower incompressibility $K_\rsat$.
This feature is also observed. 
However, it is difficult to predict which of SkAPR or APR$_\rLDP$ will 
take the longest to collapse. 
This is due to a possible trade-off between the slightly higher 
(lower) pressures for the SkAPR EOS than for the APR$_\rLDP$ EOS for 
$n\lesssim2n_\rsat$ ($n\gtrsim2n_\rsat$) and its lower nucleon effective 
masses at densities $n\gtrsim2n_\rsat$. 
In fact, what we observe is that near $500\unit{ms}$ after bounce SkAPR EOS 
predicts lower densities and temperatures at the core of the PNS than the 
APR$_\rLDP$ EOS. 
At that time, the density at the core is approximately $2.5n_\rsat$, a region 
where the effective mass for the SkAPR EOS has deviated from  its 
APR$_\rLDP$ counterpart. 
From then on the core temperature computed with the SkAPR is slightly higher 
than that for the APR$_\rLDP$ EOS. 
However, in the same region the pressure obtained with the APR$_\rLDP$ EOS is 
slightly higher. 
The competition between both effects leads to both EOSs predicting an almost 
identical collapse time to BH, Tab. \ref{tab:40M}.
We also see, as observed by O'Connor and Ott, that there is a correlation 
between the time to collapse into a BH and its initial mass. 
This is due to the accretion rate being only dependent on the low density part 
of the EOS, which was set as the same for all four EOSs.

\begin{table}[htbp]
\caption{\label{tab:40M} Black hole formation times and their gravitational 
mass at the time of collapse. The time to bounce for all EOSs is 
$t_{\rm{bounce}}=0.472\unit{s}$.}
\begin{ruledtabular}
\begin{tabular}{l | c c c}
\multicolumn{1}{c}{EOS} &
\multicolumn{1}{c}{$t_{\rm{BH}}-t_{\rm{bounce}}$} &
\multicolumn{1}{c}{$t_{\rm{BH}}$} &
\multicolumn{1}{c}{$M_{\rm{grav}}\,[M_\odot]$} \\
\hline
APR         & 1.252 & 0.780 & 2.580 \\
NRAPR       & 1.304 & 0.832 & 2.611 \\
APR$_\rLDP$ & 1.403 & 0.931 & 2.670 \\
SkAPR       & 1.405 & 0.933 & 2.672 \\
\hline
\end{tabular}
\end{ruledtabular}
\end{table}

\begin{figure*}[htb]
\centering
\includegraphics[width=0.9\textwidth]{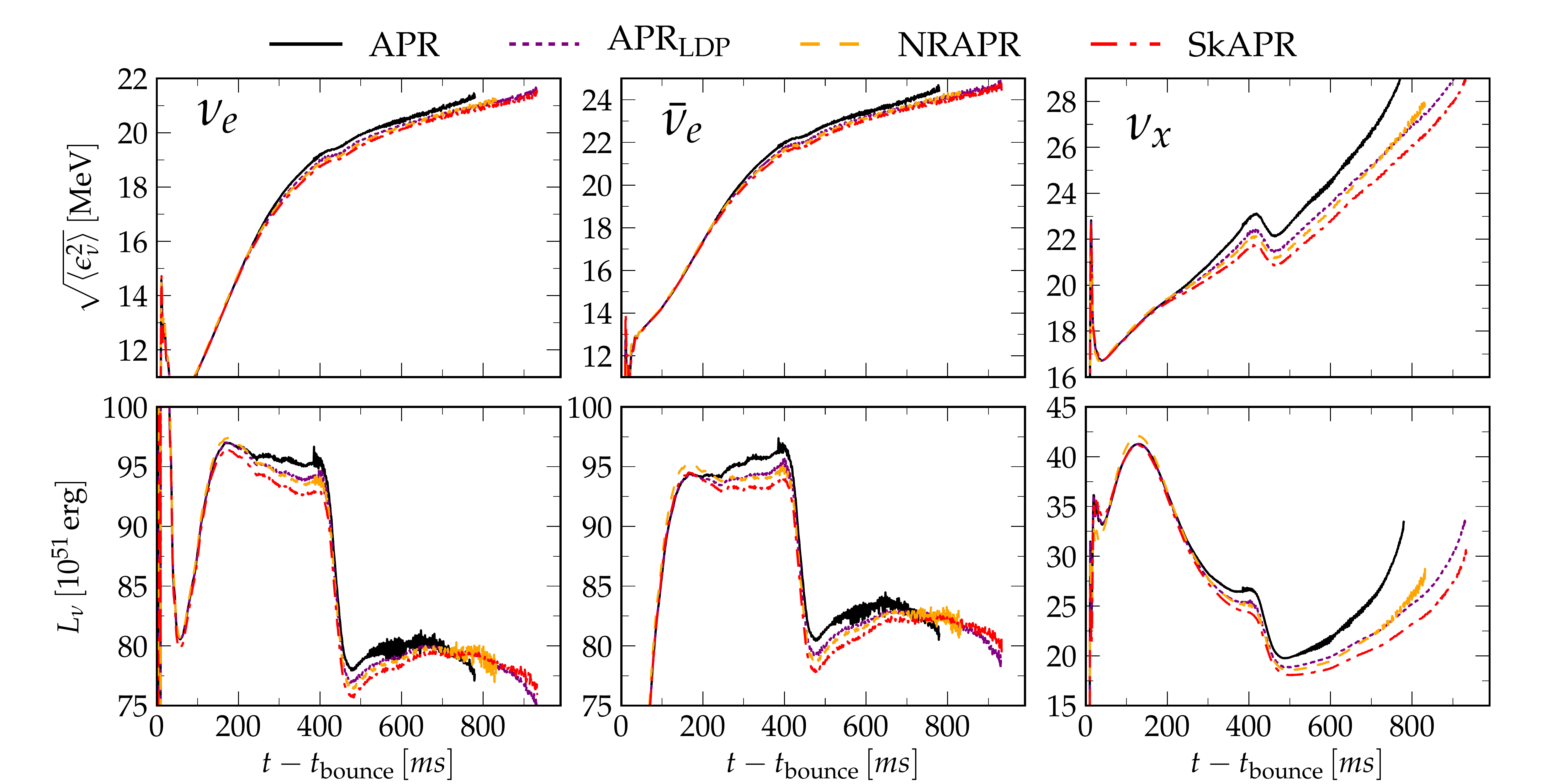}
\caption{\label{fig:40M_nu} (Color online) Root mean square energy 
$\sqrt{\langle\epsilon^2_\nu\rangle}$ (top) and luminosity $L_\nu$ (bottom) 
for electron neutrinos $\nu_e$ (left), electron anti-neutrinos $\nu_{\bar{e}}$ 
(center), and one of heavy neutrinos $\nu_x$ (right)  for the core-collapse 
of the $40\,M_\odot$ pre-supernova progenitor of Woosley and Heger
\cite{woosley:07}. The results for the heavy neutrinos $\nu_x$ have been 
divided by 4 as it includes four neutrino types. We have added $6\times10^{52}\unit{erg}$ to the 
luminosity $L_{\nu_x}$  so that it fits with the same 
scale as the luminosity for the other two species.}
\end{figure*}

As for the $15\,M_\odot$ case, differences in the inner regions of the PNS do 
not lead to significant changes in either shock or the PNS radius, bottom panels of 
Fig. \ref{fig:40M_nu}. 
However, both neutrino energies and luminosities, especially for heavy $\nu_x$ 
neutrinos, are enhanced for the EOSs that predict faster collapse to BH, Fig. 
\ref{fig:40M_nu}. 
Another feature of the neutrino spectrum is the sharp decrease in the luminosity 
for all four EOSs and neutrino species near $400\unit{ms}$ after bounce. 
This is due to the rapid change in the accretion rate as the density discontinuity 
of the Si/Si-O shell of the star passes the stalled shock front, see Fig. 4 of 
O'Connor and Ott \cite{oconnor:11}. 
For  3D simulations, Ott \etal have shown that the high neutrino luminosities 
and energies lead to a shock explosion even before the Si/Si-O shell crosses 
the shock radius \cite{ott:18}, although the hot PNS left behind is massive 
enough that it will subside into a BH once it cools down.
Unlike for the lower mass progenitor studied here, the neutrino 
luminosities show significant differences at late times due to the phase 
transition present in the APR EOS. 
Thus, it is likely that in multi-dimensional simulations the phase transition 
in the APR EOS leads to faster shock revival and expansion. 
Such a future study is indicated by results of this work.

\section{Summary and Conclusions}
\label{sec:conclusions}

Our primary objective in this work has been to build an equation of state (EOS) for simulations of supernovae, neutron stars and binary mergers based on the Akmanl, Pandharipand and Ravenhall (APR) Hamiltonian density devised to reproduce the results of the microscopic potential model calculations of Akmal and Pandharipande (AP) for nucleonic matter with varying isospin asymmetry.   
Toward this end, 
we have developed a code that takes advantage of the structure of the SRO EOS code which was devised to compute EOSs for Skyrme parametrizations of the nuclear force \cite{schneider:17}.
Here, the SRO EOS code was adapted to compute EOSs using the more intricate APR potentials \cite{akmal:98}.  
The APR potential has some distinct differences compared  to Skyrme-type potentials. 
Skyrme parameters are fit to reproduce properties of finite nuclei or empirical parameters of the expansion of energy density of nuclear matter around saturation density.
In contrast, APR has been fit to reproduce results of  variational calculations based on a microscopic potential model  for both symmetric nuclear matter (SNM) and pure neutron matter (PNM). 
These variational calculations include two- and three-body interactions as well as relativistic boost corrections. 
Furthermore, APR contains a phase transition to a neutral pion condensate that softens the EOS at high densities while still predicting cold beta-equilibrated neutron star  (NS) masses  and radii in agreement with current observations \cite{antoniadis:13, fonseca:16, most:18, nattila:16, soumi:18}.

In addition to the APR EOS,  we have developed three other EOSs: (1)  APR$_\rLDP$, an APR variant which does not include a transition to a neutral pion condensate at high densities, (2)  a finite temperature version of the non-relativistic APR model of Steiner \etaln, NRAPR \cite{steiner:05, schneider:17}, and (3) SkAPR \cite{schneider:18, margueron:18a}, 
a Skyrme-type version of APR computed with the SRO code which was fit to reproduce some of the properties.

In our calculations of the EOS of APR, we pay special attention to the surface properties of nuclear matter and its inhomogeneous phases. 
The APR model allows for more complex behavior of the effective masses of nucleons when compared to Skyrme EOSs. 
In addition, it allows for asymmetries between neutron-neutron and proton-proton gradient terms in the surface component of the Hamiltonian which is generally not present in the commonly used Skyrme EOSs. 
This allows the APR EOS to predict non-uniform nuclear matter up to higher densities, temperatures, and lower proton fractions than the Skyrme-type EOSs allow.

Using the above four EOSs, we simulated spherically symmetric core-collapse of two massive stars, a $15M_\odot$ pre-supernova progenitor \cite{woosley:02} and a $40M_\odot$ pre-supernova progenitor \cite{woosley:07}. 
We followed the evolution of the $15M_\odot$ progenitor for one second after core-bounce and the formation of a proton NS (PNS). 
Although there are some significant differences observed  across EOSs for the inner configuration of the star, neither the outer regions of the collapsing star nor the neutrino spectra seem to be significantly affected by either the phase transition included in the APR EOS or by the Skyrme or APR description of the EOS. 
Besides the development of a new EOS, one of our main goals was to determine whether the phase transition that includes a high density neutral pion condensate alters the neutrino spectrum of a collapsing star and leads to a second peak in neutrino signal, as observed by Sagert \etal for the hadron-to-quark phase transition \cite{sagert:09}. 
Note that one of the  progenitors in Ref. \cite{sagert:09} is the same as the $15\,M_\odot$ pre-supernova progenitor used here. 
However, we do not observe a second burst in neutrino luminosity and root mean square energy in our simulation with the APR EOS. 
This difference between our result and that of Sagert \etal is attributed to the lack of a second shock wave traveling through the PNS that results from the transition from hadron-to-quark matter in Sagert \etaln.
The softening in the APR EOS due to the existence of a pion condensate is not as extreme as that of a transition from hadron-to-quark matter and, thus, no second shock wave forms and thus second peak in the neutrino signal is not observed. 
We recall that the phase transition in the APR EOS as treated in \cite{constantinou:14} is almost independent of temperature. 
Therefore, it is likely that the addition of a temperature dependent phase transition facilitates the formation of a second shock wave due to the large temperatures achieved in the inner regions of the PNS and due to the low proton fractions and even higher temperatures that exist in the PNS mantle.

The $40M_\odot$ progenitor evolution was followed  until black hole (BH) formation. 
In this case, the differences across EOSs affect the BH formation time and its initial mass. 
Particularly, the softening of the APR EOS due to its prediction of a neutral pion condensate at high densities facilitates the contraction of the PNS and, thus, speeds up the NS subsidence into a BH as well as lowers its initial mass and hardens the neutrino spectrum, especially for the heavier neutrinos. 
The other three EOSs predict similar evolutions and neutrino spectra until a BH forms, which happens earlier for NRAPR as it is the softest EOS at high densities. 
We expect  differences between the EOSs to be amplified in multi-dimensional simulations.

Directions for future work suggested by the first stage of the development of the EOS of APR performed here include (1) incorporating extensions of the excluded volume approach that includes $^2\rm{H}, ^3{\rm H}$ and $^3{\rm He}$ in addition $\alpha$-particles as in Ref. \cite{lalit:18}, (2) exploring consequences for PNS evolution and (3) performing simulations of binary mergers  of neutron stars.    
More than in the evolutions of core-collapse supernovae and proto-neutron stars, the evolution of the compact object following the merger is influenced by the dense matter EOS. This is because  higher densities and temperatures are achieved in the post-merger remnant than in the case of a SN or a PNS.  The possible outcomes for the compact object include a massive stable neutron star, a hyper massive neutron star that can collapse to a black hole owing to deleptonization through loss of trapped neutrinos and rigidization of rotation, or, a prompt black hole. Future generation gravity wave detectors can inform on the possible outcomes from post-merger signals. For the post-merger evolution, time evolving effects of rotation, magnetic fields and temperature also become crucially important.

\begin{acknowledgments}
 
We acknowledge helpful discussions with Jim Lattimer. 
A.\,S.\,S. was supported in part by the National Science Foundation under award No. AST-1333520 and CAREER PHY-1151197. 
C.C., B. M. and M. P. acknowledge  
research support from the U.S. DOE grant. No. DE-FG02-93ER-40756.  C. C. also acknowledges travel support from the National Science Foundation under award Nos. PHY-1430152 (JINA Center for the Evolution of the Elements).  
%List JINA grant and other funding. 
This work benefited from discussions at the 2018 INT-JINA Symposium on ``First multi-messenger observation of a neutron star merger and its implications for nuclear physics'' supported by the National Science Foundation under Grant No. PHY-1430152 (JINA Center for the Evolution of the Elements) as also from discussions at the 2018 N3AS collaboration meeting of the
``Research Hub for Fundamental Symmetries, Neutrinos, and Applications to Nuclear Astrophysics'' supported by the
National Science Foundation, Grant PHY-1630782, and the Heising-Simons Foundation, Grant 2017-228.

\end{acknowledgments}

\appendix

\section{Equilibrium conditions}
\label{app:free}

For the most part, we follow the scheme outlined by 
Lattimer and Swesty (LS) \cite{lattimer:91}  
to determine the set of equations that determines 
equilibrium between nucleons, electrons, positrons 
and photons. Departures from the LS approach will be noted 
as the discussion proceeds. Depending on the density, temperature and net electron fraction,  
nucleons can cluster into alpha particles (proxy for light nuclei) and into heavy nuclei, 
both of which are treated using an excluded volume approach.
The total free energy of the system is 
\begin{equation}\label{app_eq:F}
 F = F_o+F_\alpha+F_h+F_e+F_\gamma\,. 
\end{equation}
Terms on the right hand side above 
are the free energies of unbound nucleons outside of 
%(\textit{outside}  
alpha particles and  heavy nuclei, alpha particles, heavy nuclei, 
leptons, and photons, respectively. 
Leptons and photons are treated as non-interacting 
relativistic uniform gases. 
Their free energies and thermodynamic properties are standard, and 
computed using the Timmes and Arnett equation of state 
(EOS) \cite{timmes:99}.  The system as a whole is  in thermal 
equilibrium at a temperature $T$ and 
electrically neutral, \ie the lepton density 
$n_{e^-}-n_{e^+}$ and the proton density $n_p=ny$ ($n$ is the total baryon density
and $y$ is the proton fraction) are related by 
$n_p =n_{e^+}-n_{e^-}$.

Because leptons and photons are assumed to form a uniform background
and are non-interacting, their free energies do not interfere 
with the overall state of nucleons inside and out of nuclei. 
Thus, for a given nucleon number density $n$, 
proton fraction $y$, and temperature $T$ we compute the 
properties of nucleons that minimizes the free energy
of the system. 
Two types of system are possible: (1) uniform matter, 
which refers to a liquid of nucleons and alpha particles, 
and (2) non-uniform matter, which includes heavy nuclei. 
The system is assumed uniform unless its temperature is 
lower than the critical temperature $T\lesssim T_c$, 
and nucleon density lower than nuclear saturation density, $n<n_\rsat\simeq0.16\unit{fm}^{-3}$. 
In the latter cases, we solve for both uniform and non-uniform 
matter. 
If only one type of matter minimizes the free 
energy of the system, then that is set as its true solution. 
However, if both solutions are possible, then we set the 
true state of the system as the one with the lowest free energy. 
We update often the possibility of finding non-uniform 
matter based on previously found solutions.

In Appendices \ref{app:apr} to \ref{app:heavy}, 
we discuss the different terms in Eq. \eqref{app_eq:F}. 
Appendix \ref{app:uniform} contains a description how to compute the 
solution for uniform matter. 
Appendix \ref{app:nonuniform}  describes how 
the solution to non-uniform matter is obtained.

\section{Derivative notations}
\label{app:der}

To simplify the notation used throughout the Appendices, 
we define the density derivatives of functions 
$F\equiv F(n_n,n_p,T)$ with respect to a nucleon density $n_t$ 
keeping $n_{-t}$ fixed as 
\begin{align}\label{app_eq:partial_t}
 \partial_{n_t} F = 
 \left.\frac{\partial F(n_n,n_p,T)}{\partial n_t}\right|_{n_{-t},T}\,.
\end{align}
Note that if $t=n$, then $-t=p$ and vice versa. 
We often interchangeably use $F(n_n,n_p)=F(n,y)$ 
making the replacements $n_n=(1-y)n$ and $n_p=yn$ 
where the number density is $n=n_n+n_p$ 
and the proton fraction $y=n_p/n$.
In similar fashion, second derivatives are denoted by
\begin{align}\label{app_eq:partial_rt}
 \partial_{n_rn_t} F = 
 \left.\frac{\partial^2 F(n_n,n_p)}{\partial n_t \partial n_r}\right|_{n_{-t},n_{-r}}\,.
\end{align}
%

% If $F\equiv F(n_n,n_p,T)$ also depends on the temperature $T$,
% its temperature derivative is written as
% %
% \begin{align}\label{app_eq:partial_T}
%  \partial_T F=\left.\frac{\partial F(n_t,n_p,T)}{\partial T}\right|_{n_n,n_p}\,.
% \end{align}
% %

If $F'\equiv F'(\eta_n,\eta_p,T)$, 
the derivatives with respect to the degeneracy 
parameters $\eta_t$ are denoted by
\begin{align}\label{app_eq:partialp_t}
 \partial_{\eta_t} F' =
 \left.\frac{\partial F'(\eta_n,\eta_p,T)}{\partial \eta_t}\right|_{\eta_{-t},T}\,.
\end{align}
Whenever we take a temperature derivative and choose to 
keep the degeneracy parameters constant instead of the 
nucleon densities, we add a prime to the $\partial$ sign, 
\ie 
\begin{align}\label{app_eq:partialp_T}
 \partial'_T F' = 
 \left.\frac{\partial F'(\eta_t,\eta_p,T)}{\partial T}\right|_{\eta_n,\eta_p}\,.
\end{align}

We also switch between derivatives 
where a set of variables such as $\xi=(n_n,n_p,T)$ 
or $\xi'=(\eta_n,\eta_p,T)$ is used to derivatives
with respect to the independent variables $\zeta=(n,y,T)$. 
In the latter case, the transformation between 
derivatives is
\begin{subequations}
 \begin{align}
 d_T F  & = \partial_T F + \sum_r(\partial_{n_r} F)(\partial_T n_r)\,,\\
 d_n F  & =                \sum_r(\partial_{n_r} F)(\partial_n n_r)\,,\\
 d_y F  & =                \sum_r(\partial_{n_r} F)(\partial_y n_r)\,,
\end{align}
\end{subequations}
for $F=F(\xi)$ and 
\begin{subequations}\label{app_eq:dF'dzeta}
 \begin{align}
 d_T F' & = \partial'_T F' + \sum_r(\partial_{\eta_r} F)(\partial_T \eta_r)\,,\\
 d_n F' &=                   \sum_r(\partial_{\eta_r} F)(\partial_n \eta_r)\,,\\
 d_y F' &=                   \sum_r(\partial_{\eta_r} F)(\partial_y \eta_r)\,.
 \end{align}
\end{subequations}
for $F'=F'(\xi')$. Above
\begin{equation}\label{app_eq:dFdzeta}
 d_T F = \left. \frac{d F}{d T}\right|_{n,y}
\end{equation}
and similarly for $F'$ and permutations of $T$, $n$, and $y$. 
% We show how the derivatives in Eq. \eqref{app_eq:dFdzeta} 
% are computed in Appendix \ref{}. 
We further define the derivative
\begin{equation}\label{app_eq:dFdzeta}
 d'_T F = \left. \frac{d F}{d T}\right|_{n,y,\zeta'}\,,
\end{equation}
where $\zeta'$ is a set of internal variables of the system. 
This will be useful when changing from derivatives with 
respect to $\zeta'$ to derivatives with respect to $\xi$.

\section{The APR model}
\label{app:apr}

In this Appendix, we collect various formulas and 
numerical notes employed in the development of the EOS of APR.

\subsection{The free energy of nucleons}
\label{app:fbulk}

The free energy of a uniform system of nucleons is computed from 
the thermodynamical relation
\begin{equation}\label{app_eq:f}
 F_\rbulk = E_\rbulk - TS_\rbulk\,.
\end{equation}
For a given density $n$, proton fraction $y$, and temperature $T$, 
the internal energy $E_\rbulk$ is computed from 
Eq. \eqref{eq:Hamiltonian}, $U_\rbulk \rightarrow \mathcal{H}(n,y,T)$.   
$E_\rbulk$ depends on the kinetic energy densities $\tau_t$, 
effective masses $m_t^\star$, and the APR potential $\mathcal{U}(n,y)$.
The entropy $S_\rbulk$ has the form 
\begin{equation}\label{app_eq:sbulk}
 S_\rbulk = \sum_t\left[ 
 \frac{5}{3}\frac{\hbar^2\tau_t}{2m^\star_t T} 
 -Tn_t\eta_t \right]\,.
\end{equation}
Note that the entropy depends also on the degeneracy parameters 
of nucleons $\eta_t$ discussed in Eq. \eqref{eq:etat} 
in Sec. \ref{ssec:bulk}.
These expressions enable the determination of the free
energies of unbound nucleons in uniform matter as well as those of
bound and unbound nucleons in non-uniform matter.

In what follows, we use capital letters for quantities per volume 
and lower case letters for specific (per baryon or per mass) quantities.
Thus, the specific free energy of the nucleon system $a$ is related 
to its internal free energy density by $f_a=F_a/n_a$. 
Here $a=i$ stands for nucleons bound inside heavy nuclei,
and $a=o$ for unbound nucleons outside heavy nuclei. 
Similarly, the specific entropy is written as $s_a=S_a/n_a$ 
and the specific internal energy as $\epsilon_a=E_a/n_a$.

\subsection{The nuclear potential}
\label{app:apr_u}

We now turn our attention the the nucleon-nucleon potential 
in the APR model given by 
\begin{equation}\label{app_eq:uapr}
 \mathcal{U}(n,y) = g_1(n)\left[1-\delta^2(y)\right] + g_2(n)\delta^2(y) \,,
\end{equation}
which may also be written in the form
\begin{equation}
\mathcal{U}(n_n,n_p)=4\frac{g_1}{n^2}n_tn_{-t}+\frac{g_2}{n^2}\left(n_t-n_{-t}
\right)^2 \,.
\end{equation}
%
%where $n_n=(1-y)n$ and $n_p=yn$ 
%with $n=n_n+n_p$ the number density 
%and $y=n_p/n$ the proton fraction.
%The isospin $t$ is defined such that 
%if $t=n$ then $-t=p$ and vice versa. 
Unless otherwise explicit, we omit the functional 
dependences after they have been shown once. 
In order to simplify expressions throughout, 
we define the auxiliary functions
%
% \begin{subequations}
\begin{align}\label{app_eq:aux}
 \phi_{i,j} =p_i+p_jn,\quad
 \psi_i     =n-p_i,\quad
 \kappa_i   = e^{-p_i^2n^2} \,.
 \end{align}
% \end{subequations}
%
Primes are used to denote total derivatives with respect 
to the total nucleon number density $n$;  thus,
%
% \begin{subequations}
\begin{align}\label{app_eq:aux_tau}
 \kappa'_i  = -2p_i^2n\kappa_i, \quad
 \kappa''_i = \kappa'_i\left[\frac{1}{n}+\frac{\kappa'_i}{\kappa_i}\right] \,.
 \end{align}
% \end{subequations}

For the low density phase (LDP), \ie 
for densities below those for which a neutral pion condensate forms, 
\begin{equation}\label{app_eq:uaprl}
\mathcal{U}\rightarrow \mathcal{U}_L 
= g_{1L}\left[1-\delta^2\right] + g_{2L}\delta^2 \,.
\end{equation}
The functions $g_{iL}$ are given by 
\begin{subequations}\label{app_eq:ldp_g}
\begin{align}
 g_{1L} &= -n^2\left[p_1+n\phi_{2,6}+\phi_{10,11}\kappa_9\right] \\
 g_{2L} &= -n^2\left[\frac{p_{12}}{n}+\phi_{7,8}+p_{13}\kappa_9\right].
\end{align} 
\end{subequations}
In the high density phase (HDP), $\mathcal{U}\rightarrow\mathcal{U}_{H}$, and
$g_{iH}$ are related to $g_{iL}$ by 
%
% \begin{subequations}
\begin{align}\label{app_eq:hdp_g}
 g_{1H} &=g_{1L}-n^2\Delta_1\,, \quad 
 g_{2H} =g_{2L}-n^2\Delta_2\,,
\end{align}
% \end{subequations}
%
where, for simplicity we write
\begin{subequations}\label{app_eq:hdp_D}
\begin{align}
\Delta_1&=\left[p_{17}\psi_{19}+p_{21}\psi_{19}^2\right]e^{p_{18}\psi_{19}}\\
\Delta_2&=\left[p_{15}\psi_{20}+p_{14}\psi_{20}^2\right]e^{p_{16}\psi_{20}}\,.
\end{align}
\end{subequations}

\subsection{Density derivatives of the nuclear potential}
\label{app:apr_dudn}

From Eq. \eqref{app_eq:uapr}, the density derivatives of the potential are given by
\begin{align}\label{app_eq:dudn}
n^2 \partial_{n_t}\mathcal{U} =
& 4f_1n_tn_{-t}+f_2(n_t-n_{-t})^2 \nonumber\\
&+4g_1n_{-t}+2g_2(n_t-n_{-t}) \,,
\end{align}
where $\mathcal{U}$ can be either $\mathcal{U}_L$  or $\mathcal{U}_H$.
If $\mathcal{U}\rightarrow\mathcal{U}_L$, then $f_i\rightarrow f_{iL}$, and so on.
We define
\begin{equation}
 f_i = \left[\frac{dg_i}{dn}-\frac{2g_i}{n}\right]\,
\end{equation}
for $i=1L$, $2L$, $1H$, and $2H$. 
Thus, we obtain
\begin{subequations}\label{app_eq:ldp_f}
\begin{align}
 f_{1L}
 &= - n^2 \left[\phi_{2,6} + p_6n + p_{11}\kappa_9 + \phi_{10,11}\kappa'_9\right] \\
 f_{2L} 
 &= - n^2 \left[-\frac{p_{12}}{n^2} + p_8 + p_{13}\kappa'_9\right]\,
\end{align}
\end{subequations}
for the low density phase, and 
%
% \begin{subequations}
\begin{align}\label{app_eq:hdp_f}
 f_{1H} = f_{1L} - n^2 \Delta'_1 \,, \quad
 f_{2H} = f_{2L} - n^2 \Delta'_2\,
\end{align} 
% \end{subequations}
%
for the high density phase, where
\begin{subequations}\label{app_eq:hdp_D}
\begin{align}
\Delta'_1&=\left[p_{17} + 2p_{21}\psi_{19}\right]e^{p_{18}\psi_{19}}+p_{18}\Delta_1\\
\Delta'_2&=\left[p_{15} + 2p_{14}\psi_{20}\right]e^{p_{16}\psi_{20}}+p_{16}\Delta_2\,.
\end{align}
\end{subequations}

The second order derivatives are expressed through
\begin{align}\label{app_eq:d2udn2}
n^2 \partial_{n_rn_t}\mathcal{U} = & 
4h_1n_tn_{-t}+h_2(n_t-n_{-t})^2 \nonumber\\
&+4f_1n_{-t}+2f_2(n_t-n_{-t})   \nonumber\\
&+4f_1n_{-r}+2f_2(n_t-n_{-t})\zeta_{rt} \nonumber\\
&+4g_1\delta_{-rt}+2g_2\zeta_{rt}\,,
\end{align}
where $\delta_{rt}=+1$ and $\zeta_{rt}=+1$ if $r=t$ while
$\delta_{rt}=0$ and $\zeta_{rt}=-1$ if $r\neq t$ and we defined
\begin{equation}
 h_i = \left[\frac{df_i}{dn}-\frac{2f_i}{n}\right]\,.
\end{equation}

Above,
\begin{subequations}\label{app_eq:ldp_h}
\begin{align}
 h_{1L}
 &= - n^2 \left[2p_6 + 2p_{11}\kappa'_9 + \phi_{10,11}\kappa''_9\right] \\
 h_{2L} 
 &= - n^2 \left[2\frac{p_{12}}{n^3} + p_{13}\kappa''_9\right]\,.
\end{align}
\end{subequations}
if we are treating the low density phase, whereas
%
% \begin{subequations}
\begin{align}\label{app_eq:hdp_f}
 h_{1H} &= h_{1L} - n^2 \Delta''_1 \,, \quad
 h_{2H} = h_{2L} - n^2 \Delta''_2\,.
\end{align} 
% \end{subequations}
%
with 
\begin{subequations}\label{app_eq:hdp_D}
\begin{align}
\Delta''_1&=2p_{21}e^{p_{18}\psi_{19}} + 2p_{18}\Delta'_1 - p^2_{18}\Delta_1\\
\Delta''_2&=2p_{14}e^{p_{16}\psi_{20}} + 2p_{16}\Delta'_2 - p^2_{16}\Delta_2\,,
\end{align}
\end{subequations}
if we are in the high density region. 
%\prak{Is this complete?}

\subsection{Nucleon effective masses and its derivatives}
\label{app:apr_meff}

The effective masses $m_t^\star$ are defined through 
\begin{equation}\label{app_eq:meff}
\frac{\hbar^2}{2m_t^\star} = 
\frac{\hbar^2}{2m_t} + 
\mathcal{M}_{t}(n_n,n_p) \,,
\end{equation}
where $m_t$ are the vacuum nucleon masses and $\mathcal{M}_{t}$ are functions of the nucleonic densities:
\begin{equation}\label{app_eq:Mt}
 \mathcal{M}_t(n,y) = (p_3 n + p_5 n_t)e^{-p_4n}\,.
\end{equation}
Thus, the density derivatives of the effective masses are 
\begin{equation}\label{app_eq:dmeffdn}
\partial_{n_r} m_t^\star = 
- \frac{2{m_t^\star}^ 2}{\hbar^2} 
\partial_{n_r} \mathcal{M}_t \,,
\end{equation}
where
\begin{equation}\label{app_eq:dMtdn}
\partial_{n_r} \mathcal{M}_t = 
(p_3 + p_5\delta_{rt})e^{-p_4n} - p_4 \mathcal{M}_t\,.
\end{equation}
The corresponding second derivatives are
\begin{align}\label{app_eq:d2meffdn2}
\partial_{n_rn_s} m_t^\star =   \frac{2{m_t^\star}^2}{\hbar^2} & \bigg[
  \frac{4m_t^\star}{\hbar^2}(\partial_{n_s} \mathcal{M}_t)
  (\partial_{n_r} \mathcal{M}_t) 
- \partial_{n_rn_s}\mathcal{M}_t\bigg] \,,
\end{align}
where
\begin{align}\label{app_eq:d2Mtdn2}
 \partial_{n_rn_s}\mathcal{M}_t = -p_4\left[
 \partial_{n_r}\mathcal{M}_t + \partial_{n_s}\mathcal{M}_t
 + p_4 \mathcal{M}_t \right].
%   - p_4e^{-p_4n} \bigg[p_3(2-p_4n) \nonumber\\
% & + p_5(\delta_{rt}+\delta_{st}-p_4n_t)  \bigg]\,.
\end{align}

\subsection{Fermi integrals}
\label{app:apr_Fermi}

We define the Fermi integrals as 
\begin{equation}\label{app_eq:Fermi}
\mathcal{F}_k(\eta)=\int\frac{u^k du}{1+\exp(u-\eta)}\,.
\end{equation} 
Their values for $k=-1/2$, $+1/2$ and $+3/2$ as well as the
inverse for $k=+1/2$ are computed using the subroutines of 
Fukushima \cite{fukushima:15a,fukushima:15b}.
The derivatives of the Fermi integrals satisfy
\begin{equation}\label{app_eq:dFermideta}
\frac{\partial\mathcal{F}_k}{\partial\eta}=k\mathcal{F}_{k-1}\,.
\end{equation} 
A useful relation used often throughout is the ratio
\begin{equation}\label{app_eq:G}
 \mathcal{G}(\eta)=2\frac{\mathcal{F}_{+1/2}(\eta)}{\mathcal{F}_{-1/2}(\eta)}\,.
\end{equation}
We will make use of the shorthand notation 
$\mathcal{G}_t=\mathcal{G}(\eta_t)$.
Whenever $\eta< -200$ we set $\eta\rightarrow-200$ to avoid
overflow and underflow in our double precision computations.
In these cases, the asymptotic forms of the Fermi integrals 
\begin{subequations}\label{app_eq:Fermi_asympt}
\begin{align}
 \lim_{\eta\rightarrow-\infty}\mathcal{F}_{-1/2}(\eta)
 &\rightarrow\sqrt{\pi}e^{\eta}\,,\\
 \lim_{\eta\rightarrow-\infty}\mathcal{F}_{+1/2}(\eta)
 &\rightarrow\tfrac{1}{2}\sqrt{\pi}e^{\eta}\,,\\
 \lim_{\eta\rightarrow-\infty}\mathcal{F}_{+3/2}(\eta)
 &\rightarrow\tfrac{3}{4}\sqrt{\pi}e^{\eta}\,.
\end{align}
\end{subequations}
can be used. Clearly, $\mathcal{G}(\eta\rightarrow-\infty)=1$.

\subsection{Degeneracy parameters}
\label{app:apr_eta}

The degeneracy parameters $\eta_t$ are computed by inverting 
Eq. \eqref{eq:nt} to obtain 
\begin{equation}\label{app_eq:etat}
\eta_t = \mathcal{F}^{-1}_{1/2} \left( \frac{2\pi^2 n_t}{\upsilon_t^{3/2}} \right)\,,
\end{equation}
where we have defined $\upsilon_t$ in Eq. \eqref{app_eq:upsilon}.
% %
% \begin{equation}
% \upsilon_t= \left( \frac{\hbar^2}{2m_t^\star T} \right)^{3/2}\,.
% \end{equation}
% %
Because we work with variables where the nucleon densities $n_t$ 
and temperatures $T$ are readily available, it is straightforward
to determine $\eta_t$. 
We use the subroutines of Fukushima to compute the above Fermi integrals 
and their inverses \cite{fukushima:15a,fukushima:15b}.
If the nucleon density is extremely low, floating point operations 
may become an issue and, thus, asymptotic limits must be used to 
compute the degeneracy parameters. 
Although such solutions do not occur in the regions of parameter 
space of interest, they do occur often when our algorithm is 
trying to determine the lowest energy state of the system. 
Therefore, for densities $\log_{10}[n_t(\mathrm{fm}^{-3})]<-100$
we set
\begin{equation}
 \lim_{n_t\rightarrow0} \eta_t =
 \ln\left(\frac{2}{\sqrt{\pi}}\frac{2\pi^2n_t}{v_t^{3/2}}\right)\,.
\end{equation}

The density derivatives of $\eta_t$ are
\begin{equation}\label{app_eq:detadn}
 \partial_{n_r}\eta_t = \frac{2\mathcal{Q}_{tr}}{\mathcal{F}_{-1/2}(\eta_t)}
\end{equation}
where we have defined
\begin{align}\label{app_eq:Qtr}
 \mathcal{Q}_{tr} = \frac{\mathcal{F}_{1/2}(\eta_t)}{n_t}
 \left(\delta_{tr} - \mathcal{R}_{tr}\right)\,
\end{align}
with 
\begin{equation}\label{app_eq:Rtr}
 \mathcal{R}_{tr}=\frac{3}{2}\frac{n_t}{m_t^\star} 
 \partial_{n_r} m_t^\star\,.
\end{equation}

% https://en.wikipedia.org/wiki/Inverse_functions_and_differentiation

% https://math.stackexchange.com/questions/619142/jacobian-of-an-inverse

\subsection{Kinetic energy density}
\label{app:apr_tau}

To compute the kinetic energy density, we 
start by defining the auxiliary function 
\begin{equation}\label{app_eq:upsilon}
\upsilon_t = \left( \frac{2m_t^\star T}{\hbar^2} \right)\,,
\end{equation}
which depends on both the nucleon densities $n_t$ and 
temperature $T$ of the system. 
Thus, the kinetic energy density becomes
\begin{equation}\label{app_eq:taut}
 \tau_t = \frac{1}{2\pi^2}\upsilon_t^{5/2}\mathcal{F}_{3/2}(\eta_t)\,.
\end{equation}
The density derivatives of $\tau_t$ are
\begin{align}\label{app_eq:dtautdn}
 \partial_{n_r} \tau_t 
 & = \frac{5}{2}\frac{\tau_t}{\upsilon_t} 
 \partial_{n_r} \upsilon_t +
 \frac{3}{2}\frac{\upsilon_t^{5/2}}{2\pi^2} 
 \mathcal{F}_{1/2}(\eta_t) \partial_{n_r}\eta_t
  \nonumber\\
 & = \frac{5}{2}\frac{\tau_t}{m_t^\star} 
 \partial_{n_r} m_t^\star +
 \frac{3}{2}\frac{\upsilon_t^{5/2}}{2\pi^2} 
 \mathcal{G}_t\mathcal{Q}_{tr}\,,
\end{align}
where derivatives of $\upsilon_t$ are computed from Eqs. 
\eqref{app_eq:dmeffdn} and \eqref{app_eq:upsilon} and 
$\partial_{n_r}\eta_t$ is defined in Eq. \eqref{app_eq:detadn}.

\subsection{Chemical and interaction potentials}
\label{app:apr_mu}

The chemical potentials are related to the degeneracy parameters
through 
\begin{equation}\label{app_eq:eta}
\eta_t=\frac{\mu_t-\mathcal{V}_t}{T} \,,
\end{equation}
where the interaction potential is 
\begin{equation}\label{app_eq:v}
\mathcal{V}_t\equiv
\left.\frac{\delta\mathcal{H}}{\delta n_t}\right|_{n_{-t},\tau_{\pm t}}\,. 
\end{equation}
Explicitly, 
\begin{align}\label{app_eq:vt}
\mathcal{V}_t
 = \tau_n (\partial_{n_t} \mathcal{M}_n) 
 + \tau_p (\partial_{n_t} \mathcal{M}_p)
 +         \partial_{n_t} \mathcal{U} \,
\end{align}
which can be computed from
Eqs. \eqref{app_eq:dMtdn} and \eqref{app_eq:dudn}.
The density derivatives are 
\begin{align}\label{app_eq:dvdn}
\partial_{n_r}\mathcal{V}_t 
   = & \left(\partial_{n_r} \tau_n\right) \left(\partial_{n_t} \mathcal{M}_n\right)
   + \tau_n \left(\partial_{n_rn_t}\mathcal{M}_n\right)
   \nonumber\\
 + & \left(\partial_{n_r} \tau_p\right) \left(\partial_{n_t} \mathcal{M}_p\right)
   + \tau_p \left(\partial_{n_rn_t}\mathcal{M}_p\right)
   \nonumber\\
 + & \partial_{n_rn_t}\mathcal{U}\,
\end{align}
which are computed using the relations in Eqs. 
\eqref{app_eq:dtautdn}, \eqref{app_eq:dMtdn}, and \eqref{app_eq:d2Mtdn2}.

Thus, we may write the chemical potential derivatives as
\begin{equation}\label{app_eq:dmut_dnr}
 \partial_{n_r}\mu_t = 
 T\partial_{n_r}\eta_t + \partial_{n_r}\mathcal{V}_t\,.
\end{equation}

\subsection{Derivatives with respect to $\eta$}
\label{app:deta}

As we will need some derivatives with respect to the degeneracy 
parameters, we calculate them here using the definition in Eq. 
\eqref{app_eq:partialp_t}. We start with the density derivatives
which are obtained from
\begin{gather}
  \begin{bmatrix}
   \partial_{n_n} \eta_n & \partial_{n_p} \eta_n \\
   \partial_{n_n} \eta_p & \partial_{n_p} \eta_p 
   \end{bmatrix}
  \begin{bmatrix}
   \partial_{\eta_n} n_n & \partial_{\eta_p} n_n \\
   \partial_{\eta_n} n_p & \partial_{\eta_p} n_p 
   \end{bmatrix}=
  \begin{bmatrix}
   1 & 0  \\
   0 & 1 
  \end{bmatrix}\,.
\end{gather}
This matrix equation leads to
\begin{equation}\label{app_eq:detardnt}
 \partial_{\eta_r} n_t = \zeta_{tr}\frac{n_r Q_{-t-r}}{\mathcal{G}_r\mathcal{O}} \,,
\end{equation}
where $\mathcal{Q}_{tr}$ was defined in Eq. \eqref{app_eq:Qtr}, 
$\mathcal{G}_t$ in Eq. \eqref{app_eq:G}, 
$\zeta_{tr}$ below Eq. \eqref{app_eq:d2udn2}, 
and 
\begin{equation}\label{app_eq:O}
 \mathcal{O} = 1 - \mathcal{R}_{nn} - \mathcal{R}_{pp} 
 - \mathcal{R}_{np}\mathcal{R}_{pn} 
 + \mathcal{R}_{nn}\mathcal{R}_{pp}\,.
\end{equation}

The $\eta_t$ derivatives of any quantity $\chi$ that 
is solely an explicit function of the nucleon 
densities $n_n$ and $n_p$ can then be computed from
\begin{equation}\label{app_eq:dchideta}
 \partial_{\eta_t}\chi = \sum_r
 \left( \partial_{n_r}\chi \right) \left( \partial_{\eta_t} n_r \right) \,.
\end{equation}

\section{Bulk observables}
\label{app:bulk_apr}

Using the results of Appendix \ref{app:apr}, the 
free energy density $F_\rbulk$ of bulk nuclear matter, 
\ie of matter composed solely of nucleons, is
\begin{equation}\label{app_eq:fbulk}
 F_\rbulk = E_\rbulk - T S_\rbulk\,.
\end{equation}
Here the energy density is
\begin{equation}\label{app_eq:ebulk}
 E_\rbulk = 
 \sum_t\frac{\hbar^2\tau_t}{2m_t^\star} + \mathcal{U}\,,
\end{equation}
while the specific entropy is
\begin{equation}\label{app_eq:sbulk}
 S_\rbulk = \sum_t \left[ 
 \frac{5}{3}\frac{\tau_t}{\upsilon_t} - n_t\eta_t \right]\,. 
\end{equation}
The pressure of the system is given by
\begin{equation}\label{app_eq:pbulk}
 P_\rbulk = \sum_t n_t\mu_t - F_\rbulk\,.
\end{equation}

\subsection{Density derivatives}
\label{app:dbulkdn}

From Eq. \eqref{app_eq:fbulk},
\begin{equation}\label{app_eq:dfbulkdn}
 \partial_{n_r} F_\rbulk = 
 \partial_{n_r} E_\rbulk - T \partial_{n_r} S_\rbulk\,,
\end{equation}
where 
\begin{equation}\label{app_eq:debulkdn}
 \partial_{n_r} E_\rbulk = \sum_t \frac{\hbar^2}{2m_t^\star}
 \left[\partial_{n_r} \tau_t - \frac{\tau_t}{m_t^\star}\partial_{n_r} m_t^\star\right] 
 - \partial_{n_r}\mathcal{U}\,,
\end{equation}
and
\begin{align}\label{app_eq:dsbulkdn}
 \partial_{n_r} S_\rbulk = \sum_t 
 \bigg[ & \frac{5}{3}\frac{\partial_{n_r} \tau_t}{\upsilon_t} 
 - \frac{5}{3}\frac{\tau_t}{\upsilon_t}\frac{\partial_{n_r} m_t^\star}{m_t^\star} 
 \nonumber\\
 & - \delta_{rt}\eta_t - n_t \partial_{n_r} \eta_t\bigg]\,.
\end{align}
For the pressure derivatives, we have
\begin{equation}\label{app_eq:dpbulkdn}
 \partial_{n_r} P_\rbulk = \sum_t \left[ n_t (\partial_{n_r} \mu_t) 
 + \delta_{rt} \mu_t \right] - \partial_{n_r}F_\rbulk\,. 
\end{equation}

\subsection{Temperature derivatives}
\label{app:dbulkdT}

Here, the temperature derivatives both at constant
nucleon densities $n_t$ and constant degeneracies $\eta_t$
are given. 
The latter will be identified with a prime in the $\partial$ sign.

\subsubsection{Constant $n_n$ and $n_p$}
\label{app:dbulkdTn}

If the densities are kept constant, 
\begin{equation}
 \partial_T n_t = 0\,.
\end{equation}
Also, 
\begin{subequations}\label{app_eq:dXdT}
\begin{align}
 \partial_T \eta_t &= - \frac{3}{2}\frac{\mathcal{G}_t}{T} \, \\
 \partial_T \tau_t &=   \frac{5}{2}\frac{\tau_t}{T} 
                     - \frac{9}{2}\frac{m_t^\star}{\hbar^2}\mathcal{G}_t n_t\, \\
 \partial_T \mathcal{V}_t  & =
   \left(\partial_T\tau_n \right) \left(\partial_{n_t}\mathcal{M}_n\right) \nonumber\\
 & \quad + \left(\partial_T\tau_p \right) \left(\partial_{n_t}\mathcal{M}_p\right)\, \\
 \label{app_eq:dXdT_mu}
 \partial_T \mu_t  &= \eta_t + T \partial_T \eta_t + 
                      \partial_T \mathcal{V}_t \,.
\end{align}
\end{subequations}
From Eq. \eqref{app_eq:dXdT} 
and Eqs. \eqref{app_eq:fbulk} to 
\eqref{app_eq:pbulk}, we obtain 
\begin{subequations}
\begin{align}
 \partial_T S_\rbulk &= \frac{1}{T}\sum_t\left[ 
   \frac{5}{2} \frac{\tau_t}{\upsilon_t} 
 - \frac{9}{4} \mathcal{G}_t n_t \right] \,,\\
 \partial_T E_\rbulk & = T \partial_T S_\rbulk \,,\\
 \partial_T F_\rbulk & = - S_\rbulk \,, \\
 \label{app_eq:dPbulkdT}
 \partial_T P_\rbulk & = \sum_t n_t (\partial_T \mu_t)
 - \partial_T F_\rbulk\,.  
\end{align}
\end{subequations}

\subsubsection{Constant $\eta_n$ and $\eta_p$}
\label{app:dbulkdTeta}

If  the degeneracy parameters are kept constant instead 
of densities, the primed derivatives $\partial'_T$ yield 
\begin{equation}
 \partial'_T \eta_t = 0 \,.
\end{equation}
This leads to the relations 
\begin{subequations}\label{app_eq:dXdTeta}
\begin{align}
\label{app_eq:dXdTeta_n}
 \partial'_T n_t &= \frac{3}{2\mathcal{O} T}
 \left(n_t(1-\mathcal{R}_{-t-t}) + n_{-t}\mathcal{R}_{t-t}\right) \, \\
 \partial'_T \tau_t &=   \frac{5}{2}\frac{\tau_t}{T}
 \left(1 + T\frac{\partial'_Tm_t^\star}{m_t^\star}\right) \, \\
 \partial'_T \mathcal{V}_t & = \sum_r \left[ 
 (\partial'_T\tau_r)(\partial_{n_t}\mathcal{M}_r) + 
 \tau_r \partial'_T(\partial_{n_t}\mathcal{M}_r) \right]\nonumber\\
 & \qquad + \partial'_T \left(\partial_{n_t}\mathcal{U}\right) \\
 \label{app_eq:dXdTeta_mu}
 \partial'_T \mu_t  &= \eta_t  +  \partial'_T \mathcal{V}_t \,.
\end{align}
\end{subequations}
Above, the $\mathcal{R}_{rt}$ were defined in Eq. \eqref{app_eq:Rtr}
and $\mathcal{O}$ in Eq. \eqref{app_eq:O}. 
For quantities not explicitly dependent on the temperature $T$, 
such as $m_t^\star$, $\mathcal{M}_t$, $\partial_{n_t}\mathcal{M}_r$, 
$\mathcal{U}$, and $\partial_{n_t}\mathcal{U}$ the $\partial'_T$ 
derivatives of are computed from
\begin{equation}\label{app_eq:dchidT}
 \partial'_T\chi = 
 \left( \partial_{n_n}\chi \right) \left( \partial'_T n_n \right)  + 
 \left( \partial_{n_p}\chi \right) \left( \partial'_T n_p \right)\,.
\end{equation}
For temperature dependent quantities, 
\begin{equation}\label{app_eq:dchipdT}
 \partial'_T\chi = \partial_T\chi + 
 \left( \partial_{n_n}\chi \right) \left( \partial'_T n_n \right)  + 
 \left( \partial_{n_p}\chi \right) \left( \partial'_T n_p \right)\,.
\end{equation}

Finally, 
\begin{subequations}
\begin{align}
\label{app_eq:dSbulk_dTp}
 \partial'_T S_\rbulk &= \sum_t\left[T\frac{\partial'_T\tau_t}{\upsilon_t}
 - \eta_t\partial'_Tn_t \right] \,,\\
 \partial'_T E_\rbulk & =\sum_t\left[\frac{\tau_t}{\upsilon_t} 
 + \frac{3}{5}T\frac{\partial'_T\tau_t}{\upsilon_t}\right]  
 + \partial'_T\mathcal{U} \,,\\
\label{app_eq:dFbulk_dTp}
 \partial'_T F_\rbulk & = \partial'_T E_\rbulk 
 - T\partial'_T S_\rbulk - S_\rbulk \,, \\
 \label{app_eq:dPbulkdTeta}
 \partial'_T P_\rbulk & = \sum_t \left[ \mu_t (\partial'_T n_t)  + 
 n_t (\partial'_T \mu_t)  \right] - \partial'_T F_\rbulk\,.  
\end{align}
\end{subequations}

\section{The nuclear surface}
\label{app:surface}

Here, we review the algorithm used in Sec. II B of SRO  to determine 
the nuclear surface tension per unit area $\sigma(y_i,T)$. 
For the purpose of this discussion, we assume two phases in equilibrium: 
the dense phase is assumed to have density $n_i$ and proton fraction $y_i$ 
whereas the dilute phase has density $n_o\leq n_i$ and proton fraction $y_o$.
The procedure described below is used to determine the parameters $\lambda$, $q$, 
and $p$ in  Eqs. \eqref{eq:sigma}, and \eqref{eq:h} and the 
coefficients of the critical temperature $T_c(y_i)$, Eq. \eqref{eq:Tc}, for 
which the dense and the dilute phases coexist.

%To obtain the parameters $\lambda$, $q$, and $p$ and a functional form for $T_c(y_i)$, 
We follow \cite{lim:12,lattimer:85,steiner:05} to study
the two phase equilibrium of bulk nucleonic matter. 
For a given proton fraction $y$, there exists a critical temperature
$T_c$ and a critical density $n_c$ for which both the dense and dilute
phases have the same density $n_i=n_o$ and the same proton fraction $y_i=y_o$.  
The quantities $n_c$ and $T_c$ are obtained by simultaneously solving 
%\cite{lim:12}
%
\begin{equation}\label{app_eq:coexistence}
 \left.\frac{\partial P_\rbulk}{\partial n}\right|_T=0
 \qquad \mathrm{and} \qquad
 \left.\frac{\partial^2P_\rbulk}{\partial n^2}\right|_T=0\,,
\end{equation}
for proton fractions $y\leq0.50$ \footnote{Because we ignore Coulomb
contributions to the surface tension, the formalism presented in this
section is almost symmetric under the $y\rightarrow1-y$ transformation.
The symmetry is only slightly broken by the small difference $\Delta$
in the neutron and proton rest masses, $m_n=m_p+\Delta$, which we
ignore here when considering $y > 0.5$. }. 
Here, $P_\rbulk$ is the bulk pressure given by Eq. \eqref{app_eq:pbulk}. 
Once the critical temperature $T_c$ has been determined for a range of 
proton fractions $y$, the fit using Eq. \eqref{eq:Tc} is performed.

After determining $T_c(y)$, we compute the properties of semi-infinite
nucleonic matter for which the density varies along
the $z$ axis and is constant in the remaining two.
Ignoring Coulomb effects, we assume that in the limits
$z\rightarrow\pm\infty$ matter saturates at densities $n_i$ and $n_o$
and proton fractions $y_i$ and $y_o$.
These two phases are in equilibrium if their pressures as well as their 
neutron and proton chemical potentials are the same, \ie
\begin{align}\label{eq:phaseequilibrium}
 P_{\rbulk,i}=P_{\rbulk,o}\,, \quad \mu_{ni}=\mu_{no}\,, \quad\mathrm{and}\quad \mu_{pi}=\mu_{po}\,.
\end{align}
Here, the pressures $P_{\rbulk,i}=P_{\rbulk}(n_i,y_i)$ and 
$P_{\rbulk,o}=P_{\rbulk}(n_o,y_o)$ are computed from Eq. \eqref{app_eq:pbulk} 
and the chemical potentials $\mu_{ta}$ from Eqs. \eqref{app_eq:eta} and 
\eqref{app_eq:etat}.

Equations \eqref{eq:phaseequilibrium} are solved simultaneously with
\begin{equation}\label{eq:yi}
 y_i=\frac{n_{pi}}{n_{ni}+n_{pi}}
\end{equation}
to obtain the neutron and proton densities $n_{ni}$, $n_{pi}$, $n_{no}$, and $n_{po}$ 
of the high and low density phases , respectively.

Once the neutron and proton densities of the two coexisting phases have been
calculated, we determine the surface shape that minimizes $\sigma(y_i,T)$.
Since we assume the system to be homogeneous across two dimensions, the surface
tension per unit area is given by
\cite{ravenhall:83,steiner:05}
\begin{align}\label{eq:tension}
 \sigma(y_i,T)=
 \int_{-\infty}^{+\infty}\bigg[&F_\rbulk(z)+E_S(z)+P_{\rbulk,o}\nonumber\\
 &-\mu_{no}n_n(z)-\mu_{po}n_p(z)\bigg]dz\,,
\end{align}
where, $P_{\rbulk,o}$, $\mu_{no}$, and $\mu_{po}$ or, alternatively, 
$P_{\rbulk,i}$, $\mu_{ni}$, and $\mu_{pi}$ are solutions to Eqs. 
\eqref{eq:phaseequilibrium}. 
The quantity $F_\rbulk(z)=F_\rbulk,(n(z),y(z),T)$ is the bulk free energy 
density across the $z$ axis, whereas $E_S(z)$ is the spatially-varying 
contribution to the energy density of the Hamiltonian in Eq. \eqref{eq:HS}.

To minimize Eq. \eqref{eq:tension}, we assume that the neutron and
proton densities have a Woods-Saxon form, \ie
\begin{equation}\label{eq:WS}
 n_t(z)=n_{to}+\frac{n_{ti}-n_{to}}{1+\exp\left((z-z_t)/a_t\right)}\,,
\end{equation}
where $z_n$ and $a_n$ ($z_p$ and $a_p$) are the neutron
(proton) half-density radius and its diffuseness \cite{woods:54}, respectively.
This form has the desired limits
 $\lim_{z\rightarrow-\infty}n_t(z)=n_{ti}$ and
$\lim_{z\rightarrow+\infty}n_t(z)=n_{to}$. 
Following Refs.  \cite{steiner:05,lattimer:85,ravenhall:83}, we set 
the proton half-density radius $z_p$ at $z=0$ and minimize the surface 
tension per unit area with respect to the three other variables $z_n$, 
$a_n$, and $a_p$.  This allows us to tabulate values of the surface 
tension per unit area $\sigma(y_i,T)$ as a function of the proton 
fraction $y_i$ of the dense phase and the temperature $T$ of the 
semi-infinite system.  
This is used to determine the parameters $\lambda$ and $q$ in
Eq. \eqref{eq:sigma} and $p$ in Eq. \eqref{eq:h} by
performing a least squares fit.

It is worth mentioning that the surface free energy density should, in
general, include a contribution from the neutron skin
$\sigma\rightarrow\sigma+\mu_n\nu_n$, where $\nu_n$ is the neutron
excess \cite{ravenhall:83,lim:12}.  However, we follow \LSn, and
neglect this term.  In future work, this term should be included since
its effects are important for very neutron rich matter \cite{lim:12}.

\section{Alpha particles}
\label{app:alpha}

In this section,  quantities related to the alpha particles 
that appear in the uniform phase are collected. 
Alpha particles are treated as hard spheres with volume 
$v_\alpha$ and its number density is related to its chemical 
potential through
\begin{equation}\label{app_eq:mualpha2}
 n_\alpha = 8 n_Q e^{\mu_\alpha/T} \,
\end{equation}
where $n_Q = (m_n T/ 2\pi^2\hbar^2)^{3/2}$ is the quantum concentration 
with $m_n$ denoting the neutron mass. 
For alpha particles in equilibrium with a nucleon gas 
with neutron and proton chemical potentials,
$\mu_{no}$ and $\mu_{po}$, respectively, and pressure
$P_{\rbulk,o}$, the alpha particle chemical potential satisfies
\begin{equation}\label{app_eq:mualpha3}
 \mu_\alpha = 2(\mu_{no}+\mu_{po}) + B_\alpha - v_\alpha P_{\rbulk,o} \,,
\end{equation}
where $B_\alpha=28$ MeV is the binding energy of alpha particles.
Unbound nucleons are treated as in Appendix
\ref{app:apr}.

\subsection{Thermodynamical properties}
\label{ssec:alpha_prop}

Since alpha particles are treated in the excluded volume approach, 
their internal energy, entropy, free energy, and pressure are, 
respectively, 
\begin{subequations}\label{app_eq:alpha_prop}
\begin{align}
 E_\alpha & = \left(\frac{3}{2}T - B_\alpha \right)n_\alpha \,,\quad
 S_\alpha   = \left(\frac{5}{2} - \frac{\mu_\alpha}{T}\right)n_\alpha \,,\\
 F_\alpha & = \left(\mu_\alpha-B_\alpha-T\right)n_\alpha\,,\quad
 P_\alpha   = n_\alpha T \,.
\end{align}
\end{subequations}

\subsection{Derivatives of alpha particle thermal variables}
\label{ssec:alpha_prop_der}

Derivatives of the  alpha particle density 
with respect to the neutron and proton densities 
are given by
\begin{equation}\label{app_eq:dna_dnt}
  \partial_{n_t} n_\alpha = 
  \frac{n_\alpha}{T} \partial_{n_t}\mu_\alpha\,,
\end{equation}
with $t=no$ for neutrons and $t=po$ for protons.  
The chemical potential derivatives are given by
\begin{equation}\label{app_eq:dmua_dnt}
 \partial_{n_t}\mu_\alpha = 
   2(\partial_{n_t}\mu_{no} + \partial_{n_t}\mu_{po}) 
 - v_\alpha \partial_{n_t} P_{\rbulk,o}\,.
\end{equation}
Nucleon chemical potential and pressure derivatives are 
obtained from Eqs. \eqref{app_eq:dmut_dnr} and 
\eqref{app_eq:dpbulkdn}, respectively.

From Eqs. \eqref{app_eq:dna_dnt} and \eqref{app_eq:dmua_dnt}, 
density derivatives of the alpha particle
thermodynamical quantities are
\begin{subequations}\label{app_eq:dadnt}
\begin{align}
 \label{app_eq:dadnt_S}
 \partial_{n_t}S_\alpha
 & = \frac{S_\alpha}{T} \partial_{n_t}\mu_\alpha \,,\quad
 \partial_{n_t}E_\alpha 
  = \frac{E_\alpha}{T} \partial_{n_t}\mu_\alpha \,,\\
  \label{app_eq:dadnt_F}
 \partial_{n_t}F_\alpha
 & = \left[n_\alpha + \frac{F_\alpha}{T}\right]\partial_{n_t}\mu_\alpha \,,\quad
 \partial_{n_t}P_\alpha
  = n_\alpha \partial_{n_t}\mu_\alpha \,.
\end{align}
\end{subequations}

Temperature derivatives at constant densities are
\begin{equation}\label{app_eq:dna_dT}
  \partial_T n_\alpha = \frac{n_\alpha}{T} 
   \left(\frac{3}{2}-\frac{\mu_\alpha}{T}+\partial_T\mu_\alpha\right)\,,
\end{equation}
where 
\begin{equation}\label{app_eq:dmua_dT}
 \partial_T\mu_\alpha = 
   2(\partial_T\mu_{no} + \partial_T\mu_{po}) 
 - v_\alpha \partial_T P_o\,,
\end{equation}
and, thus, 
\begin{subequations}\label{app_eq:dadT}
\begin{align}
 \partial_T S_\alpha
 \label{app_eq:dSa_DT}
 & = S_\alpha\frac{\partial_T n_\alpha}{n_\alpha} + \frac{\mu_\alpha}{T^2} n_\alpha \,,\\
 \partial_T E_\alpha 
 & = E_\alpha\frac{\partial_T n_\alpha}{n_\alpha} + \frac{3}{2}n_\alpha \,,\\
 \label{app_eq:dFa_DT}
 \partial_T F_\alpha
 & = F_\alpha\frac{\partial_T n_\alpha}{n_\alpha} + 
 \left(\partial_T \mu_\alpha - 1 \right)n_\alpha \,,\\
 \label{app_eq:dadT_P}
 \partial_T P_\alpha
 & = T \frac{\partial_T n_\alpha}{n_\alpha} \,.
\end{align}
\end{subequations}

Derivatives with respect to $\eta_t$ are straightforwardly obtained by
using Eqs. \eqref{app_eq:dchideta} while derivatives with respect to 
temperature $T$ keeping $\eta_t$ constant are computed using Eq. 
\eqref{app_eq:dchipdT} and results in Eqs. \eqref{app_eq:dXdTeta_n} 
and \eqref{app_eq:dna_dnt} through \eqref{app_eq:dadnt}.

\section{Heavy nuclei}
\label{app:heavy}

In the LS approach,  the free energy $F_h$ of the 
representative heavy nucleus has contributions 
from four terms: 
\begin{equation}
 F_h = F_i + F_{TR} + F_S + F_C \,,
\end{equation}
where the various terms are, respectively, the free
energy $F_i$ of bulk nucleons inside nuclei, the 
translational free energy $F_{TR}$ due to nuclear
motion inside the Wigner-Seitz cell, the surface 
free energy $F_S$, and the coulomb free energy 
$F_C$.

Nucleons inside heavy nuclei are treated as 
in Appendix \ref{app:apr}.  We assume they 
have constant density $n_i=n_{ni}+n_{pi}$
and proton fraction $y_i=n_{pi}/n_i$, where
$n_{ni}$ ($n_{pi}$) is the neutron (proton) 
density.

\subsection{Surface and Coulomb contributions}
\label{app:virial}

The surface and coulomb free energies are given by 
\cite{lattimer:91}
\begin{subequations}
 \begin{align}
  F_S & = \frac{3s(u)\sigma}{r}\,,\\
  F_C & = \frac{4\pi\alpha_C}{5}(n_i y_i r)^2c(u)\,,
 \end{align}
\end{subequations}
where $\alpha_C$ is the fine structure constant, 
and $s(u)$ and $c(u)$ are shape functions chosen 
to satisfy physical limits. 
The function $\sigma\equiv\sigma(y_i,T)$ was defined 
in Eq. \eqref{eq:sigma}. 
The quantities $F_S$, $F_C$, and $F_{TR}$ all 
depend on the generalized radius $r$. 
However, in most of the parameter space $F_{TR}$ is 
small compared to  $F_S$ and $F_C$. 
Furthermore, in regions where $F_{TR}$ is comparable
to $F_S$ and $F_C$, \ie near the transition from uniform
to non-uniform matter at high temperatures, their 
contributions to the total free energy are unimportant
when compared to contributions of nucleons, photons, 
and electrons. 
Thus, when minimizing the total nuclear free energy
with respect to the generalized radius $r$, 
$F_{TR}$ may be ignored to obtain  
\begin{equation}\label{app_eq:virial}
 \frac{\partial F_N}{\partial r}=0
 \quad\Leftrightarrow\quad
 F_S=2F_C\,.
\end{equation}
%
%Eq. \eqref{app_eq:virial} 
This result is known as the nuclear virial 
theorem and is generally valid at $T=0$. 
In this model, it implies that 
\begin{equation}\label{app_eq:r}
 r = \frac{9\sigma}{2\beta}\left[\frac{s(u)}{c(u)}\right]^{1/3}\,.
\end{equation}
where $\beta\equiv\beta(n_i,y_i,T)$ is given by
\begin{equation}\label{app_eq:beta}
 \beta = 9\left[\frac{\pi\alpha_C}{15}\right]^{1/3}(n_iy_i\sigma)^{2/3}\,. 
\end{equation}
We may thus combine $F_S$ and $F_C$ into a single term
\begin{equation}
 F_{SC} = F_S + F_C = \beta \mathcal{D}(u)\,.
\end{equation}

As discussed in LS \cite{lattimer:91} 
and SRO \cite{schneider:17}, 
the shape functions have the forms
%
% \begin{subequations}
 \begin{align}
  s(u) & = uv\,,\quad
  c(u)  = \mathcal{D}(u)^3/s(u)^2 \,,
 \end{align}
% \end{subequations}
%
where, for simplicity, $v=(1-u)$ and 
$\mathcal{D}(u)$ is well approximated by 
\cite{lattimer:91, lim:17}
\begin{equation}\label{app_eq:Del}
 \mathcal{D}(u)=uv
 \frac{vD(u)^{1/3}+uD(v)^{1/3}}{u^2+v^2+0.6u^2v^2}
\end{equation}
where $D(u)=1-\tfrac{3}{2}u^{1/3}+\tfrac{1}{2}u$.

\subsubsection{The shape function $\mathcal{D}$}
\label{app:Del}

Derivatives of the function $\mathcal{D}(u)$ 
introduced in Eq. \eqref{app_eq:Del} are 
\begin{subequations}
 \begin{align}
  \partial_u \mathcal{D} & = \mathcal{D}
  \left[\frac{1}{u}-\frac{1}{v} + \frac{P'}{P}-\frac{Q'}{Q}\right]\,,\\
  \partial_{uu} \mathcal{D} & = \frac{\left(\partial_u\mathcal{D}\right)^2}{\mathcal{D}} 
  + \mathcal{D}\bigg[-\frac{1}{u^2}+\frac{1}{v^2}\nonumber\\
  & \quad +\frac{P''}{P}-\frac{P'^2}{P^2}-\frac{Q''}{Q}+\frac{Q'^2}{Q^2}\bigg]\,,
 \end{align}
\end{subequations}
where
\begin{subequations}\label{app_eq:PQ}
 \begin{align}
   P &= vD(u)^{1/3}+uD(v)^{1/3}\,,\\
   Q &= u^2+v^2+0.6u^2v^2\,,\\
   P' &= \frac{1}{3} \left[\frac{vD'(u)}{D(u)^{2/3}} 
        + \frac{uD'(v)}{D(v)^{2/3}}\right] \nonumber\\ 
      & \quad - D(u)^{1/3} + D(v)^{1/3}\,,\\
   Q' &= 2\left(u-v+0.6uv(v-u)\right)\,,\\
   P'' &=  \frac{2}{3} \left[-\frac{D'(u)}{D(u)^{2/3}} 
         + \frac{D'(v)}{D(v)^{2/3}}\right] \nonumber\\ 
     & \quad   + \frac{1}{3} \left[\frac{vD''(u)}{D(u)^{2/3}} 
         + \frac{uD''(v)}{D(v)^{2/3}}\right] \nonumber\\ 
     & \quad   - \frac{2}{9} \left[\frac{vD'(u)^2}{D(u)^{5/3}} 
        + \frac{uD'(v)^2}{D(v)^{5/3}}\right]\,,\\
   Q'' &= 4+1.2(u^2+v^2)-4.8uv\,,
 \end{align}
\end{subequations}
and
\begin{subequations}
 \begin{align}
  D'(u) & = \frac{1}{2}(1-u^{-2/3})\,,\quad
  D'(v)  = - \frac{1}{2}(1-v^{-2/3})\,,\\
  D''(u) & = \frac{1}{3}(u^{-5/3})\,,\quad
  D''(v)  = \frac{1}{3}(u^{-5/3})\,.
 \end{align}
\end{subequations}

\subsubsection{The surface tension $\sigma$}
\label{app:sigma}

The surface and coulomb free energies depend on the 
surface tension $\sigma\equiv\sigma(y_i,T)$ defined in 
Eq. \eqref{eq:sigma}.
Its first order derivatives are
%
% \begin{subequations}
 \begin{align}
  \partial_{y_i}\sigma & = \sigma\left[\frac{\partial_{y_i}h}{h}-\frac{R'}{R}\right]\,,\quad
  \partial_T    \sigma  = \sigma\left[\frac{\partial_T    h}{h}\right]\,,
 \end{align}
% \end{subequations}
%
while the second order ones are 
\begin{subequations}
 \begin{align}
  \partial_{y_iy_i}\sigma 
  & = \frac{(\partial_{y_i}\sigma)^2}{\sigma}
    + \sigma\bigg[\frac{\partial_{y_iy_i}h}{h} \nonumber\\
  & \quad - \frac{(\partial_{y_i}h)^2}{h^2}
  -\frac{R''}{R}+\frac{R'^2}{R^2}\bigg]\,,\\ 
  \partial_{Tz}\sigma 
  & = \frac{(\partial_T\sigma)(\partial_z\sigma)}{\sigma} 
  + \sigma \left[\frac{\partial_{Tz} h}{h} 
  -\frac{(\partial_zh)(\partial_T h)}{h^2}\right]\,,
 \end{align}
\end{subequations}
where $z=y_i$ or $T$. 
Derivatives of $h$ are computed in 
Appendix \ref{app:h} where we have used the notation
\begin{subequations}
 \begin{align}
  R   & = y_i^{-\lambda} + q + (1-y_i)^{-\lambda}\,,\\
  R'  & = - \lambda \left(y_i^{-\lambda-1} - (1-y_i)^{-\lambda-1}\right)\,,\\ 
  R'' & = \lambda(\lambda+1) \left(y_i^{-\lambda-2} + (1-y_i)^{-\lambda-2}\right)\,.
 \end{align}
\end{subequations}

\subsubsection{The function $\beta$}
\label{app:beta}

In Eq. \eqref{app_eq:beta}, the function $\beta=\beta_0 (n_i y_i \sigma)^{2/3}$, 
where $\beta_0$ is a constant.
Its first order derivatives are 
\begin{subequations}
 \begin{align}
   \partial_{n_i}\beta & = \frac{2}{3}\frac{\beta}{n_i}\,,\quad
   \partial_{y_i}\beta  = \frac{2}{3}\beta\left[\frac{1}{y_i} 
   + \frac{\partial_{y_i}\sigma}{\sigma}\right]\,,\\
   \partial_T\beta & = \frac{2}{3}\beta\left[\frac{\partial_{T}\sigma}{\sigma}\right]\,,
  \end{align}
\end{subequations}
whereas the second order derivatives are
\begin{subequations}
 \begin{align}
  \partial_{n_in_i}\beta & = 
  \frac{(\partial_{n_i}\beta)^2}{\beta} - \frac{\partial_{n_i}\beta}{n_i}\,,\\
  \partial_{y_iy_i}\beta & = \frac{(\partial_{y_i}\beta)^2}{\beta} 
  + \frac{2}{3}\beta\bigg[-\frac{1}{y_i^2} \nonumber\\
  & \quad + \frac{\partial_{y_iy_i}\sigma}{\sigma} 
  - \frac{(\partial_{y_i}\sigma)^2}{\sigma^2}\bigg]\,,\\
   \partial_{n_iz}\beta & = \frac{(\partial_{n_i}\beta)(\partial_z\beta)}{\beta}\,,\\
   \partial_{Tz}\beta   & = \frac{(\partial_T\beta)(\partial_z\beta)}{\beta} + 
   \frac{2}{3}\beta\left[\frac{\partial_{Tz}\sigma}{\sigma} - 
   \frac{(\partial_T\sigma)(\partial_z\sigma)}{\sigma^2}\right]\,,
  \end{align}
\end{subequations}
where $z=y_i$ or $T$.

\subsubsection{The radius $r$ of heavy nuclei}
\label{app:r}

The nuclear radius defined in Eq. \eqref{app_eq:r} 
can be written as 
\begin{equation}\label{app_eq:r1} 
 r=\frac{9\sigma}{2\beta}\frac{Q}{P} \,,
\end{equation}
where $P$ and $Q$ are functions solely of the occupied 
volume fraction $u$ defined, together with their derivatives,
in Eqs. \eqref{app_eq:PQ}.
Thus, $r\equiv r(u,n_i,y_i,T)$ and its derivatives are 
\begin{subequations}
 \begin{align}
  \partial_u r & = r\left[\frac{Q'}{Q}-\frac{P'}{P}\right] \,,\quad
  \partial_{n_i} r  = r\left[-\frac{\partial_{n_i}\beta}{\beta}\right] \,,\\
  \partial_z r & = r\left[-\frac{\partial_z\beta}{\beta}                             + \frac{\partial_z\sigma}{\sigma}\right]\,.
 \end{align}
\end{subequations}
where $z=y_i$ or $T$, and
\begin{subequations}
 \begin{align}
  \partial_{uu} r & = \frac{(\partial_ur)^2}{r} \nonumber\\
  & \quad + r \left[\frac{Q''}{Q} - \frac{Q'^2}{Q^2} - 
  \frac{P''}{P} + \frac{P'^2}{P^2}\right] \,,\\
  \partial_{uw} r & = \frac{(\partial_ur)(\partial_wr)}{r} \,,\\ 
  \partial_{n_iw} r & = \frac{(\partial_{n_i} r)(\partial_{w} r)}{r} 
  + r\left[\frac{(\partial_{n_i}\beta)(\partial_{w}\beta)}{\beta^2}
  - \frac{\partial_{n_iw}\beta}{\beta}\right] \,,\\
  \partial_{zz'} r & = \frac{(\partial_{z} r)(\partial_{z'} r)}{r} 
  + r \bigg[\frac{(\partial_{z}\beta)(\partial_{z'}\beta)}{\beta^2} \nonumber\\
  & \quad - \frac{\partial_{zz'}\beta}{\beta^2} + \frac{\partial_{zz'}\sigma}{\sigma}
  - \frac{(\partial_{z}\sigma)(\partial_{z'}\sigma)}{\sigma^2}\bigg]\,,
 \end{align}
\end{subequations}
where $w=n_i$, $y_i$, or $T$, 
and $z$ and $z'$ are either $y_i$ or $T$.

\subsubsection{The mass number $A$ of heavy nuclei}
\label{app:A}

The mass number $\bar{A}$ of the representative
heavy nucleus in the single nucleus approximation (SNA) is 
\begin{equation}\label{app_eq:A}
 \bar{A}=\frac{4\pi n_i r^3}{3}\,.
\end{equation}
Thus, $\bar{A}\equiv\bar{A}(u,n_i,y_i,T)$ and its
first order derivatives are
\begin{subequations}
\begin{align}
 \partial_w\bar{A} & =4\pi n_i r^2\partial_w r\,,\\
 \partial_{n_i}\bar{A} & =4\pi n_i r^2\partial_{n_i} r + \frac{4\pi r^3}{3}\,.
\end{align}
\end{subequations}
for $w=u$, $y_i$ or $T$. 
The second order derivatives are
\begin{subequations}
 \begin{align}
  \partial_{ww'}\bar{A} & = \partial_w\bar{A}\left[
  \frac{2\partial_{w'}r}{r} + 
  \frac{\partial_{ww'}r}{\partial_w r}\right] \,,\\
  \partial_{n_iw}\bar{A} & = \partial_w\bar{A}
  \left[\frac{\partial_{n_iw}r}{\partial_wr} +
  \frac{2\partial_wr}{r} + \frac{1}{n_i}\right] \,,\\
  \partial_{n_in_i}\bar{A} & = 4\pi n_i r^2 \bigg[
  \partial_{n_in_i} r 
   + \frac{2(\partial_{n_i}r)^2}{r} + 
  \frac{2\partial_{n_i}r}{n_i} \bigg]\,,
 \end{align}
\end{subequations}
for $w$ and $w'$ one of $u$, $y_i$ or $T$.

\subsubsection{Surface and Coulomb free energies}
\label{app:sc_free}

The combined free energy of the surface and coulomb
terms is $F_{SC}=\beta\mathcal{D}(u)$.
The associated free energy derivatives are
%
% \begin{subequations}
 \begin{align}\label{app_eq:dFSC_dw}
  \partial_u F_{SC} &= \beta (\partial_u\mathcal{D})\,,\quad
  \partial_w F_{SC} = (\partial_w\beta) \mathcal{D}\,,
 \end{align}
% \end{subequations}
%
for $w=n_i$, $y_i$, or $T$. 
The second order derivatives are
\begin{subequations}\label{app_eq:d2FSC_dww}
 \begin{align}
  \partial_{uu} F_{SC}  &= \beta (\partial_{uu}\mathcal{D})\,,\quad
  \partial_{uw} F_{SC}  = (\partial_w\beta) (\partial_{u}\mathcal{D})\,,\\
  \partial_{ww'} F_{SC} &= (\partial_{ww'}\beta) \mathcal{D}\,,
 \end{align}
\end{subequations}
for $w$ and $w'$ one of $n_i$, $y_i$, or $T$.

\subsection{Contribution from translational motion}
\label{app:fH}

The translational free energy is $F_{TR}=uvn_if_{TR}$
where \cite{lattimer:91}
\begin{equation}
f_{TR} = \frac{h}{\bar{A}}\left[\mu_{TR}-T\right]\,.
\end{equation}
The function $h\equiv h(y_i,T)$ was defined in Eq. 
\eqref{eq:h} (see also Eq. \eqref{app_eq:h}), and 
$\bar{A}$, the mass number of the representative 
heavy nucleus in SNA, was defined in 
Eq. \eqref{app_eq:A} and
%
% \begin{equation}\label{app_eq:A}
%  \bar{A}=\frac{4\pi n_i r^3}{3}\,
% \end{equation}
%
% where $r$ is the generalized nuclear size discussed 
% in Appendix \ref{app:r} and
%
\begin{equation}\label{app_eq:mutr}
\mu_{TR}=T\ln\left(\frac{uvn_i}{n_Q \bar{A}^{5/2}}\right)\,,
\end{equation}
where $n_Q = (m_n T/ 2\pi^2\hbar^2)^{3/2}$. 
Recall that $v=(1-u)$. 

\subsubsection{The function $h$}
\label{app:h}

We now compute derivatives of auxiliary functions 
needed later. We start with the function $h(y_i,T)$
defined as 
\begin{equation}\label{app_eq:h}
 h\left(y_i,T\right)=
 \begin{dcases}
    g^p\,,& \mathrm{if\, } T\leq T_c(y_i)\,;\\
    0\,,              & \mathrm{otherwise}\quad,
\end{dcases}
\end{equation}
where $g(y_i,T)=[1-({T}/{T_c})^2]$,
$p$ is a parameter to be determined, and 
$T_c\equiv T_c(y_i)$ has the form
\begin{equation}\label{eq:Tc}
 T_c(y_i) = T_{c0}\left[a_c+b_c\delta^2
 +c_c\delta^4+d_c\delta^6\right]
\end{equation} 
with $\delta\equiv\delta(y_i)=1-2y_i$.

To compute derivatives of the auxiliary 
function $h$, we first determine the derivatives
of $T_c$:
\begin{subequations}
 \begin{align}
  \partial_{y_i} T_c & = 
  - 4T_{c0}\left[b_c\delta+2c_c\delta^3+3d_c\delta^5\right]\,,\\
  \partial_{y_iy_i} T_c & = 
  8T_{c0}\left[b_c+6c_c\delta^2+15d_c\delta^4\right]\,.
  \end{align}
\end{subequations}

Next, we compute the derivatives of $g$: 
\begin{subequations}
 \begin{align}
  \partial_{y_i} g      & = (\partial_{T_c}g)(\partial_{y_i}T_c) \,,\quad
  \partial_{T}   g       = -\frac{2T}{T_c^2}\,,\\
  \partial_{T  T}   g   & = -\frac{2}{T_c^2}\,,\quad
  \partial_{y_iT}   g    = 4\frac{T}{T_c^3}\partial_{y_i}T_c\,,\\
  \partial_{y_iy_i} g   & = (\partial_{T_cT_c}g)(\partial_{y_i}T_c)^2 + (\partial_{T_c}g)(\partial_{y_iy_i}T_c) \,,
  \end{align}
\end{subequations}
where
%
% \begin{subequations}
 \begin{align}
  \partial_{T_c} g    & = -2\frac{T^2}{T_c^3}\,,\quad
  \partial_{T_cT_c} g  = 6\frac{T^2}{T_c^4}\,.
 \end{align}
% \end{subequations}
%

The first order derivatives of $h$ then become 
%
% \begin{subequations}
 \begin{align}
  \partial_{T}   h & = g'\partial_{T}  g \,,\quad
  \partial_{y_i} h  = g'\partial_{y_i}g \,,
 \end{align}
% \end{subequations}
%
where $g'=pg^{p-1}$. 
The second order derivatives are 
\begin{subequations}
 \begin{align}
  \partial_{TT}     h & = 
  g'\partial_{TT} g + g''(\partial_{T} g)^2 \,,\\
  \partial_{y_iT}   h & = 
  g'\partial_{Ty_i}g + g''(\partial_{T} g)(\partial_{y_i}g) \,,\\
  \partial_{y_iy_i} h & = 
  g'\partial_{y_iy_i} g + g''(\partial_{y_iy_i} g)^2 \,,
 \end{align}
\end{subequations}
where $g''=p(p-1)g^{p-2}$.

\subsubsection{Translational chemical potential $\mu_{TR}$}
\label{app:mutr}

The translational chemical potential 
$\mu_{TR}\equiv\mu_{TR}(u,n_i,y_i,T)$ 
defined in Eq. \eqref{app_eq:mutr}
has the first order derivatives 
\begin{align}
 \partial_w \mu_{TR} & = - \frac{5T}{2\bar{A}}\partial_w\bar{A} + \mu_w \,,
\end{align}
where $w$ is one of $u$, $n_i$, $y_i$, or $T$, and 
\begin{align}\label{app_eq:mu_w}
 \mu_w = 
  \begin{dcases}
     T\left[\frac{1}{u}-\frac{1}{v}\right]\,,&\text{if $w=u$}\,;\\
     \frac{T}{n_i}\,,&\text{if $w=n_i$}\,;\\
     0\,,&\text{if $w=y_i$}\,;\\
     \frac{\mu_{TR}}{T} - \frac{3}{2}\,,&\text{if $w=T$}\,,
\end{dcases}
\end{align}
with $v=1-u$.

The second order derivatives are
\begin{align}
 \partial_{ww'} \mu_{TR} & =  \nu_{ww'} - \frac{5T}{2\bar{A}} \left[ \partial_{ww'}\bar{A}-\frac{(\partial_{w}\bar{A})(\partial_{w'}\bar{A})}{\bar{A}}
  \right] \,,
\end{align}
where $w$ and $w'$ are one of $u$, $n_i$, $y_i$, or $T$ and 
\begin{align}\label{app_eq:nu_ww'}
 \nu_{ww'} = 
  \begin{dcases}
     \partial_{w'}\mu_w \,,&\text{if $w'\neq T$}\,;\\
     \partial_{w'}\mu_w - \frac{5}{2\bar{A}}\partial_w\bar{A} \,,&\text{if $w'=T$}\,,
\end{dcases}
\end{align}
which are readily computed from Eqs. \eqref{app_eq:mu_w}.

%
% \begin{align}
%  \nu_{ww'} = 
%   \begin{dcases}
%      \partia_{w'}\mu_w
%      T\left[\frac{1}{u^2}+\frac{1}{v^2}\right]\,,&\text{if $w=w'=u$}\,;\\
%      -\frac{T}{n_i^2}\,,&\text{if $w=w'=n_i$}\,;\\
%      \frac{\mu_w}{T}\,,&\text{if $w\ne T$ and $w'=T$}\,;\\
%      \frac{\mu_{TR}}{T} - \frac{3}{2}\,,&\text{if $w=T$}\,;\\
%      0,\,,&\text{otherwise}\,.
% \end{dcases}
% \end{align}
% 

\subsubsection{Translational free energy}
\label{app:tr_obs}

In explicit form, the translational free energy is
\begin{equation}
 F_{TR}=\frac{uvn_i}{\bar{A}}h\left[\mu_{TR}-T\right]\,,
\end{equation}
where $v=1-u$ and, $\bar{A}$ and $\mu_{TR}$ are given 
in Eqs. \eqref{app_eq:A} and \eqref{app_eq:mutr},
respectively.

Its derivatives are
%
% \begin{subequations}
 \begin{align}\label{app_eq:dFTR_dw}
   \partial_w F_{TR} &= F_{TR}\left[\omega_w
   - \frac{\partial_w\bar{A}}{\bar{A}}
   + \frac{\partial_w(\mu_{TR}-T)}{\mu_{TR}-T}\right]\,,%\\
%   \partial_u F_{TR} &= F_{TR}\left[\frac{1}{u}-\frac{1}{v}
%    -\frac{\partial_u\bar{A}}{\bar{A}}
%    +\frac{\partial_u\mu_{TR}}{\mu_{TR}-T}\right]\,,\\
%   \partial_{n_i} F_{TR} &= F_{TR}\left[\frac{1}{n_i}
%   - \frac{\partial_{n_i}\bar{A}}{\bar{A}} 
%   + \frac{\partial_{n_i}\mu_{TR}}{\mu_{TR}-T}\right]\,,\\
%   \partial_{y_i} F_{TR} &= F_{TR}\left[\frac{\partial_{y_i}h}{h}
%   - \frac{\partial_{y_i}\bar{A}}{\bar{A}} 
%   + \frac{\partial_{y_i}\mu_{TR}}{\mu_{TR}-T}\right]\,,\\
%   \partial_T F_{TR} &= F_{TR}\left[\frac{\partial_T h}{h}
%   - \frac{\partial_T\bar{A}}{\bar{A}} 
%   + \frac{\partial_T\mu_{TR}-1}{\mu_{TR}-T}\right]\,,
 \end{align}
% \end{subequations}
%
for $w$ one of $u$, $n_i$, $y_i$, or $T$, and 
\begin{align}\label{app_eq:omega_w}
 \omega_w = 
  \begin{dcases}
     \left[\frac{1}{u}-\frac{1}{v}\right]\,,&\text{if $w=u$}\,;\\
     \frac{1}{n_i}\,,&\text{if $w=n_i$}\,;\\
     \frac{\partial_{y_i}h}{h}\,,&\text{if $w=y_i$}\,;\\
     \frac{\partial_T h}{h}\,,&\text{if $w=T$}\,,
\end{dcases}
\end{align}

The second order derivatives are
%
% \begin{subequations}
 \begin{align}\label{app_eq:d2FTR_dww}
  \partial_{ww'} F_{TR} &= 
  \frac{(\partial_{w} F_{TR})(\partial_{w'} F_{TR})}{F_{TR}} 
  + F_{TR}\Omega_{ww'}\,,
 \end{align}
% \end{subequations}
%
with
\begin{align}
 \Omega_{ww'} & = \partial_{w'}\omega_w 
  - \frac{\partial_{ww'}\bar{A}}{\bar{A}}
  + \frac{(\partial_w\bar{A})(\partial_{w'}\bar{A})}{\bar{A}^2}
  \nonumber\\ & \qquad 
  + \frac{\partial_{ww'}(\mu_{TR}-T)}{\mu_{TR}-T}
  \nonumber\\ & \qquad 
  - \frac{\partial_w(\mu_{TR}-T)\partial_{w'}(\mu_{TR}-T)}{(\mu_{TR}-T)^2}
\end{align}
for $w$ and $w'$ one of $u$, $n_i$, $y_i$, or $T$. 
The values of $\partial_{w'}\omega_w$ are readily computed
from Eqs. \eqref{app_eq:omega_w}.

\section{Uniform matter}
\label{app:uniform}

For uniform matter, the nuclear part of the 
free energy of the system is 
\begin{equation}\label{app_eq:Fu}
 F_u = F_o + F_\alpha\,,
\end{equation}
where the free energy of unbound nucleons is 
\begin{equation}\label{app_eq:fo}
 F_o = u_\alpha F_{\rbulk,o}\,,
\end{equation}
Above, $F_{\rbulk,o}=F_\rbulk(n_o,y_o,T)$, see Appendix \ref{app:apr},
where $n_o=n_{no}+n_{po}$ is the nucleon number density in the 
uniform phase while $n_{no}$ and $n_{po}$ are the neutron and 
proton number densities, respectively. 
The proton fraction of unbound nucleons is  $y_o=n_{po}/n_o$. 
The index $o$ refers to nucleons outside of heavy nuclei. 
The term $u_\alpha = (1-n_\alpha v_\alpha)$ 
represents the excluded volume fraction by alpha particles, 
which are treated as hard spheres with number density 
$n_\alpha$ and volume $v_\alpha$. 
As in LS, we set $v_\alpha=24\unit{fm}^3$.

The free energy of alpha particles is 
\begin{equation}\label{app_eq:fu}
 F_\alpha = n_\alpha f_\alpha \,
\end{equation}
where $f_\alpha = (\mu_\alpha-B_\alpha-T)$, 
with $\mu_\alpha$ and $B_\alpha$ the chemical potential 
and binding energy of alpha particles, respectively. 
The relationship between the chemical potential and 
number density of alpha particles has been defined in 
Eq. \eqref{app_eq:mualpha2} in Appendix \ref{app:alpha}.

%We set as independent variables in our system the par
%We minimize the free energy $F_N$ with respect to the 
%alpha particle number density $n_\alpha$. 

The conservation equations for baryon 
number and charge are
\begin{subequations}\label{app_eq:conservation_u}
 \begin{align}
 \label{app_eq:conservation_un}
  n & = 4n_\alpha + u_\alpha n_o  \\
 \label{app_eq:conservation_uny}
  ny & = 2n_\alpha + u_\alpha n_{po}\,. 
 \end{align}
\end{subequations}
Minimizing $F_N$ with respect to the 
alpha particle number density $n_\alpha$ yields the 
the chemical potential of alpha particles:
\begin{equation}\label{app_eq:mualpha1}
 \partial_{n_\alpha} F_u=0 \Leftrightarrow 
 \mu_\alpha = 2(\mu_{no}+\mu_{po}) + B_\alpha - P_{\rbulk,o} v_\alpha\,.
\end{equation}
%
%Here $\partial_{n_\alpha}$ is the derivative with respect to $n_\alpha$. 
%Eq. \eqref{app_eq:mualpha1} 
As expected, the alpha particle chemical potential 
depends on the chemical potentials of the protons and 
neutrons, $\mu_{no}$ and $\mu_{po}$, respectively, which 
are given in Appendix \ref{app:apr_mu}. 
The pressure $P_{\rbulk,o}$ due to nucleons outside of 
alpha particles is given in Eq. \eqref{app_eq:pbulk}.

\subsection{Solution of the uniform system}
\label{app:u_sol}

To solve the system of Eqs. \eqref{app_eq:conservation_u} and 
\eqref{app_eq:mualpha1},  we choose 
$X_p=n_{po}/n$ if $y\leq0.5$ and 
$X_n=n_{no}/n$ if $y>0.5$ as independent variables. 
 As in LS, these choices are used to eliminate $n_\alpha$ from 
Eqs. \eqref{app_eq:conservation_u} and yield 
\begin{subequations}
 \begin{align}
  n_{no}&=\frac{-X_p n(1-y)v_\alpha + 2(1-2y+X_p)}{2-nyv_\alpha}~n\\
  n_{po}&=X_pn,
 \end{align}
for $y\leq0.5$. In the case  $y>0.5$, 
 \begin{align}
  n_{po}&=\frac{-X_n ny v_\alpha - 2(1-2y-X_n)}{2-n(1-y)v_\alpha}~n\\
  n_{no}&=X_nn \,.
 \end{align}
\end{subequations}
Once an initial guess for $X_p$ or $X_n$ is obtained, the 
nucleon densities $n_{no}$ and $n_{po}$ as well as their
chemical potentials $\mu_{no}$ and $\mu_{po}$,  and the nucleon 
pressure $P_o$ are readily computed. 
The chemical potential of alpha particles $\mu_\alpha$ is then 
determined from Eq. \eqref{app_eq:mualpha1} and its density 
$n_\alpha$ from Eq. \eqref{app_eq:mualpha2}. 
These are then used to check if one of the equalities in Eq. 
\eqref{app_eq:conservation_u} is satisfied.  If not, an iterative 
procedure is employed to satisfy the conservation equations. 
We choose the equality in Eq. \eqref{app_eq:conservation_un} 
as it is more easily solved by the root finding routines 
\texttt{nleqslv} of Hasselman \cite{hasselman:16} in the 
$y\rightarrow0$ limit.

We note that in the limiting cases where alpha 
particles disappear, $n_\alpha\rightarrow0$, 
$y_o\rightarrow y$ which leads to 
$X_p\rightarrow y$ if $y\leq0.5$ and 
$X_n\rightarrow (1-y)$ if $y>0.5$.

\subsection{Change of variables}
\label{app:change_u}

Once a solution for the uniform system has been determined,
we use results of Appendix \ref{app:apr} and 
\ref{app:alpha} to compute derivatives of the chosen set of 
internal variables, here $\xi'=(\eta_{no},\eta_{po})$, with 
respect to the independent variables $\zeta=(n,y,T)$. 
To do this, we rely on the conservation equations Eqs.
\eqref{app_eq:conservation_u} rewritten as 
\begin{subequations}\label{app_eq:conservation_uA}
 \begin{align}
 \label{app_eq:conservation_unA}
  \mathcal{A}_1 & = n - 4n_\alpha - u_\alpha n_o = 0\\
 \label{app_eq:conservation_unyA}
  \mathcal{A}_2& = ny - 2n_\alpha - u_\alpha n_{po} = 0 \,. 
 \end{align}
\end{subequations}
where $u_\alpha = (1-n_\alpha v_\alpha)$.

Explicitly, we solve the systems
\begin{gather}\label{app_eq:M_u}
  \begin{bmatrix}
   \partial_{\eta_n} \mathcal{A}_1 & \partial_{\eta_p} \mathcal{A}_1 \\
   \partial_{\eta_n} \mathcal{A}_2 & \partial_{\eta_p} \mathcal{A}_2 
   \end{bmatrix}
  \begin{bmatrix}
   d_\zeta \eta_n \\
   d_\zeta \eta_p 
   \end{bmatrix} =
  \begin{bmatrix}
   d'_\zeta\mathcal{A}_1 \\
   d'_\zeta\mathcal{A}_2 
  \end{bmatrix}\,
\end{gather}
to compute $d_\zeta \eta_t$, for $\zeta=(n,y,T)$. 
The derivative notations are the same as in Appendix \ref{app:der}. 
This allows us to compute the derivatives of the thermodynamical
properties as shown below. 
For completeness we write the full expression appearing in Eqs. 
\eqref{app_eq:M_u} in Appendix \ref{app:B}. 
% using the LU decomposition code available with the 
% open-source LS code \cite{lattimer:91}, 

\subsection{Thermodynamics of uniform matter}
\label{app:thermo_u}

We write the free energy and entropy densities of uniform 
matter as
\begin{subequations}
 \begin{align} 
  \label{app_eq:Fu}
  F_u & = F_\alpha + u_\alpha F_{\rbulk,o} \,,\\
  \label{app_eq:Su}
  S_u & = S_\alpha + u_\alpha S_{\rbulk,o} \,,
 \end{align}
\end{subequations}
where $u_\alpha=(1-n_\alpha v_\alpha)$ is the 
volume fraction excluded by the alpha particles.

We then compute derivatives with respect to the 
independent variables $\zeta=(n,y,T)$: 
\begin{subequations}\label{app_eq:dFSu_dzeta}
 \begin{align}
  \label{app_eq:dFu_dzeta}
  d_\zeta F_u & = d_\zeta F_\alpha + u_\alpha (d_\zeta F_{\rbulk,o}) 
  - v_\alpha (d_\zeta n_\alpha) F_{\rbulk,o} \,,\\
  \label{app_eq:dSu_dzeta}
  d_\zeta S_u & = d_\zeta S_\alpha + u_\alpha (d_\zeta S_{\rbulk,o}) 
  - v_\alpha (d_\zeta n_\alpha) S_{\rbulk,o} \,.
 \end{align}
\end{subequations}
The derivatives of 
$\chi=(n_\alpha,F_\alpha,S_\alpha,F_{\rbulk,o},S_{\rbulk,o})$, 
with respect to $\zeta$ are obtained from Eqs. 
\eqref{app_eq:dF'dzeta}, \ie
\begin{subequations}\label{app_eq:dchi_dzeta}
 \begin{align}
  d_n \chi & = \sum_t (\partial_{\eta_t} \chi) (d_n \eta_t) \,,\\
  d_y \chi & = \sum_t (\partial_{\eta_t} \chi) (d_y \eta_t) \,,\\
  d_T \chi & = \partial'_T \chi + \sum_t 
   (\partial_{\eta_t} \chi) (d_T \eta_t) \,.
 \end{align}
\end{subequations}
Here, $d_\xi \eta_t$ are determined from the solutions of 
Eqs. \eqref{app_eq:M_u}. 
The $\partial_{\eta_t}\chi$ terms are computed from Eq. 
\eqref{app_eq:dchideta}, using Eq. \eqref{app_eq:detardnt} as 
well as Eqs. \eqref{app_eq:dfbulkdn}, \eqref{app_eq:dsbulkdn}, 
\eqref{app_eq:dna_dnt}, \eqref{app_eq:dadnt_F}, and 
\eqref{app_eq:dadnt_S} for $F_{\rbulk,o}$, $S_{\rbulk,o}$, 
$n_\alpha$, $F_\alpha$, and $S_\alpha$, respectively.

The $\partial'_T \chi$ terms are computed from Eqs. 
\eqref{app_eq:dFbulk_dTp}, and \eqref{app_eq:dSbulk_dTp} 
for $\chi = F_{\rbulk,o}$ and $S_{\rbulk,o}$, respectively. 
For alpha particle related quantities, $\partial'_T \chi$
is computed with help from Eq. \eqref{app_eq:dchipdT}, 
and Eqs. \eqref{app_eq:dna_dT}, \eqref{app_eq:dFa_DT}, 
\eqref{app_eq:dSa_DT}, respectively, for 
$\chi = n_\alpha,\,F_\alpha$, and $S_\alpha$. 
The terms in $\partial'_T n_t$ and $\partial_{n_t}\chi$ 
in Eq. \eqref{app_eq:dchipdT} are determined using 
Eq. \eqref{app_eq:dXdTeta_n} and 
Eqs. \eqref{app_eq:dna_dnt}, \eqref{app_eq:dadnt_F}, 
and \eqref{app_eq:dadnt_S} with $n_t\rightarrow n_{to}$. 
% Meanwhile, ...
%
% \begin{subequations}
% \label{app_eq:example}
%  \begin{align}
%   \label{app_eq:dFa_dzeta}
%   d_T F_a & = 
%   \partial'_T F_\alpha + \sum_t 
%   (\partial_{\eta_t} F_\alpha) (d_T \eta_t) \,.
%  \end{align}
% \end{subequations}
%
% Here, $\partial'_T F_\alpha$ can be computed from Eqs. 
% \eqref{app_eq:dadnt_F}, \eqref{app_eq:dFa_DT},  
% \eqref{app_eq:dchipdT}, and \eqref{app_eq:dXdTeta_n}.
% The derivative $\partial_{\eta_t} F_\alpha$ is obtained 
% from Eqs. \eqref{app_eq:dchideta}, \label{app_eq:detardnt} 
% and \eqref{app_eq:dadnt_F}, while $d_T \eta_t$ 
% is determined solving Eqs. \eqref{app_eq:M_u}. 
% The derivatives of the nucleon chemical potentials 
% $\mu_{to}$ with respect to the independent variables $\zeta$, 
% $d_\zeta \mu_{to}$, can also be computed from Eqs. 
% \eqref{app_eq:dF'dzeta}. 

The internal energy $E_u$ and its derivatives 
are directly obtained from the relation
\begin{equation}\label{app_eq:E_u}
 E_u = F_u + T S_u\,,
\end{equation}
which leads to 
\begin{subequations}
 \begin{align}\label{app_eq:dEu_dzeta}
  d_n E_u & = d_n F_u + T d_n S_u\,,\\
  d_y E_u & = d_y F_u + T d_y S_u\,,\\
  d_T E_u & = d_T F_u + T d_T S_u + S_u \,.
 \end{align}
\end{subequations}

The pressure ensues from the relation 
\begin{equation}\label{app_eq:P_u}
 P_u = n (d_n F_u) - F_u\,.
\end{equation}
Pressure derivatives are computed using the thermodynamical 
relations  Eqs. (B1) and (B2) of LS, \ie
\begin{subequations}
 \begin{align}\label{app_eq:dPu_dzeta}
  d_n P_u & = n d_{nn} F_u \,,\\
  d_y P_u & = n \left(\mu_{no} - \mu_{po} + d_{ny} F_u \right)\,,\\
  d_T P_u & = S_u + n d_{nT} F_u \,, 
 \end{align}
\end{subequations}
where
\begin{subequations}
 \begin{align}\label{app_eq:thermo}
  d_{TT} F_u & = - d_T S_u\,,\\
  d_{Tn} F_u & = (1-y) d_T \mu_{no} + y d_T \mu_{po} \,,\\
  d_{Ty} F_u & = -n (d_T\mu_{no} - d_T\mu_{po}) \,,\\
  d_{yy} F_u & = -n (d_y\mu_{no} - d_y\mu_{po})\,,\\
  d_{yn} F_u & = - (\mu_{no} - \mu_{po}) - n (d_n\mu_{no} - d_n\mu_{po})\,,\\
  d_{nn} F_u & = (1-y)d_n\mu_{no} + y d_n\mu_{po}\,.
 \end{align}
\end{subequations}

The nucleon chemical potential derivatives are readily 
obtained from the previously derived results, see
Eqs. \eqref{app_eq:dmut_dnr}, \eqref{app_eq:detardnt}, 
\eqref{app_eq:dchideta}, and \eqref{app_eq:M_u},  
and from 
\begin{subequations}
 \begin{align}\label{app_eq:dmuu_dzeta}
  d_n \mu_{to} & = \sum_t (\partial_{\eta_{to}}\mu_{to})(d_n \eta_{to}) \,,\\
  d_y \mu_{to} & = \sum_t (\partial_{\eta_{to}}\mu_{to})(d_n \eta_{to}) \,,\\
  d_T \mu_{to} & = \partial'_T \mu_{to} 
                 + \sum_t (\partial_{\eta_{to}}\mu_{to})(d_n \eta_{to}) \,.
 \end{align}
\end{subequations}

\section{Non-uniform matter}
\label{app:nonuniform}

In this case, the total free energy $F_{nu}$ of nucleons is 
\begin{equation}\label{app_eq:FNU}
 F_{nu} = F_o + F_\alpha + (F_i + F_{TR} + F_{SC}) \,,
\end{equation}
where the various terms are, respectively, the free energy 
of nucleons outside nuclei, of alpha particles, and of 
heavy nuclei. 
The free energy of heavy nuclei has contributions from 
nucleons inside heavy nuclei, $F_i$, as well as translational, 
$F_{TR}$, and the sum of surface and coulomb parts, $F_{SC}$. 
The terms in Eq. \eqref{app_eq:FNU} are given by 
\begin{subequations}\label{app_eq:FNUexp}
\begin{align}
 F_o&=v u_\alpha F_{\rbulk,o}\,,\quad
 F_\alpha=v n_\alpha f_\alpha\,,\\
 F_i&=u F_{\rbulk,i}\,,\quad
 F_{TR}=u v n_if_{TR}\,,\\
 F_{SC}&=\beta\mathcal{D}(u)\,.
\end{align} 
\end{subequations}
In Eqs. \eqref{app_eq:FNUexp}, $u$ is the volume fraction
occupied by heavy nuclei, $v=(1-u)$, $n_\alpha$ ($v_\alpha$) 
is the number density (volume) of alpha particles from Eq. 
\eqref{app_eq:mualpha2}. 
The terms $F_{\rbulk,o} = F_\rbulk(n_o,y_o,T)$, 
$F_{\rbulk,i} = F_\rbulk(n_i,y_i,T)$, and $f_\alpha$ are, 
respectively, the bulk free energy densities 
of nucleons outside and inside of heavy nuclei, discussed in 
Appendix \ref{app:bulk_apr}, and the free energy of alpha 
particles discussed in \ref{app:alpha}. 
Similar to how we defined $n_o$ and $y_o$ for uniform matter 
before,  $n_i=n_{pi}+n_{ni}$ ($y_i=n_{pi}/n_i$) refer to the 
density (proton fraction) of nucleons inside of heavy nuclei.

\subsection*{Solution of the non-uniform system}
\label{app:nu_sol}

Here we describe the procedure for minimizing  
the total free energy $F_{nu}$ of nucleons  with respect 
to appropriately chosen internal variables of the system. 
We choose the variables, 
$y_i$, $n_i$, $u$, $r$, $n_{no}$, $n_{po}$ and $n_\alpha$, 
which are constrained by 
the conservation equations of mass and charge 
\begin{subequations}\label{app_eq:conservation_nu} 
 \begin{align}
  n&=un_i+v[4n_\alpha+n_{o}u_\alpha]\,,\\
  ny&=un_iy_i+v[2n_\alpha+n_{po}u_\alpha]\,.
 \end{align}
 \end{subequations}

Two other constraints stem from minimizing 
$F_{nu}$ with respect to $r$ and $n_\alpha$ and lead to the
Eqs. \eqref{app_eq:r} and \eqref{app_eq:mualpha1}.
%
%\begin{align}
%  r & = \frac{9\sigma}{2\beta}\left[\frac{s(u)}{c(u)}\right]^{1/3}\,,\\
%  \mu_\alpha & = 2(\mu_{no} + \mu_{po}) + B_\alpha - P_ov_\alpha\,.
%\end{align}
%\end{subequations}

Thus, the system of equations to be solved is reduced to three 
equations obtained by computing the derivatives of $F_{nu}$ with 
respect to $n_i$, $y_i$, and $u$. 
The resulting equations can be rearranged to read as 
\begin{subequations}\label{app_eq:solve_nu}
 \begin{align}
   A_1 &= P_{\rbulk,i}-B_1-P_{\rbulk,o}-P_\alpha=0\,,\\
   A_2 &= \mu_{ni}-B_2-\mu_{no}=0\,,\\
   A_3 &= \mu_{pi}-B_3-\mu_{po}=0\,.
 \end{align}
\end{subequations}
%
%Eqs. \eqref{app_eq:solve_nu} dictate 
These equations establish the pressure and chemical 
equilibrium between nucleons inside heavy nuclei and in the 
uniform liquid of free nucleons and alpha particles surrounding 
heavy nuclei. 
Here, $P_{\rbulk,i}$, $P_{\rbulk,o}$, and $P_\alpha$ are the 
pressures of nucleons inside and outside heavy nuclei and of 
alpha particles, while $\mu_{ta}$ are the chemical potentials 
of neutrons, $t=n$, and protons, $t=p$, 
inside, $a=i$, and outside, $a=o$, heavy nuclei. 
% The form of the $B_i \equiv B_i(u,n_i,y_i,T)$ terms, 
% which are due to the translational, surface, and coulomb 
% contributions to the the total free energy
% of nucleons is discussed in Appendix \ref{app:B}.

The terms $B_i$ in Eq. \eqref{app_eq:solve_nu},
which determine the  equilibrium between heavy nuclei immersed 
in a uniform liquid of nucleons and alpha particles, are 
computed from the derivatives of $\hat{F} = F_{TR} + F_{SC}$. 
%which includes the translational and coulomb plus surface free energies. 
Explicitly, 
\begin{subequations}\label{app_eq:B}
 \begin{align}
 B_1 &=\partial_u\hat{F}-\frac{n_i}{u}\partial_{n_i} \hat{F}\,,\\
 B_2 &=\frac{1}{u}\left[\frac{y_i}{n_i}\partial_{y_i}\hat{F}
       -\partial_{n_i}\hat{F}\right]\,,\\
 B_3 &=-\frac{1}{u}\left[\frac{1-y_i}{n_i} \partial_{y_i}\hat{F}
       +\partial_{n_i}\hat{F}\right]\,,
 \end{align}
\end{subequations}
where $\partial_w$ is a partial derivative with respect to the
internal variable $w=u$, $n_i$, or $y_i$, keeping the other 
ones constant. 
Their forms were given in Eqs. \eqref{app_eq:dFSC_dw} and 
\eqref{app_eq:dFTR_dw}.

As in SRO, we solve Eqs. \eqref{app_eq:solve_nu} for the
three independent variables $\vartheta = (\log_{10} n_{no}, 
\log_{10} n_{po},\log_{10} u)$ using the root finding 
routines \texttt{nleqslv} of Hasselman \cite{hasselman:16}. 
In solving Eqs. \eqref{app_eq:solve_nu}, we find that 
numerical computations of the Jacobian matrix is, in most 
cases, as accurate as direct computations of 
$\partial_{\vartheta} A_j$.

Regardless of whether the Jacobian is computed numerically or 
semi-analytically, quite often an initial guess of $\vartheta$ 
does not result in a solution being found unless the root 
finding algorithm is implemented with quadruple precision. 
However, this choice renders the code extremely slow and, 
is thus impractical. 
Furthermore, matters become more complicated near the 
phase transition from uniform to non-uniform matter where 
it is unclear if a solution exists. 
Thus, we sometimes resort to computing the free 
energy of nucleons for millions of sets $\vartheta$. 
These are sorted to form a set of increasing total 
free energy and up to a thousand may be used as initial 
guesses to solve Eqs. \eqref{app_eq:solve_nu}. 
Once a solution is found, we check for unphysical situations
such as 
(1) negative number densities for any of the particles, 
(2) negative adiabatic index 
$\Gamma = d \ln P/d \ln n\vert_s$, 
(3) charge of the heavy nucleus is too small, usually $Z\leq6$, 
(4) the nucleon number density inside heavy nuclei is lower than 
that in the uniform nucleon liquid, and  
(5) unrealistic volume fraction occupied heavy nuclei , 
$u<0$ or $u>1$, etc.

Once a solution in the nonuniform case is deemed physical, 
its free energy is compared to that of the uniform system. 
The solution that has the lowest free energy is then taken 
as the true solution of the system. 
As in the uniform system case, once a solution $\vartheta$ 
for the the non-uniform is found we use it and its derivatives 
to improve initial guesses when moving to a nearby point 
$\zeta=(n,y,T)$ in the parameter space.

\subsection{Change of variables}
\label{app:change_nu}

Once a solution for the non-uniform system has been determined,
we use results of Appendices \ref{app:apr},  \ref{app:alpha}, 
and \ref{app:heavy} to compute derivatives of the chosen set of 
internal variables, here $\xi=(u,n_i,y_i,\eta_{no},\eta_{po})$, 
with respect to the independent variables $\zeta=(n,y,T)$. 
To do this, we rely on Eqs.\eqref{app_eq:conservation_nu} and 
\eqref{app_eq:solve_nu} rewritten as 
\begin{subequations}\label{app_eq:conservation_nuA}
 \begin{align}
 \label{app_eq:conservation_nunA}
  \mathcal{B}_1 & = n - un_i - v\left[4n_\alpha - n_ou_\alpha\right] = 0\,,\\
 \label{app_eq:conservation_nunyA}
  \mathcal{B}_2 & = ny - un_iy_i - v\left[2n_\alpha - n_{po}u_\alpha\right] = 0 \,,\\
  \mathcal{B}_3 & = \mu_{pi}-B_3-\mu_{po}=0\,,\\
  \mathcal{B}_4 & = \mu_{ni}-B_2-\mu_{no}=0\,,\\
  \mathcal{B}_5 & = P_i-B_1-P_o-P_\alpha=0\,,
 \end{align}
\end{subequations}
Then, using the LU decomposition code available with the 
open-source LS code \cite{lattimer:91}, we solve the systems
\begin{equation}\label{app_eq:M_nu}
 (\partial_\xi \mathcal{B})(d_\zeta \xi) = d_\zeta \mathcal{B}
\end{equation}
where 
\begin{gather}\label{app_eq:dmB_dxi}
\partial_\xi \mathcal{B} = 
  \begin{bmatrix}
   \partial_u \mathcal{B}_1 & 
   \partial_{n_i} \mathcal{B}_1 &
   \partial_{y_i} \mathcal{B}_1 &   
   \partial_{\eta_{no}} \mathcal{B}_1 & 
   \partial_{\eta_{po}} \mathcal{B}_1 \\
   \partial_u \mathcal{B}_2 & 
   \partial_{n_i} \mathcal{B}_2 &
   \partial_{y_i} \mathcal{B}_2 &   
   \partial_{\eta_{no}} \mathcal{B}_2 & 
   \partial_{\eta_{po}} \mathcal{B}_2 \\
   \partial_u \mathcal{B}_3 & 
   \partial_{n_i} \mathcal{B}_3 &
   \partial_{y_i} \mathcal{B}_3 &   
   \partial_{\eta_{no}} \mathcal{B}_3 & 
   \partial_{\eta_{po}} \mathcal{B}_3 \\
   \partial_u \mathcal{B}_4 & 
   \partial_{n_i} \mathcal{B}_4 &
   \partial_{y_i} \mathcal{B}_4 &   
   \partial_{\eta_{no}} \mathcal{B}_4 & 
   \partial_{\eta_{po}} \mathcal{B}_4 \\
   \partial_u \mathcal{B}_5 & 
   \partial_{n_i} \mathcal{B}_5 &
   \partial_{y_i} \mathcal{B}_5 &   
   \partial_{\eta_{no}} \mathcal{B}_5 & 
   \partial_{\eta_{po}} \mathcal{B}_5   
   \end{bmatrix}\,,
\end{gather}
\begin{gather}\label{app_eq:dxi_dzeta}
d_\zeta \xi = 
  \begin{bmatrix}
   d_\zeta u \\
   d_\zeta n_i \\
   d_\zeta y_i \\
   d_\zeta \eta_{no} \\
   d_\zeta \eta_{po} 
   \end{bmatrix} 
\quad \text{and} \quad d_\zeta \mathcal{B} = 
  \begin{bmatrix}
   d'_\zeta\mathcal{B}_1 \\
   d'_\zeta\mathcal{B}_2 \\
   d'_\zeta\mathcal{B}_3 \\
   d'_\zeta\mathcal{B}_4 \\
   d'_\zeta\mathcal{B}_5 
  \end{bmatrix}\,,
\end{gather}
%
%for $d_\zeta \xi$, 
for $\zeta=(n,y,T)$. 
The resulting  expressions are given explicitly in 
Appendix \ref{app:B}. 
%It is sufficient for now that 
The solutions to Eqs. \eqref{app_eq:M_nu}
allow us to compute  derivatives of the 
thermodynamical properties as shown below.

\subsection{Thermodynamics of non-uniform matter}
\label{app:thermo_u}

We write the free energy and entropy densities of 
non-uniform matter as
%
% \begin{subequations}
 \begin{align} 
  F_{nu} & = F_h + vF_u \,,\quad
  S_{nu}  = S_h + vS_u \,,
 \end{align}
% \end{subequations}
% 
where $F_u$ and $S_u$ are as in Eqs. \eqref{app_eq:Fu} and 
\eqref{app_eq:Su}, respectively, and 
\begin{subequations}
 \begin{align}
 F_h & = u F_{\rbulk,i} + F_{TR} + F_{SC}\,,\\
 S_h & = u S_{\rbulk,i} + S_{TR} + S_{SC}\,,
\end{align}
\end{subequations}
with $F_{\rbulk,i} \equiv F_\rbulk (n_i,y_i,T)$ defined in Eq. 
\eqref{app_eq:fbulk} and $S_{\rbulk,i} \equiv S_\rbulk (n_i,y_i,T)$
in Eq. \eqref{app_eq:sbulk}.

The derivatives of $F_{nu}$ and $S_{nu}$ with respect to 
the independent variables $\zeta=(n,y,T)$ are
\begin{subequations}\label{app_eq:dFSnu_dzeta}
 \begin{align}
  \label{app_eq:dFnu_dzeta}
  d_\zeta F_{nu} & = d_\zeta F_h + v (d_\zeta F_u) - (d_\zeta u) F_u \,,\\
  \label{app_eq:dSnu_dzeta}
  d_\zeta S_{nu} & = d_\zeta S_h + v (d_\zeta F_u) - (d_\zeta u) F_u \,.
 \end{align}
\end{subequations}
The derivatives $d_\zeta F_u$ and $d_\zeta S_u$ were
computed in Eqs. \eqref{app_eq:dFSu_dzeta}, while 
$d_\zeta u$ are obtained from solving the system of 
Eqs. \eqref{app_eq:M_nu}. 
We are left with evaluating $d_\zeta F_h$ and $d_\zeta S_h$. 
These are readily computed from the results obtained in 
Appendices \ref{app:apr} and \ref{app:heavy}: 
\begin{subequations}\label{app_eq:dFSh_dzeta}
 \begin{align}
  \label{app_eq:dFh_dzeta}
  d_\zeta F_h 
  & = (d_\zeta u) F_{\rbulk,i} + u (d_\zeta F_{\rbulk,i}) \nonumber\\
  & \qquad + d_\zeta F_{SC} + d_\zeta F_{TR} \,,\\
  \label{app_eq:dSh_dzeta}
  d_\zeta S_h
  & = (d_\zeta u) S_{\rbulk,i} + u (d_\zeta S_{\rbulk,i}) \nonumber\\
  & \qquad + d_\zeta S_{SC} + d_\zeta S_{TR} \,,
 \end{align}
\end{subequations}
where $\zeta=(n,y,T)$. 
The derivative terms in the right hand side of 
Eqs. \eqref{app_eq:dFSh_dzeta} are given by
\begin{subequations}\label{app_eq:dFb_dzeta}
 \begin{align}
  \label{app_eq:dFbulk_dT} 
  d_T F_{\rbulk,i} & = \partial_T F_{\rbulk,i} 
  + (\partial_{n_i} F_{\rbulk,i})(d_T n_i) \nonumber\\
  & \quad + (\partial_{y_i} F_{\rbulk,i})(d_T y_i) \,,\\
   \label{app_eq:dFbulk_dn} 
  d_n F_{\rbulk,i} & =  
  (\partial_{n_i} F_{\rbulk,i})(d_n n_i) \nonumber\\
  & \quad + (\partial_{y_i} F_{\rbulk,i})(d_n y_i) \,,\\
   \label{app_eq:dFbulk_dy} 
  d_y F_{\rbulk,i} & =  
  (\partial_{n_i} F_{\rbulk,i})(d_y n_i) \nonumber\\
  & \quad + (\partial_{y_i} F_{\rbulk,i})(d_y y_i) \,,
 \end{align}
\end{subequations}
and similarly for $S_{\rbulk,i}$ by replacing $F \rightarrow S$. 
The derivatives $\partial_T F_{\rbulk,i}$ and 
$\partial_T S_{\rbulk,i}$ were computed in Eqs. 
\eqref{app_eq:dFbulk_dTp} and \eqref{app_eq:dSbulk_dTp}, 
respectively, and $d_\zeta n_i$ and $d_\zeta y_i$ were 
obtained from solving Eqs. \eqref{app_eq:M_nu}. 
The other derivatives are
\begin{subequations}\label{app_eq:dnidyi}
 \begin{align}
 \partial_{n_i} F_{\rbulk,i} & = (1-y_i) (\partial_{n_{ni}} F_{\rbulk,i})
 + y_i (\partial_{n_{pi}} F_{\rbulk,i})\,,\\
 \partial_{y_i} F_{\rbulk,i} & = n_i\left(\partial_{n_{pi}} F_{\rbulk,i}
 -\partial_{n_{ni}} F_{\rbulk,i}\right)\,
 \end{align}
\end{subequations}
and similarly so for $S_{\rbulk,i}$ 
by replacing $F \rightarrow S$.

The other derivatives to be computed in Eqs. 
\eqref{app_eq:dFSh_dzeta} involve  the translational, 
surface and Coulomb contributions.
The needed derivatives of the free energies  are
\begin{subequations}\label{app_eq:dFH_dzeta}
 \begin{align}
  d_T F_H & = \partial_T F_H + (\partial_{n_i} F_H)(d_T n_i) \nonumber\\
  & \quad + (\partial_{y_i} F_H)(d_T y_i) + (\partial_u F_H)(d_T u) \,,\\
  d_n F_H & = (\partial_{n_i} F_H)(d_T n_i) \nonumber\\
  & \quad + (\partial_{y_i} F_H)(d_n y_i) + (\partial_u F_H)(d_n u) \,,\\
  d_y F_H & = (\partial_{n_i} F_H)(d_y n_i) \nonumber\\
  & \quad + (\partial_{y_i} F_H)(d_y y_i) + (\partial_u F_H)(d_y u) \,,
 \end{align}
\end{subequations}
where $F_H$ may be either $F_{SC}$ or $F_{TR}$. 
Again, the terms $d_\zeta u$, $d_\zeta n_i$, and $d_\zeta y_i$
are computed by solving Eqs. \eqref{app_eq:M_nu}.
The derivatives $\partial_{w} F_{SC}$, for 
$w=u$, $n_i$, $y_i$, and $T$, were computed in 
Eq. \eqref{app_eq:dFSC_dw} and $\partial_{w} F_{TR}$ 
in Eqs. \eqref{app_eq:dFTR_dw}.

The entropy for translational and surface plus coulomb 
terms are computed from $S_H=-\partial_T F_H$ and,  
their derivatives in Eq. \eqref{app_eq:dFH_dzeta} are
%
% \begin{subequations}
 \begin{align}\label{app_eq:dSH_dw}
  \partial_w S_H & = - \partial_{Tw} F_H \,,
 \end{align}
% \end{subequations}
%
where if 
$S_H$ ($F_H$) is either $S_{SC}$ ($F_{SC}$) or 
$S_{TR}$ ($F_{TR}$). The second order derivatives
$\partial_{Tw} F_H$ were computed in Eqs. \eqref{app_eq:d2FSC_dww}
and \eqref{app_eq:d2FTR_dww} for $F_{SC}$ and $F_{TR}$, 
respectively.

From the free energy and entropy, the internal energy 
$E_{nu}$ for non-uniform matter and its derivatives are
\begin{equation}\label{app_eq:E_nu}
 E_{nu} = F_{nu} + T S_{nu} \,,
\end{equation}
and
\begin{subequations}
 \begin{align}\label{app_eq:dEnu_dzeta}
  d_n E_{nu} & = d_n F_{nu} + T d_n S_{nu}\,,\\
  d_y E_{nu} & = d_y F_{nu} + T d_y S_{nu}\,,\\
  d_T E_{nu} & = d_T F_{nu} + T d_T S_{nu} + S_{nu} \,.
 \end{align} 
\end{subequations}

The pressure follows from the relation 
\begin{equation}\label{app_eq:P_nu}
 P_u = n (d_n F_{nu}) - F_{nu}\,.
\end{equation}
Derivatives of pressure  are computed using thermodynamical 
relations found in Eqs. (B1) and (B2) of LS, \ie
\begin{subequations} 
 \begin{align}\label{app_eq:dPnu_dzeta}
  d_n P_{nu} & = n d_{nn} F_{nu} \,,\\
  d_y P_{nu} & = n \left(\mu_{no} - \mu_{po} + d_{ny} F_{nu} \right)\,,\\
  d_T P_{nu} & = S + n d_{nT} F_{nu} \,, 
 \end{align}
\end{subequations}
where
\begin{subequations} 
 \begin{align}\label{app_eq:thermonu}
  d_{TT} F_{nu} & = - d_T S_{nu}\,,\\
  d_{Tn} F_{nu} & = (1-y) d_T \mu_{no} + y d_T \mu_{po} \,,\\
  d_{Ty} F_{nu} & = -n (d_T\mu_{no} - d_T\mu_{po}) \,,\\
  d_{yy} F_{nu} & = -n (d_y\mu_{no} - d_y\mu_{po})\,,\\
  d_{yn} F_{nu} & = - (\mu_{no} - \mu_{po}) - n (d_n\mu_{no} - d_n\mu_{po})\,,\\
  d_{nn} F_{nu} & = (1-y)d_n\mu_{no} + y d_n\mu_{po}\,.
 \end{align}
\end{subequations}

Derivatives of the chemical potential are readily 
obtained from the previously derived results in 
Eqs. \eqref{app_eq:dchideta} and \eqref{app_eq:M_u} 
and from 
\begin{subequations} 
 \begin{align}\label{app_eq:dmunu_dzeta}
  d_n \mu_{to} & = \sum_t (\partial_{\eta_{to}}\mu_{to})(d_n \eta_{to}) \,,\\
  d_y \mu_{to} & = \sum_t (\partial_{\eta_{to}}\mu_{to})(d_n \eta_{to}) \,,\\
  d_T \mu_{to} & = \partial'_T \mu_{to} 
                 + \sum_t (\partial_{\eta_{to}}\mu_{to})(d_n \eta_{to}) \,.
 \end{align}
\end{subequations}

Note that Eqs. \eqref{app_eq:E_nu} through 
\eqref{app_eq:dmunu_dzeta} are simply Eqs. 
\eqref{app_eq:E_u} through \eqref{app_eq:dmuu_dzeta} 
with $u \rightarrow nu$.

\section{Transformations of variables}
\label{app:B}

We now show explicitly the terms in equations solved 
to change from internal variables to independent variables. 
We start with the matrices for the uniform system shown 
in Eqs. \eqref{app_eq:conservation_uA} and \eqref{app_eq:M_u}.
First, we compute the derivatives of $\mathcal{A}$ with 
respect to the independent variables, keeping 
the other independent variables as well as the internal 
variables fixed:  
\begin{subequations}
 \begin{align}
 d'_n \mathcal{A}_1 & = 1 \,, \quad
 d'_n \mathcal{A}_2  = y \,,\\
 d'_y \mathcal{A}_1  & = 0 \,,\quad
 d'_y \mathcal{A}_2  = n \,,\\
 d'_T \mathcal{A}_1 & = (v_\alpha n_o    - 4)\partial'_T n_\alpha 
 - u_\alpha \partial'_T n_o \,,\\
 d'_T \mathcal{A}_2 & = (v_\alpha n_{po} - 2)\partial'_T n_\alpha 
 - u_\alpha \partial'_T n_{po}\,,
 \end{align}
\end{subequations}
where $n_o=n_{no}+n_{po}$. The derivatives 
$\partial'_T n_{to}$ and $\partial'_T n_\alpha$
were computed in Eq. \eqref{app_eq:dXdTeta_n}
and in Eq. \eqref{app_eq:dchipdT} with help from results 
of Appendix \ref{app:alpha}, respectively. 
Derivatives with respect to the 
independent variables are
\begin{subequations}
 \begin{align}
  \partial_{\eta_{no}} \mathcal{A}_1 
  & = (v_\alpha n_o - 4)    \partial_{\eta_{no}} n_\alpha 
  - u_\alpha \partial_{\eta_{no}}n_o\,,\\
  \partial_{\eta_{po}} \mathcal{A}_1 
  & = (v_\alpha n_o - 4)    \partial_{\eta_{po}} n_\alpha 
  - u_\alpha \partial_{\eta_{po}}n_o\,,\\
  \partial_{\eta_{no}} \mathcal{A}_2 
  & = (v_\alpha n_{po} - 2) \partial_{\eta_{no}} n_\alpha 
  - u_\alpha \partial_{\eta_{no}}n_{po}\,,\\
  \partial_{\eta_{po}} \mathcal{A}_2 
  & = (v_\alpha n_{po} - 2) \partial_{\eta_{po}} n_\alpha 
  - u_\alpha \partial_{\eta_{po}}n_{po}\,.
 \end{align}
\end{subequations}
where the derivatives $\partial_{\eta_{to}} n_{ro}$ and 
$\partial_{\eta_{to}} n_\alpha$ were computed in Eq. \eqref{app_eq:detadn} 
and in Eq. \eqref{app_eq:dchideta} with help from results of 
Appendix \ref{app:alpha}, respectively.

In Appendix \ref{app:change_nu}, we showed the system 
of equations to be solved to compute the derivatives of the internal
variables $u$, $n_i$, $y_i$, $\eta_{no}$, and $\eta_{po}$
with respect to the independent variables $n$, $y$, and $T$. 
Derivatives with respect to the density $n$ required are 
\begin{subequations}
 \begin{align}
  d'_n \mathcal{B}_1 & = 1 \,,\quad
  d'_n \mathcal{B}_2  = y \,,\\
  d'_n \mathcal{B}_3 & = 0 \,,\quad
  d'_n \mathcal{B}_4  = 0 \,,\quad
  d'_n \mathcal{B}_5  = 0 \,, 
 \end{align}
\end{subequations}
whereas those with respect to the proton fraction $y$ are
\begin{subequations}
 \begin{align}
  d'_y \mathcal{B}_1 & = 0 \,,\quad
  d'_y \mathcal{B}_2  = n \,,\\
  d'_y \mathcal{B}_3 & = 0 \,,\quad
  d'_y \mathcal{B}_4  = 0 \,,\quad
  d'_y \mathcal{B}_5  = 0 \,.
 \end{align}
\end{subequations}
Derivatives with respect to the temperature $T$ are  
\begin{subequations}
 \begin{align}
  d'_T \mathcal{B}_1 & = v\left[(v_\alpha n_o-4) \partial'_T n_\alpha 
  - u_\alpha \partial'_T n_o\right] \,,\\
  d'_T \mathcal{B}_2 & = v\left[(v_\alpha n_{po}-2) \partial'_T n_\alpha 
  - u_\alpha \partial'_T n_{po}\right] \,,\\
  d'_T \mathcal{B}_3 & = \partial_T\mu_{pi} - \partial_T B_3 - \partial'_T \mu_{po} \,,\\
  d'_T \mathcal{B}_4 & = \partial_T\mu_{ni} - \partial_T B_2 - \partial'_T \mu_{no} \,,\\
  d'_T \mathcal{B}_5 & = \partial_T P_{\rbulk,i} - \partial_T B_1 
                       - \partial'_T P_{\rbulk,o} - \partial'_T P_\alpha \,.
 \end{align}
\end{subequations}
Above, the temperature derivatives $\partial_T\chi$ for 
$\chi=P_{\rbulk,i}$ and $\mu_{ti}$ were computed in Eqs. 
\eqref{app_eq:dPbulkdT} and \eqref{app_eq:dXdT_mu}, respectively.
Primed derivatives $\partial'_T\chi$ for 
$\chi=n_{to}$, $\mu_{to}$, and $P_{\rbulk,o}$ were computed in 
Eqs. \eqref{app_eq:dXdTeta_n}, \eqref{app_eq:dXdTeta_mu}, and 
\eqref{app_eq:dPbulkdTeta}.
The $\partial'_T P_\alpha$ and $\partial'_T n_\alpha$ terms are 
computed using Eq. \eqref{app_eq:dchipdT}, results in 
Eqs. \eqref{app_eq:dXdTeta_n} and Appendix \ref{app:alpha}.
The temperature derivatives of the $B_i$ terms are
\begin{subequations}
 \begin{align}
  \partial_{T}B_1 & = \partial_{uT}\hat{F} - \frac{n_i}{u}\partial_{n_iT}\hat{F}\,,\\
  \partial_{T}B_2 & = \frac{1}{u}
  \left[\frac{y_i}{n_i}\partial_{y_iT}\hat{F}-\partial_{n_iT}\hat{F}\right]\,,\\
  \partial_{T}B_3 & = - \frac{1}{u}
  \left[\frac{1-y_i}{n_i}\partial_{y_iT}\hat{F}+\partial_{n_iT}\hat{F}\right]\,.
 \end{align}
\end{subequations}

Now we record the derivatives of $\mathcal{B}$ 
with respect to the internal variables. 
We start with  derivatives with respect to 
the volume fraction occupied by heavy nuclei $u$: 
\begin{subequations}\label{app_eq:dB_dz}
 \begin{align}
  \partial_u \mathcal{B}_1 & = - n_i + (4n_\alpha + u_\alpha n_o) \,,\\
  \partial_u \mathcal{B}_2 & = - n_iy_i + (2n_\alpha + u_\alpha n_{po})  \,,\\
  \partial_u \mathcal{B}_3 & = - \partial_u B_3 \,,\quad
  \partial_u \mathcal{B}_4  = - \partial_u B_2 \,,\quad
  \partial_u \mathcal{B}_5  = - \partial_u B_1 \,.
 \end{align}
\end{subequations}
Derivatives 
with respect to the number density inside heavy nuclei $n_i$ are
\begin{subequations}
 \begin{align}
 \partial_{n_i} \mathcal{B}_1 & = -u \,,\quad
 \partial_{n_i} \mathcal{B}_2  = -uy_i \,,\\
 \partial_{n_i} \mathcal{B}_3 & = - \partial_{n_i}B_3 + \partial_{n_i}\mu_{pi} \,,\\
 \partial_{n_i} \mathcal{B}_4 & = - \partial_{n_i}B_2 + \partial_{n_i}\mu_{ni} \,,\\
 \partial_{n_i} \mathcal{B}_5 & = - \partial_{n_i}B_1 + \partial_{n_i}P_{\rbulk,i} \,,
 \end{align}
\end{subequations}
and those with respect to the proton fraction inside heavy nuclei $y_i$ are
\begin{subequations}
 \begin{align}
  \partial_{y_i} \mathcal{B}_1 & = 0 \,,\quad
  \partial_{y_i} \mathcal{B}_2  = -un_i \,,\\
  \partial_{y_i} \mathcal{B}_3 & = - \partial_{y_i}B_3 + \partial_{y_i}\mu_{pi} \,,\\
  \partial_{y_i} \mathcal{B}_4 & = - \partial_{y_i}B_2 + \partial_{y_i}\mu_{ni} \,,\\
  \partial_{y_i} \mathcal{B}_5 & = - \partial_{y_i}B_1 + \partial_{y_i}P_{\rbulk,i} \,,
 \end{align}
\end{subequations}
Derivatives of the degeneracy parameter of unbound neutrons $\eta_{no}$ are
\begin{subequations}
 \begin{align}
   \partial_{\eta_{no}} \mathcal{B}_1 & = v u_\alpha \partial_{\eta_{no}} n_o
    - v(v_\alpha n_o - 4)\partial_{\eta_{no}}n_\alpha \,,\\
   \partial_{\eta_{no}} \mathcal{B}_2 & = v u_\alpha \partial_{\eta_{no}} n_{po}
    - v(v_\alpha n_{po} - 2)\partial_{\eta_{no}}n_\alpha \,,\\
   \partial_{\eta_{no}} \mathcal{B}_3 & = 
   - \partial_{\eta_{no}}\mu_{po}  \,,\quad
   \partial_{\eta_{no}} \mathcal{B}_4  = 
   - \partial_{\eta_{no}}\mu_{no}  \,,\\
   \partial_{\eta_{no}} \mathcal{B}_5 & = 
   - \partial_{\eta_{no}}P_{\rbulk,o} - \partial_{\eta_{no}}P_\alpha \,,
 \end{align}
\end{subequations}
and those with respect to the degeneracy parameter of unbound protons $\eta_{po}$ are
\begin{subequations}
 \begin{align}
   \partial_{\eta_{po}} \mathcal{B}_1 & = v u_\alpha \partial_{\eta_{po}} n_o
    - v(v_\alpha n_o - 4)\partial_{\eta_{po}}n_\alpha \,,\\
   \partial_{\eta_{po}} \mathcal{B}_2 & = v u_\alpha \partial_{\eta_{po}} n_{po}
    - v(v_\alpha n_{po} - 2)\partial_{\eta_{po}}n_\alpha \,,\\
   \partial_{\eta_{po}} \mathcal{B}_3 & = 
    - \partial_{\eta_{po}}\mu_{po}  \,,\quad
   \partial_{\eta_{po}} \mathcal{B}_4  = 
   - \partial_{\eta_{po}}\mu_{no}  \,,\\
   \partial_{\eta_{po}} \mathcal{B}_5 & = 
   - \partial_{\eta_{po}}P_{\rbulk,o} - \partial_{\eta_{po}}P_\alpha \,.
 \end{align}
\end{subequations}

Equations \eqref{app_eq:dB_dz} make use of the results
\begin{align}\label{app_eq:dFdny}
 \partial_{n_i}F & = (1-y_i)\left(\partial_{n_{ni}}F + \partial_{n_{pi}}F\right)\,,\\
 \partial_{y_i}F & = n_i\left(\partial_{n_{pi}}F-\partial_{n_{ni}}F \right)
\end{align}
for $F=P_{\rbulk,i}$ and $\mu_{ti}$.
The derivatives with respect to $\eta_{to}$ are computed from 
\begin{equation}\label{app_eq:dFdeta}
 \partial_{\eta_{to}}F = 
 (\partial_{\eta_{to}}n_{no})(\partial_{n_{no}}F) + 
 (\partial_{\eta_{to}}n_{po})(\partial_{n_{po}}F) \,,
\end{equation}
where $\partial_{n_{ro}}F$, for $F=n_{to}$, $\mu_{to}$, 
$P_{\rbulk,o}$, $n_\alpha$, $P_\alpha$, and were computed 
in Eqs. \eqref{app_eq:detadn}, \eqref{app_eq:dmut_dnr}, 
\eqref{app_eq:dpbulkdn}, \eqref{app_eq:dna_dnt}, 
and \eqref{app_eq:dadnt_F}, respectively.

Next, we turn to derivatives of the functions $B$ defined in 
Eq. \eqref{app_eq:B} with respect to the internal variables. 
We begin with derivatives with respect to $u$: 
\begin{subequations}\label{app_eq:dBdu}
 \begin{align}
  \partial_{u}B_1 & = \partial_{uu}\hat{F} - \frac{n_i}{u}\partial_{n_iu}\hat{F}
  + \frac{n_i}{u^2}\partial_{n_i}\hat{F}\,,\\
  \partial_{u}B_2 & = -\frac{B_2}{u} + \frac{1}{u}
  \left[\frac{y_i}{n_i}\partial_{y_iu}\hat{F}-\partial_{n_iu}\hat{F}\right]\,,\\
  \partial_{u}B_3 & = -\frac{B_3}{u} - \frac{1}{u}
  \left[\frac{1-y_i}{n_i}\partial_{y_iu}\hat{F}+\partial_{n_iu}\hat{F}\right]\,,
 \end{align}
\end{subequations}
Derivatives with respect to $n_i$ are
\begin{subequations}\label{app_eq:dBdni}
 \begin{align}
  \partial_{n_i}B_1 & = \partial_{un_i}\hat{F} - \frac{n_i}{u}\partial_{n_in_i}\hat{F}
  - \frac{1}{u}\partial_{n_i}\hat{F}\,,\\
  \partial_{n_i}B_2 & = 
  \frac{y_i}{un_i}\partial_{y_in_i}\hat{F}
  -\frac{y_i}{un_i^2}\partial_{y_i}\hat{F}
  -\frac{\partial_{n_in_i}\hat{F}}{u}\,,\\
  \partial_{n_i}B_3 & = -\frac{1-y_i}{un_i}\left(\partial_{y_in_i}\hat{F} 
  - \frac{\partial_{y_i}\hat{F}}{n_i}\right) 
  - \frac{\partial_{n_in_i}\hat{F}}{u}\,,
 \end{align}
\end{subequations}
and with respect to $y_i$ are
\begin{subequations}\label{app_eq:dBdyi}
 \begin{align}
  \partial_{y_i}B_1 & = \partial_{uy_i}\hat{F} 
  - \frac{n_i}{u}\partial_{n_iy_i}\hat{F}\,,\\
  \partial_{y_i}B_2 & = \frac{y_i}{un_i}\partial_{y_iy_i}\hat{F} 
  +\frac{1}{un_i}\partial_{y_i}\hat{F}
  -\frac{\partial_{n_iy_i}\hat{F}}{u}\,,\\
  \partial_{y_i}B_3 & = -\frac{1-y_i}{un_i}\left(\partial_{y_iy_i}\hat{F} 
  - \frac{\partial_{y_i}\hat{F}}{1-y_i}\right) 
  - \frac{\partial_{n_iy_i}\hat{F}}{u}\,.
 \end{align}
\end{subequations}
Recall that $\hat{F}=F_{TR}+F_{SC}$ and that the second 
derivatives of $F_{SC}$ and $F_{TR}$ were computed in Eqs. 
\eqref{app_eq:d2FSC_dww} and \eqref{app_eq:d2FTR_dww}, 
respectively.

For completeness, we write elements of the Jacobian of the system of equations
being solved, \ie Eqs. \eqref{app_eq:solve_nu}. As in LS, we write 
the system as 
\begin{equation}
 A_k = A_{ki}(x_i,n_i)-B_k(x_i,n_i,u)-A_{ko}(n_{no},n_{po})\,,
\end{equation}
where $A_{ko}=(P_{\rbulk,o}+P_\alpha,\mu_{no},\mu_{po})$,  
$A_{ki}=(P_{\rbulk,i},\mu_{ni},\mu_{pi})$ and $B_i$ 
are as in Eq. \eqref{app_eq:B}. 
Since we are solving for 
$\vartheta = \log_{10}\theta$ where $\theta = (n_{no}, n_{po}, u)$, 
\begin{equation}
 d_{\vartheta_{i}} A 
 = \left.\frac{dA}{d\vartheta_i}\right|_{\vartheta_j,\vartheta_k} 
 = \left.\frac{dA}{d\theta_i}\right|_{\theta_j,\theta_k} \ln(10)\theta_i 
\end{equation}
where $i$, $j$, and $k$ denote permutations of the elements of 
$\vartheta$ and $\theta$.
Thus, the elements of the Jacobian matrix may be computed from 
the relations 
\begin{subequations}
 \begin{align}
  d_\theta A_{ki} & = \partial_{n_i} A_{ki} \partial_\theta n_i 
                    + \partial_{y_i} A_{ki} \partial_\theta y_i\,,\\
  d_\theta A_{ko} & = \partial_{\theta} A_{ko} \,,\\
  d_\theta B_{k}  & = \partial_{\theta} B_k 
                    + \partial_{n_i} B_k \partial_\theta n_i 
                    + \partial_{y_i} B_k \partial_\theta y_i\,.
 \end{align}
\end{subequations}
The elements $\partial_{n_i} A_{ki}$ and $\partial_{y_i} A_{ki}$
were computed in Eqs. \eqref{app_eq:dFdny}. 
Elements $\partial_{\theta} A_{ko}$ were computed in Eqs. 
\eqref{app_eq:dpbulkdn} and \eqref{app_eq:dmut_dnr}. 
Note that, $\partial_u A_{ko}=0$. 
The terms $\partial_{\xi} B_k$, for $\xi=(u,n_i,y_i)$ were 
determined in Eqs. \eqref{app_eq:dBdu}, \eqref{app_eq:dBdni}, 
and \eqref{app_eq:dBdyi}, while $\partial_{n_t} B_k=0$.
Finally, the derivatives $\partial_\theta \xi$ are determined
from 
\begin{subequations}
 \begin{align}
  \partial_u n_i & = -\frac{n_1}{u} \,,\quad
  \partial_u y_i  = -\frac{n_2-y_in_1}{un_i} \,,\\
  \partial_{n_{to}} n_i & = - \frac{v}{u} 
  \left[u_\alpha + (4-n_ov_\alpha)\partial_{n_{to}}n_\alpha\right]\,,\\
  \partial_{n_{to}} y_i & = -\frac{y_i}{n_i}\partial_{n_{to}} n_i - \frac{v}{un_i} M_t \,,
 \end{align}
\end{subequations}
with $n_1 = (n_i - u_\alpha n_o - 4n_\alpha)$, 
$n_2 = (y_in_i - u_\alpha n_{po} - 2n_\alpha)$, and 
$M_t=(2-n_{po}v_\alpha)\partial_{n_t}n_\alpha + \delta_{tp} u_\alpha$.
As before, we have used the notation $v=1-u$, $u_\alpha=(1-n_\alpha v_\alpha)$. 
The derivatives $\partial_{n_t}n_\alpha$ are computed from 
Eq. \eqref{app_eq:dna_dnt}.

\bibliography{apreos}

\end{document}